\documentclass[]{pasj01}
\usepackage{lineno}
\begin{document} 
\Received{2021/12/17}
\Accepted{2022/05/30}
%\Published{yyyy/mm/dd}

\title{Decades' long-term variations in NS-LMXBs observed with MAXI/GSC, RXTE/ASM and Ginga/ASM}

%%% begin:list of authors
% Do NOT capitalize all letters in "textsc".

\author{Kazumi \textsc{Asai}, Tatehiro \textsc{Mihara}, and Masaru \textsc{Matsuoka}}%
\altaffiltext{}{MAXI team, RIKEN, 2-1 Hirosawa, Wako, Saitama 351-0198, Japan }
\email{kazumi@crab.riken.jp}


\maketitle

\begin{abstract}
We investigated the decades' long-term X-ray variations in bright low-mass X-ray binaries
containing a neutron star (NS-LMXB).
The light curves of
MAXI/GSC and RXTE/ASM
covers $\sim$ \textcolor{black}{26}~yr, and high-quality X-ray light curves are obtained from 33 NS-LMXBs.
Among them, together with Ginga/ASM, two sources
(GX~3$+$1 and GX~9$+$1) showed an apparent sinusoidal variation with the period of $\sim 5$~yr and $\sim 10$~yr
in the \textcolor{black}{34}~yr light curve.
Their X-ray luminosities were
$(1-4)\times10^{37}$ erg s$^{-1}$
in the middle of the luminosity distribution of the NS-LMXB.
Other seven sources 
(Ser~X-1, 4U~1735--444, GX~9$+$9, 4U~1746$-$37, 4U~1708$-$40, 4U~1822$-$000, and 1A~1246$-$588)
have also similar sinusoidal variation,
although the profiles (amplitude, period, and phase) are variable.
Compering the 21 sources with known orbital periods,
a possible cause of 
the long-term sinusoidal variation might be 
the mass transfer cycles induced by the irradiation to the donor star.
\end{abstract}

%\linenumbers

\section{Introduction}
%LMXB intro
Many of the bright X-ray sources are low mass X-ray binaries 
with a weakly magnetized neutron star 
(NS-LMXB: see \cite{Barret2001} for a review).
Based on the temporal activity, 
NS-LMXBs are divided into two types: persistent type and transient type.
Furthermore, many persistent NS-LMXBs and bright phases of transient NS-LMXBs
are divided into two groups: Z sources and Atoll sources, 
based on their behavior on the color--color
diagram and the hardness--intensity diagram \citep{Hasinger1989}.
Z sources are very bright, and the luminosities sometimes become
close to the Eddington luminosity ($L_{\rm E}$).
On the other hand, Atoll sources are generally less bright
(\,$\ltsim$ 0.5 $L_{\rm E}$).

%LMXB orbital period and long-term period
The orbital periods of NS-LMXBs range 
from minutes to $\sim 20$ d \citep{Liu2007}.
Moreover, long-term variations on the time scale beyond the orbital period are known.
These super-orbital periods range in tens to hundreds of days,
which are thought to be related to the properties of the accretion disk, such as
radiation-induced warping and precession (see \cite{Charles2008} for a review).
On the other hand, several NS-LMXBs display
very long-term quasi-periodic modulations
(approximately several to tens of years).
The variations are thought to have a different origin.
Kotze and Charles (2010) (hereafter KC10)
suggest that 
the long-term variations are due to the variation
of the mass-transfer rate from the donor,
which is a consequence of solar-like magnetic cycles
(\cite{Applegate1987}, \cite{Warner1988}).
Solar-like cycles of $\sim$ 10~yr
are observed from many late type stars
\citep{Baliunas1995}.

%irradiation
It is pointed out the importance of irradiation in NS-LMXBs 
for the outburst properties and their long-term evolution \citep{Ritter2008}.
The former is relevant to irradiating the accretion disk.
Irradiation during an outburst leads to drastic changes
in the outburst properties because 
the irradiation changes the conditions
for the occurrence of disk instabilities.
The latter is relevant to the irradiating donor star.
The irradiation of the donor star can destabilize mass transfer
and lead to irradiation-driven mass transfer cycles,
i.e., to a secular evolution.
However, it is not clear to 
estimate the effect of irradiation
for the secular evolution because of several unclear factors.

%GX3+1
GX~3$+$1 shows the long-term variation
on the time scale of years
superimposed with the short-term variations
on the time scale of hours
\citep{Seifina2012}.
The short-term variations are due to the transitions
between branches in terms of its color--color diagram,
which are independent of the long-term variation.
The spectral index is constant during the long-term variations.

In this paper,
we pay attention to the sinusoidal variation and
report the analysis of the long-term variations
during 1996--\textcolor{black}{2021} for 41 NS-LMXBs.
In section 2, we present the details of X-ray light curve analysis.
We show the results in section 3. 
We discuss the cause of the sinusoidal variation in section 4.

\section{Observations and data analysis}

\renewcommand{\arraystretch}{0.75}
\begin{table*}
  \tbl{List of NS-LMXBs observed with MAXI/GSC and RXTE/ASM.}
  {
  \begin{tabular}{lccccccc}
      \hline
        Name & Type\footnotemark[$*$]
        & \textcolor{black}{L$_{\rm ave}$ \footnotemark[$\dagger$]}
& Distance & P$_{\rm orb}$ \footnotemark[$\ddag$]  &
\textcolor{black}{ASM/GSC\footnotemark[$\S$]}  &
Comment \footnotemark[$\|$] &
Reference\footnotemark[$\sharp$] \\
 & & (10$^{36}$ erg s$^{-1}$) & (kpc) & (hr) & & & \\
      \hline
Sco~X-1       & Z          & 222 &  2.8 & 18.90 & 0.95 & NP  &  (1) \\ 
GX~17$+$2     & Z          & 209 & 12.6 & -- & 0.97 & NP  & (2) \\
GX~5$-$1      & Z          & 173 &    9 & -- & 0.94 & NP  & (3) \\  
Cyg~X-2       & Z          & 129 &   11 & 235.2 & 0.97 & NP  & (4) \\
GX~349$+$2    & Z          & 126 &   5  & 22.5 & 0.88 & NP  & (3) \\    
LMC~X-2       & Z          & 99 &   50 & 8.16 & 1.02 & NP  & (5) \\     
GX~340$+$0    & Z          & 96 &   11 & -- & 0.98 & NP  & (3) \\   
GX~13$+$1     & Z, A        & 35  &    7 & 577.6 & 0.92 & NP & (1) \\        
4U~1820$-$303 & A, UCXB     & 34  &  7.6 & 0.19 & 1.02 & FV  & (1) \\       
Ser~X-1       & A          & 33  &  8.4 & 2 & 0.96 & MP & (1) \\        
GX~9$+$1      & A          & 29  &  5.0 & -- & 0.96 & CP & (1) \\         
4U~1705$-$440 & A          & 21  &  7.4 & -- & 1.00 &  FV  & (1) \\ 
4U~1735$-$444 & A          & 22  &  8.5 & 4.65 & 1.00 & MP & (4) \\         
4U~1624$-$490 & ADC        & 21  &   15 & 20.89 & 1.06 &  NP  & (7) \\      
SAX~J1747.0$-$2853 & T     & 20  &    9 & -- & --   & --  & (8)\\       
GX~9$+$9      & A          & 12  &  5.0 & 4.20 & 0.98 & NP  & (3) \\         
4U~1254$-$690 & A          & 11  &   13 & 3.93 & 0.99 & NP  & (10) \\          
Cir~X-1       & Z, A        & 12  &  7.8 & 398.4 & 1.04 & LV  & (6) \\    
GX~3$+$1      & A          & 10 &  4.5 & -- & 1.14 & CP  & (9) \\         
GS~1826$-$238 & T          & 8.4   &  7.0 & 2.088 & 1.41 & LV  & (11) \\           
4U~1746$-$37  & A          & 6.8  & 11.0 & 5.16 & 0.76 & MP & (1) \\   
4U~1708$-$40  & -          & 5.3   &  8   & -- & 0.83 & MP  & (12) \\            
4U~1724$-$307 & -          & 4.1   &  7.4 & -- & --   & --  & (1) \\       
4U~1543$-$624 & UCXB       & 3.9   &  9.2 & 0.303 & 1.02 & NP & (1) \\ 
4U~1636$-$536 & A          & 3.8   &  6.0 & 3.80 & 0.99 & LV  & (1) \\      
Aql~X-1       & T          & 2.7   & 5.0  & 18.95 & --   & --  & (1) \\ 
4U~2127$+$119 (M15 X-2) & UCXB       & 2.4   & 10.3 & 0.376 & 0.90 & NP  & (13) \\              
4U~0513$-$40  & UCXB       & 2.3   & 12   & -- & 1.03 &  NP  & (14) \\  
EXO~1745$-$248& T          & 2.0   & 5.9  & -- & --   & --  & (15) \\ 
4U~1608$-$522 & T          & 1.9   & 4.1  & 12.89 & --   & --  & (1) \\ 
4U~1822$-$000 & -       & 1.8   & 6.3  & 3.2 & 0.92 & MP & (16)\\ 
XTE~J1709$-$267 & T        & 1.4   & 8.5  & -- & --   & --  & (17) \\ 
4U~1916$-$053 & UCXB       & 1.2 & 8.9  & 0.83 & 1.40 & NP  & (4)\\    
4U~1745$-$203 & T          & 1.1 & 8.5  & -- & --   & --  & (15)\\ 
4U~0614$+$091 & UCXB       & 1.1 & 3.2  & -- & 0.94 & NP & (18)\\  
SLX~1735$-$269& -          & 1.0 & 7.3  & -- & 2.64 &  NP  & (4)\\
HETE~J1900.1$-$2455 & T    & 0.9 & 5    & 1.39 & -- & --  & (4)\\      
1H~0918$-$548 & UCXB       & 0.5 & 4.8  & -- & 0.94 & NP  & (19)\\  
1A~1246$-$588 & UCXB       & 0.4 &  5   & -- & 0.83 &  MP  & (20)\\                 
4U~1323$-$619 & -          & 0.3 & 4.2  & 2.93 & 0.64 &  NP  & (21) \\ 
1H~1556$-$605 & -          & 0.3 & 4    & 9.1 & 1.58 & NP & (3) \\            
\hline
\end{tabular}
}
\label{tab1}
\begin{tabnote}
\footnotemark[$*$] The source types are indicated by 'Z'-Z source, 'A'-Atoll source, 'ADC'-ADC source, 'UCXB'-Ultra Compact X-ray Binary, 'T'-transient.\\   
\footnotemark[$\dag$] Luminosity in the 2--10~keV band in the unit of $10^{36}$~erg~s$^{-1}$ 
from MJD=55100 to 
\textcolor{black}{
MJD=59662.} \\ 
\footnotemark[$\ddag$] Orbital periods of systems are adopted from \citet{Liu2007}
except fot Ser~X-1 \citep{Cornelisse2013}.\\ 
\footnotemark[$\S$] ASM/GSC for
MJD=55100--
\textcolor{black}{55460}.\\ 
%\footnotemark[$\|$] Type of variation 
%(see text for explanation).\\ 
\footnotemark[$\|$] Type of variation:
%%%%old
%'LP' show a possible long-term period beyond the observation period, 
%%
'CP' show a clear periodic variation, 
'MP' show a modified periodic variation,
'NP' show no periodic variation,
'FV' show fast variability,
'LV' show large variability,
'--' are Transient and Contamination sources.\\
\footnotemark[$\sharp$] Reference of distance. 
(1) \citet{Liu2007},
(2) \citet{Lin2012},
(3) \citet{Christian1997},
(4) \citet{Galloway2008},
(5) \citet{Freedman2001}
(6) \citet{DAi2012},
(7) \citet{Xiang2007},
(8) \citet{Natalucci2000}
(9) \citet{Kuulkers2000},
(10) \citet{Zand2003},
(11) \citet{Barret2000},
(12) \citet{Revnivtsev2011},
(13) \citet{White2001},
(14) \citet{Harris1996},
(15) \citet{Valenti2007},
(16) \citet{Shahbaz2007},
(17) \citet{Ludlam2017},
(18) \citet{Kuulkers2010},
(19) \citet{Jonker2004},
(20) \citet{Jonker2007},
(21) \citet{Gambino2016}.\\
%\footnotemark[$\sharp$]  ... \\  
%\footnotemark[$**$]  ... \\ 
%\footnotemark[$\dag\dag$]  ... \\ 
\end{tabnote}
\end{table*}
\renewcommand{\arraystretch}{1} 

The long-term X-ray activity is continuously
monitored with MAXI
(Monitor of All-sky X-ray Image: \cite{Matsuoka2009})
since 2009 August.
We obtained long-term one-day bin light curves of
MAXI/GSC (Gas Slit Camera: \cite{Mihara2011}; \cite{Sugizaki2011})
\footnote{$<$http://maxi.riken.jp/pubdata/v7l/$>$.} for 41 NS-LMXBs
from 2009 August to \textcolor{black}{2021} December.
We also analyzed the data of the same sources observed 
with the ASM (All Sky Monitor: \cite{Levine1996}) onboard 
RXTE (Rossi X-ray Timing Explorer: \cite{Bradt1993}) 
in 2--10~keV from 1996 February to 2011 December.
The ASM data are obtained from the archived results provided by the RXTE-ASM teams at MIT and NASA/GSFC. 
\footnote{$<$http://xte.mit.edu/$>$.}

The obtained count rates of GSC and ASM were 
converted to luminosities by assuming a Crab-like spectrum
\citep{Kirsch2005} and the distance listed in table~\ref{tab1}.
ASM data are converted to a flux in Crab unit using the nominal
relation of 1~Crab = 75~counts s$^{-1}$ for ASM.
\footnote{$<$http://xte.mit.edu/XTE/ASM\_lc.html$>$.}
GSC data are converted to a flux in Crab unit using the nominal
relation of 1~Crab = 
\textcolor{black}{3.45}~photons s$^{-1}$ cm$^{-2}$
in 2--10~keV for GSC.
\footnote{$<$http://maxi.riken.jp/top/readme.html$>$.}
The assumption of a Crab-like spectrum is acceptable
in the hard state (low luminosity $\leq 5\times10^{36}$ erg s$^{-1}$)
because the energy spectrum is approximated
by a power law with a photon index of 1--2.
On the other hand, in the soft state (high luminosity $\geq 10^{37}$ erg s$^{-1}$), 
the energy spectrum is dominated by the thermal emission, and the luminosity obtained
by assuming Crab-like spectrum is underestimated in the 2--10~keV band \citep{Asai2015}.
In this paper, we do not care about it
because we handle only the relative difference.

We excluded following data in each source.

\begin{description}

\item[GX~3$+$1:]
Data of MJD = 56187--56287 and 56652--56997
for contamination by Swift~J174510.8$-$262411.
Data of MJD = 55916--55922
for another south-west source.

\item[SLX~1735$-$269:]
Data of MJD = 56187--56287 and 56652--56997
for contamination by Swift~J174510.8$-$262411.

\item[4U~1624$-$490:]
Data of MJD = 58460--58580
for contamination by MAXI~J1631$-$479.

\item[4U~1708$-$40:]
Data of the count rate above 1 photons cm$^{-2}$ s$^{-1}$ by solar X-ray leakage. 

\item[4U~1916$-$053:]
Data of MJD = 55945--55951
and MJD = 56311--56315 by solar X-ray leakage.

\item[4U~1323$-$619:]
Data of MJD = 58509--58610 for contamination by MAXI~J1348$-$630.

\item[Ser~X-1:]
Data with the errors above 0.012 Crab in 2--10~keV.

\end{description}

% Source selection
We analyzed ASM and GSC data of the 41 NS-LMXBs
and investigated the long-term variability
during \textcolor{black}{$\sim 26$ yr} (see table~1).
The X-ray light curves are shown in the Appendix.
Here we used GSC data for the overlapped period
during MJD=55100--55800.
Table~1 shows the flux ratio of ASM and GSC during the overlapped period.

We estimated the luminosity in the 2--10~keV energy band
from MJD=55100 to MJD=\textcolor{black}{59662} for the 41 NS-LMXBs 
using GSC data and listed in $L_{\rm ave}$.
%41-8=33
High-quality persistent light curves are obtained from 33 sources.
The remaining eight sources are excluded from the following analyses.
%6
Six of excluded eight sources are transients sources.
The active period of HETE~J1900.1$-$2455
is too short ($ \leq $10~yr) to investigate the long-term variability.
Other sources
(EXO~1745$-$248, Aql~X-1, XTE~J1709-267, 4U~1608$-$522, and 4U~1745$-$203)
are below the GSC detection limit during quiescence.
%2
The other excluded two sources, SAX~J1747.0$-$2853 and 4U~1724$-$307, have contamination from nearby sources.

%\citet{Kotze2010}
%Kotze and Charles (2010) 
KC10 indicated that all the 20 sources in their paper 
were considered to be better fitted
with a single sine wave than with a constant value.
We also focus on the properties of the sinusoidal variation.
Here, we tried to classify observed variations
into 
five types as follows. 
The types are described in the comment column of table~1. 

\begin{description}
\item[CP (clear periodic variation):]
Two bright Atoll sources (GX~3$+$1 and GX~9$+$1)
show clear sinusoidal variations.
We define CP sources as can be fitted with a single sinusoidal curve and a tilted line.
We focus on these sources in the next session.

\item[MP (modified periodic variation):]
Seven sources 
(Ser~X-1, 4U~1735$-$444, GX~9$+$9, 4U~1746$-$37,
4U~1708$-$40, 4U~1822$-$000, and 1A~1246$-$588) 
show the modified periodic variation.
It is difficult to fit their light curves with 
the periodic model functions.
We also focus on these sources in the next session.

\item[NP (no periodic variation):]
Nineteen sources show no periodic variation.
Eight sources (Z sources and GX~13$+$1) of them 
show almost constant baseline although
there are small variations around baseline.
Other eight sources
(X-ray luminosities $<$ $\sim 3\times10^{37}$ erg s$^{-1}$)
of them show almost constant.
The remaining three sources 
(4U~1543$-$624, 4U1916$-$053, and 1H~1556$-$605)
show a decreasing trend in
luminosity.

\item[FV (fast variation):]
Two sources (4U1820$-$303 and 4U1705$-$440) have
the luminosity change
with a shorter variability than 1~yr.
The variation seems to have a different origin 
from that of the sinusoidal variation
investigated in this paper.

\item[LV (large variation):]
Three sources (Cir~X-1, 4U~1636$-$536, and GS~1826$-$238)
show the large luminosity change of 1--2 orders of magnitude.
Again, the large luminosity change seems to
have a different origin from that of
the sinusoidal variation investigated in this paper.
\end{description}

\begin{figure}
\includegraphics[width=88mm]{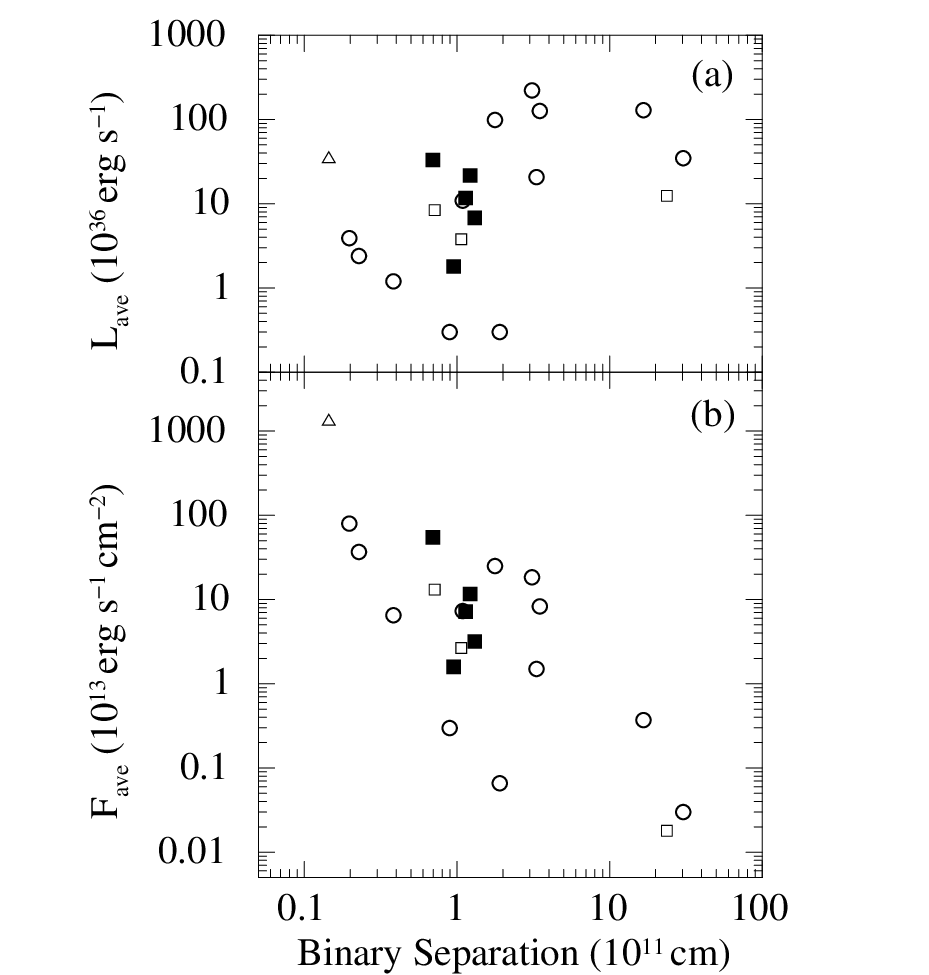}
\caption{(a) Average luminosity in the 2--10~keV
against binary separation.
(b) Average irradiating flux on the donor star.
We assumed the donor star mass is 0.5 $M_\odot$.
Filled squares MP (five sources),
open circles NP,
open triangles FV,
and open squares LV.
Two CP sources (GX~3$+$1 and GX~9$+$1)
are not plotted since the orbital period is not known. 
}
\label{fig1}
\end{figure}

Figure~\ref{fig1}a
shows the average luminosity against
the binary separation
for 21 sources.
The  orbital periods of 21 sources are known among 33 sources.
The binary separation was estimated by Kepler's third law
assuming a neutron star mass of 1.4~$\Mo$ and 
the mass of the donor star of 0.5~$\Mo$
although their actual masses are uncertain.
We also display the 
types in the figure.
Especially, filled marks indicate periodic variation,
which are relevant to the
long-term variation investigated in this paper.

Figure~\ref{fig1}b
shows the irradiating flux on the donor star
against the binary separation.
Here we treated the donor star as a point source
and calculated simply irradiation flux
$F=L/4 \pi d^{2}$,
where $L$ and $d$ denote the average luminosity and
the binary separation, respectively.

Next, we pick up CP and MP sources
with periodic and modified periodic variation.
To confirm the property of variation,
we also analyzed the archive data of
the X-ray All Sky Monitor (ASM: \cite{Tsunemi1989}) 
on board the Ginga satellite \citep{Makino1987}.
The Ginga/ASM data from 1987 to 1991
are obtained from the archived site of DARTS.
\footnote{
$<$https://data.darts.isas.jaxa.jp/pub/ginga/GINGA-ASM-1.2/$>$.
}
The 1--6~keV counts s$^{-1}$ are converted to the 2--10~keV flux
assuming the spectrum of Crab nebula.
For 1A1246$-$588 (MP), there were no data of Ginga/ASM.  

\section{Results}
Figure~2 shows the light curves of
two
CP sources 
(GX~3$+$1 and GX~9$+$1)
with the apparent sinusoidal variation from 1987 to
\textcolor{black}{2021.}
First, we fitted the light curves of GX~3$+$1 and GX~9$+$1
with a sinusoidal and linear function model.
The fitting parameters are shown in table~2.

\begin{figure*}
  \includegraphics[width=85mm]{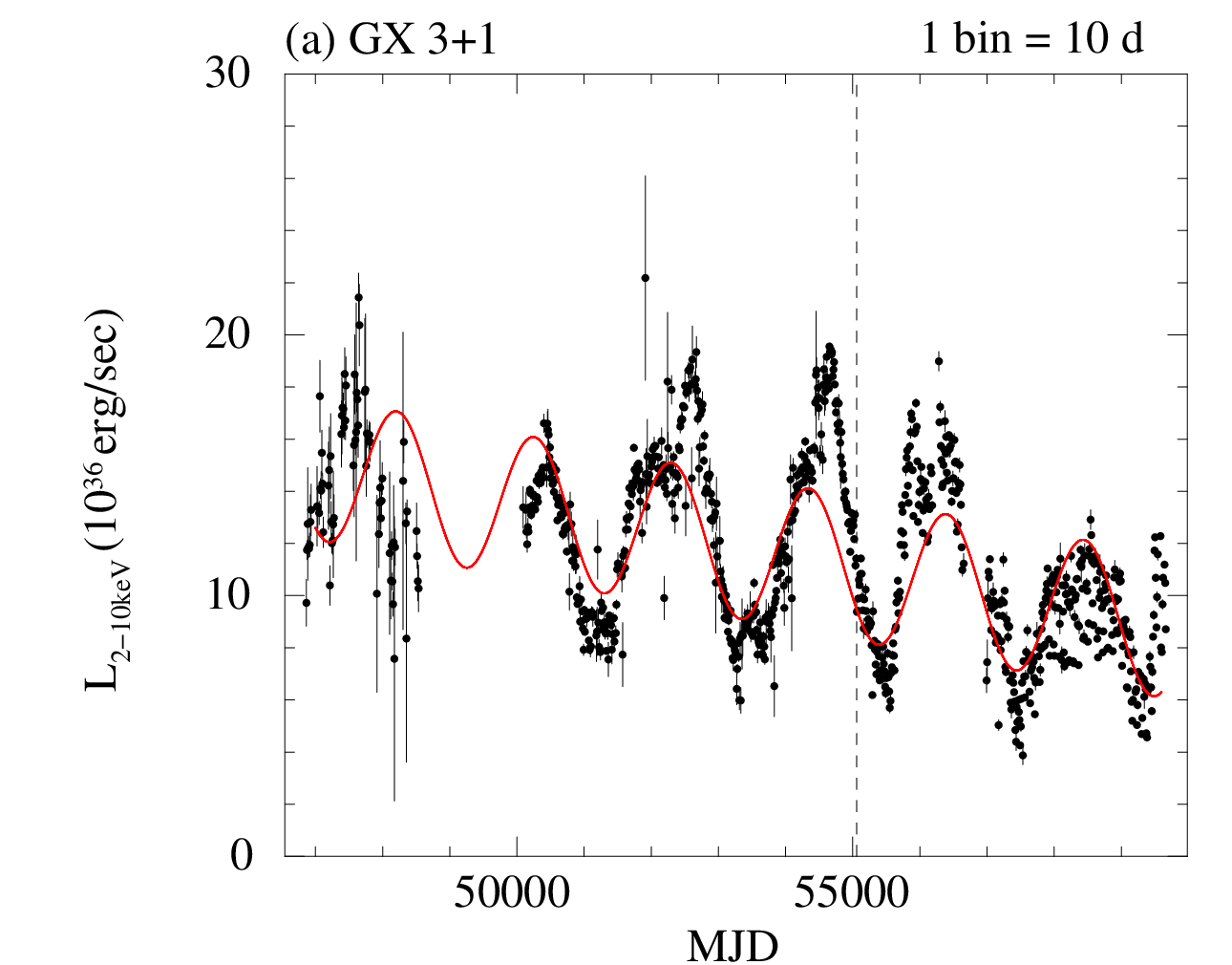}
  \includegraphics[width=85mm]{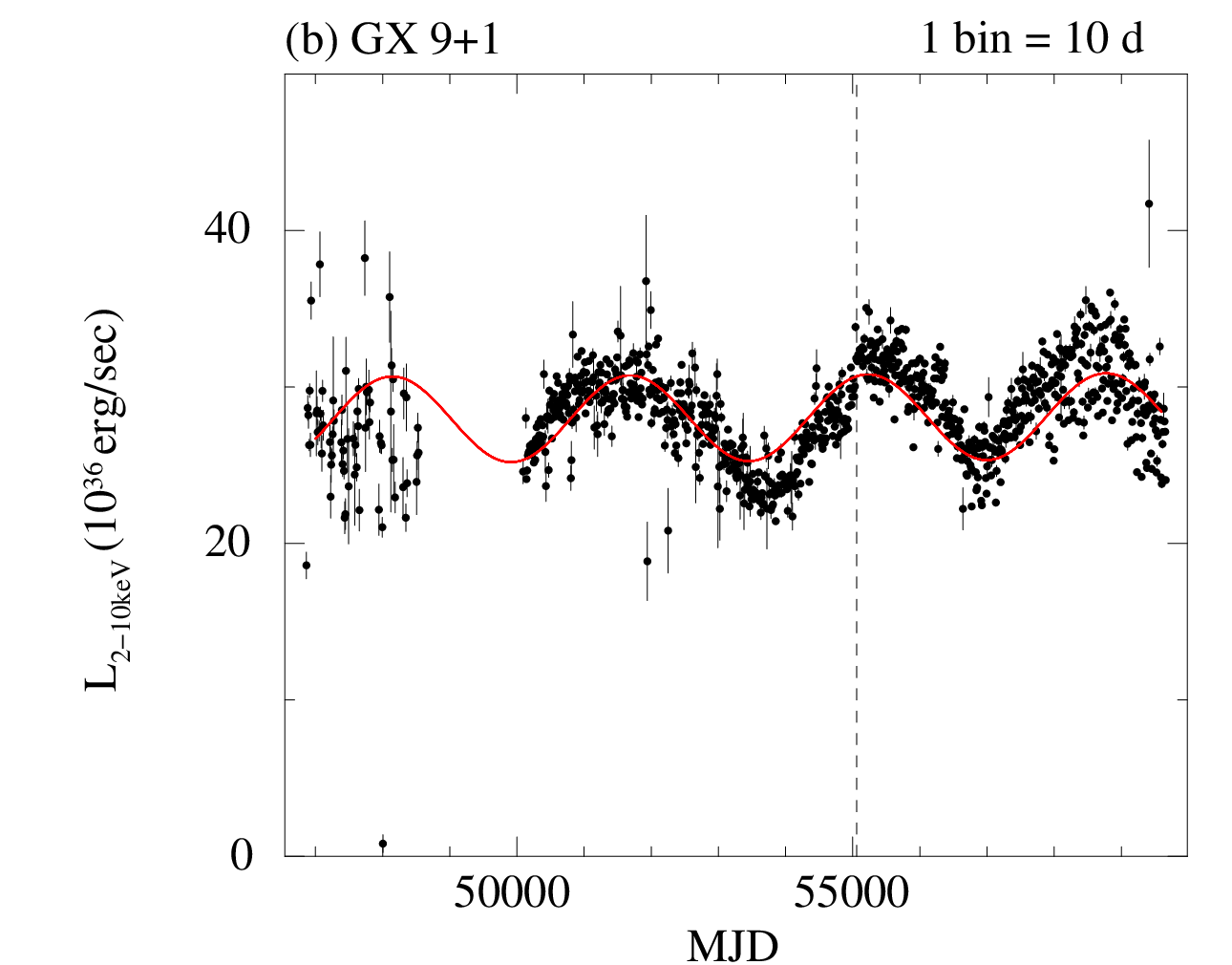}
\caption{
Light curves observed by Ginga/ASM, RXTE/ASM, and MAXI/GSC.
For GX~3$+$1 and GX~9$+$1, we show the fitted sinusoidal curves in red (Color online).
The model function and parameters are shown in table~2.
The data of GX~9$+$1 has a discrepancy between ASM and GSC fluxes.
MAXI data are processed in a regular way as for other sources. 
There is no contamination source nor
background uncertainty by the ridge emission.
So we plotted the data as they are.
}
\label{fig2}
\end{figure*}  

\begin{table*}
  \tbl{Fitting results of a sinusoidal and linear function model.
  Plots in figure~2.}
{
 \begin{tabular}{lccccc}
      \hline
        Name & Model\footnotemark[$*$]
        & Average luminosity  
        & \multicolumn{2}{c}{Sin}
        & Slope\footnotemark[$\S$]\\
        & 
        & ($10^{36}$ erg s$^{-1}$)
        & $A$\footnotemark[$\dag$]
        & $P$\footnotemark[$\ddagger$] (d) 
        & ($10^{36}$ erg s$^{-1}$ d$^{-1}$)  \\
      \hline
GX~3$+$1      & CONS$+$LINR$+$SIN
              & 9.7
              & 0.28 & 2046 & $-4.81 \times 10^{-4}$  \\ 
GX~9$+$1      & CONS$+$LINR$+$SIN
              & 29.5 
              & 0.09 & 3542 & $+1.90 \times 10^{-5}$  \\
\hline
    \end{tabular}
    }
    \label{tab2}
\begin{tabnote}
\footnotemark[$*$] Model components. CONS: constant component,
LINR: linear component,
SIN: sinusoidal component. 
\\   
\footnotemark[$\dag$] Amplitude of sinusoidal component against average luminosity. \\ 
\footnotemark[$\ddag$] Period of sinusoidal component.\\ 
\footnotemark[$\S$] Slope of baseline. \\
\end{tabnote}
\end{table*}

Next, in order to investigate the variability of the sinusoidal profile
of two CP soures and seven MP sources,
we fit each peak with a Gaussian profile (figure~3 and 4).
The parameters are
the width (sigma) of Gaussian profile (GW) and 
the peak luminosity (GN). Those correspond to 
the period and amplitude of the sinusoidal variation,
respectively.
In order to see variability of the peak luminosity and width,
we plotted the GN against the GW in figure~5.
The amplitude and periods of two CP sources
(GX~3$+$1 and GX~9$+$9)
are more stable than those of MP sources.
The average GWs for each source are $\sim$ 300--2300~d,
which correspond approximately to
a half sinusoidal period of GX~3$+$1 ($\sim$~2.5~yr)
and a period of GX~9$+$1 ($\sim$~10~yr), respectively.

\begin{figure*}
  \includegraphics[width=85mm]{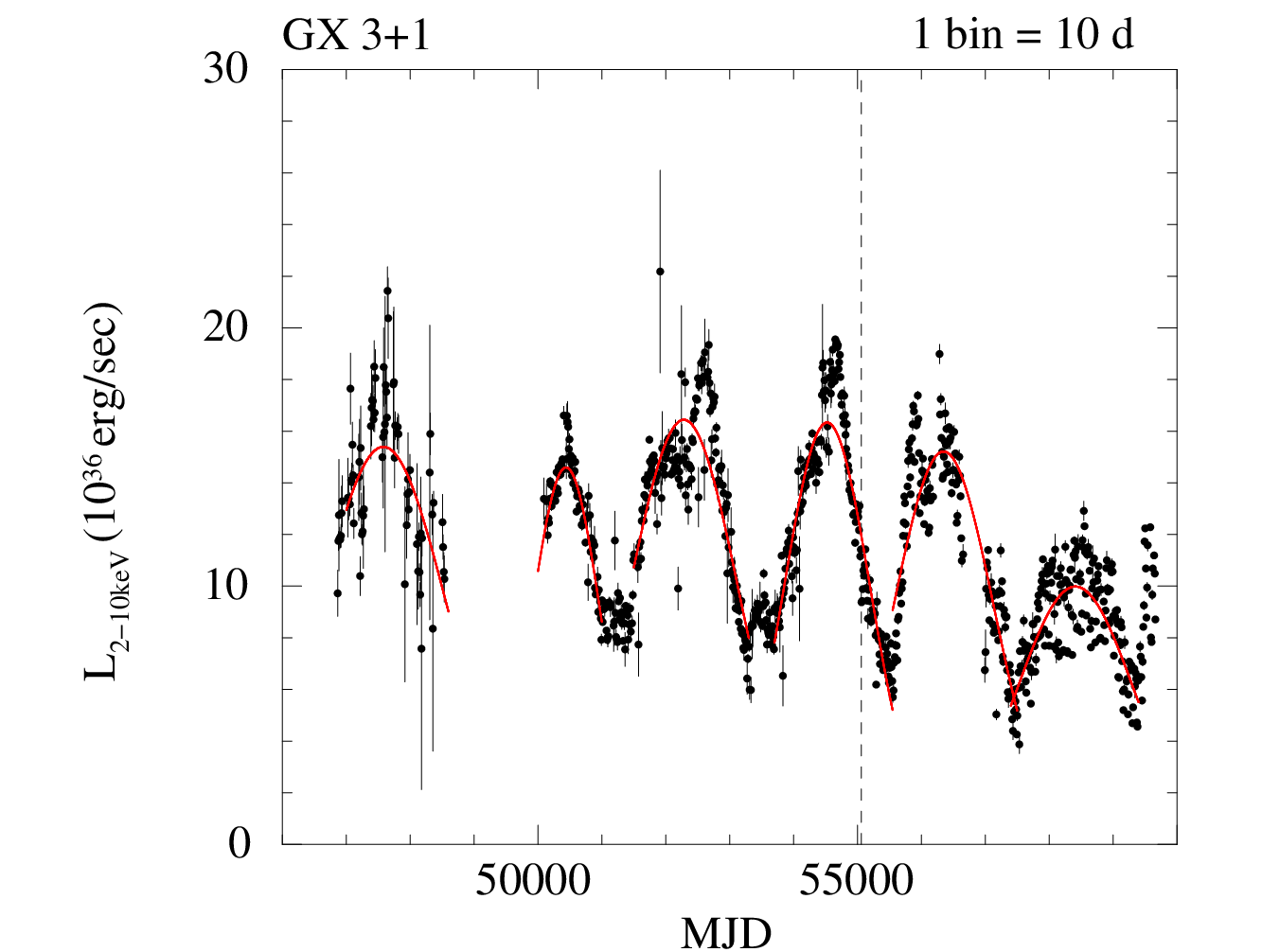}
  \includegraphics[width=85mm]{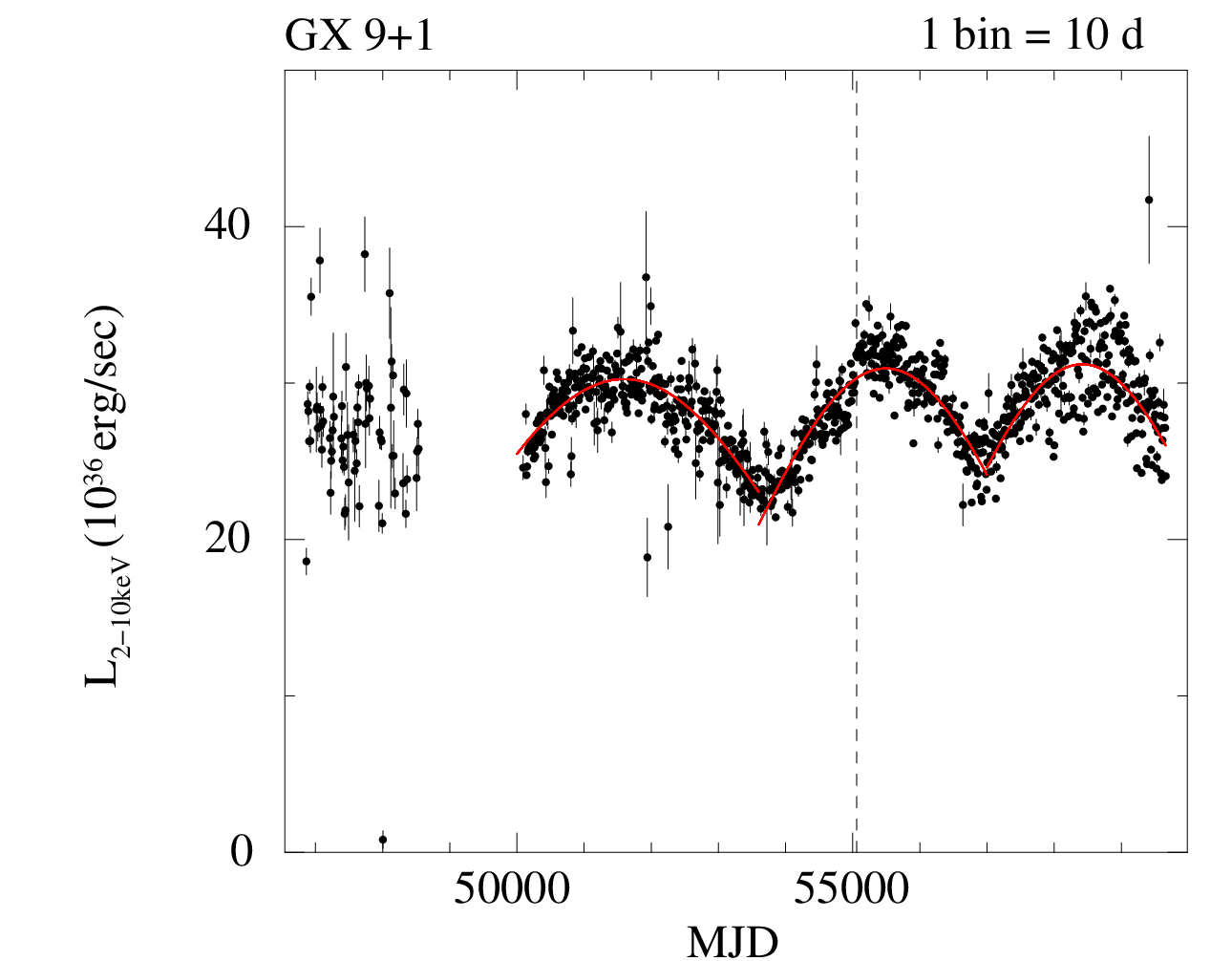}
\caption{Light curves of two CP sources fitted the gaussian peaks. Parameters are shown in table~3.
The fitting results were shown in red color curve.
(Color online)}
\label{fig3}
\end{figure*}  

\begin{figure*}
  \includegraphics[width=75mm]{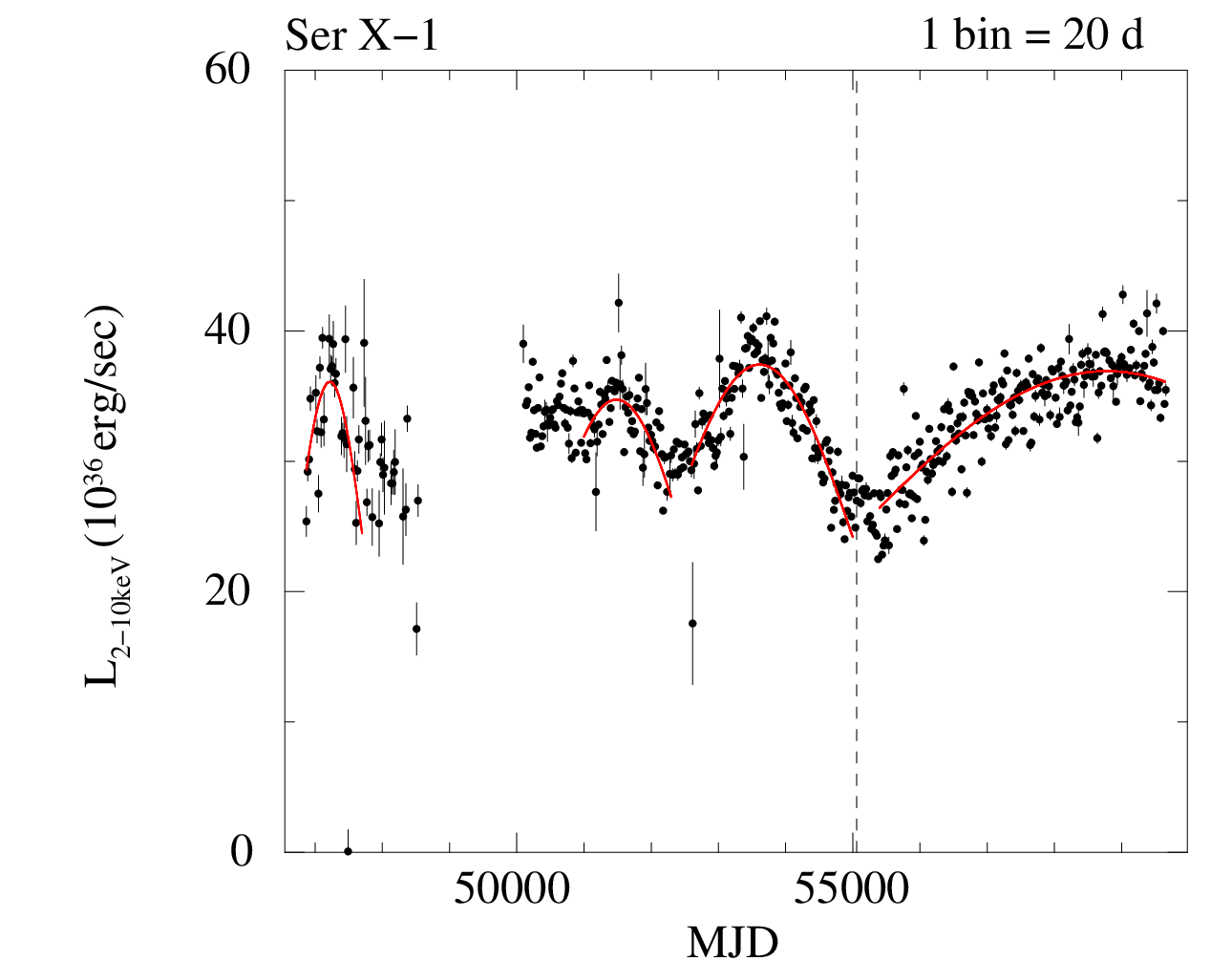}
  \includegraphics[width=75mm]{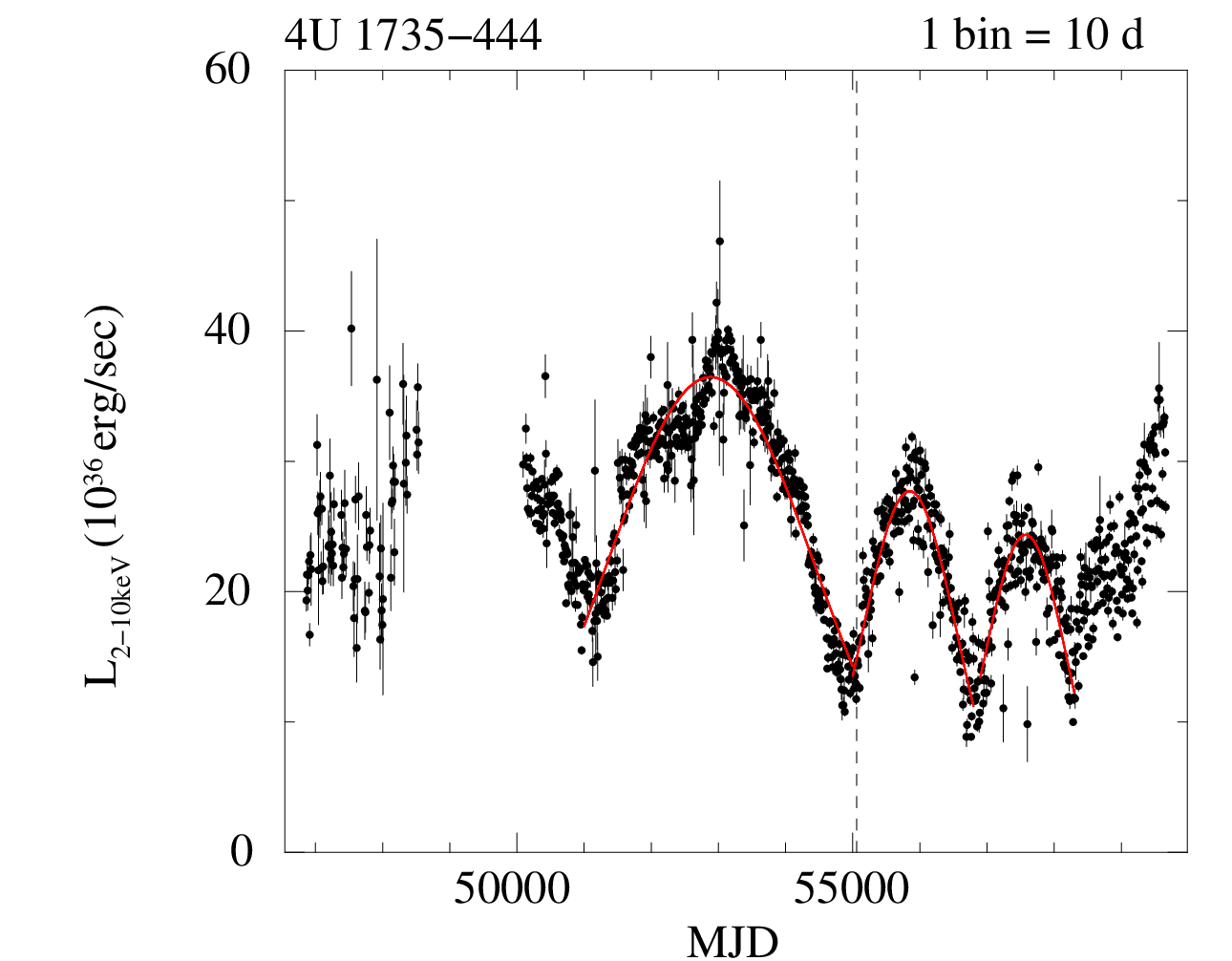}
  \includegraphics[width=75mm]{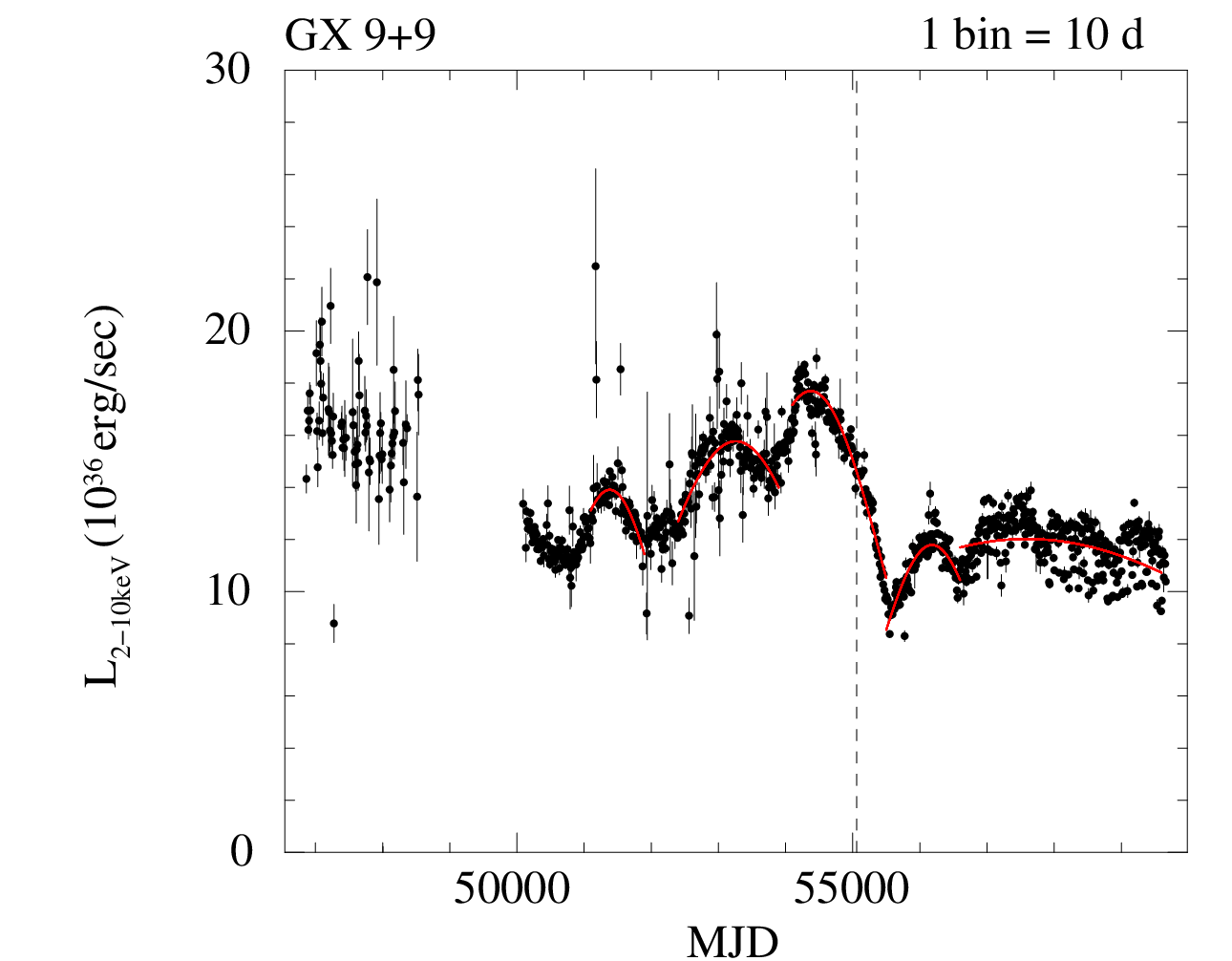}
  \includegraphics[width=75mm]{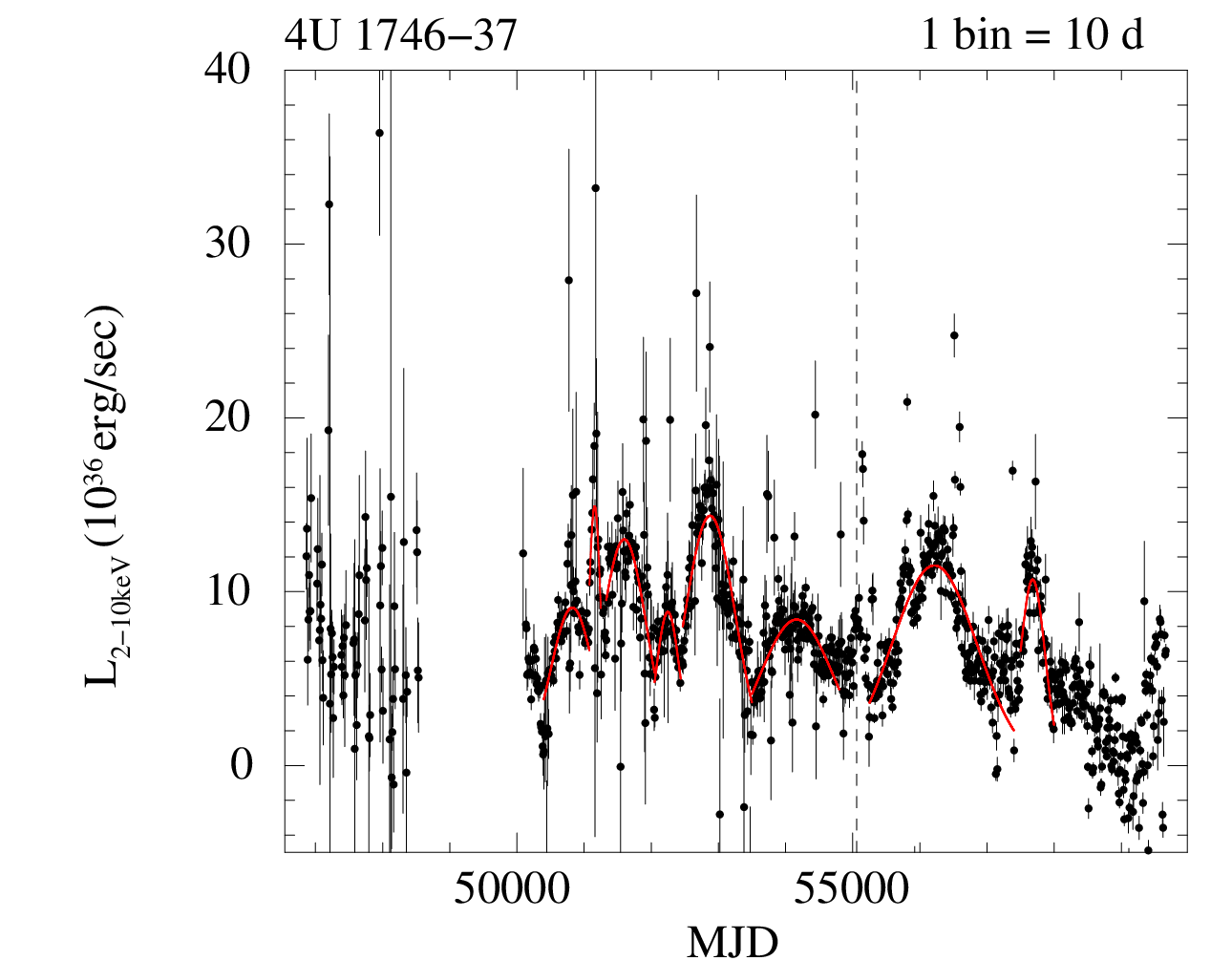}
  \includegraphics[width=75mm]{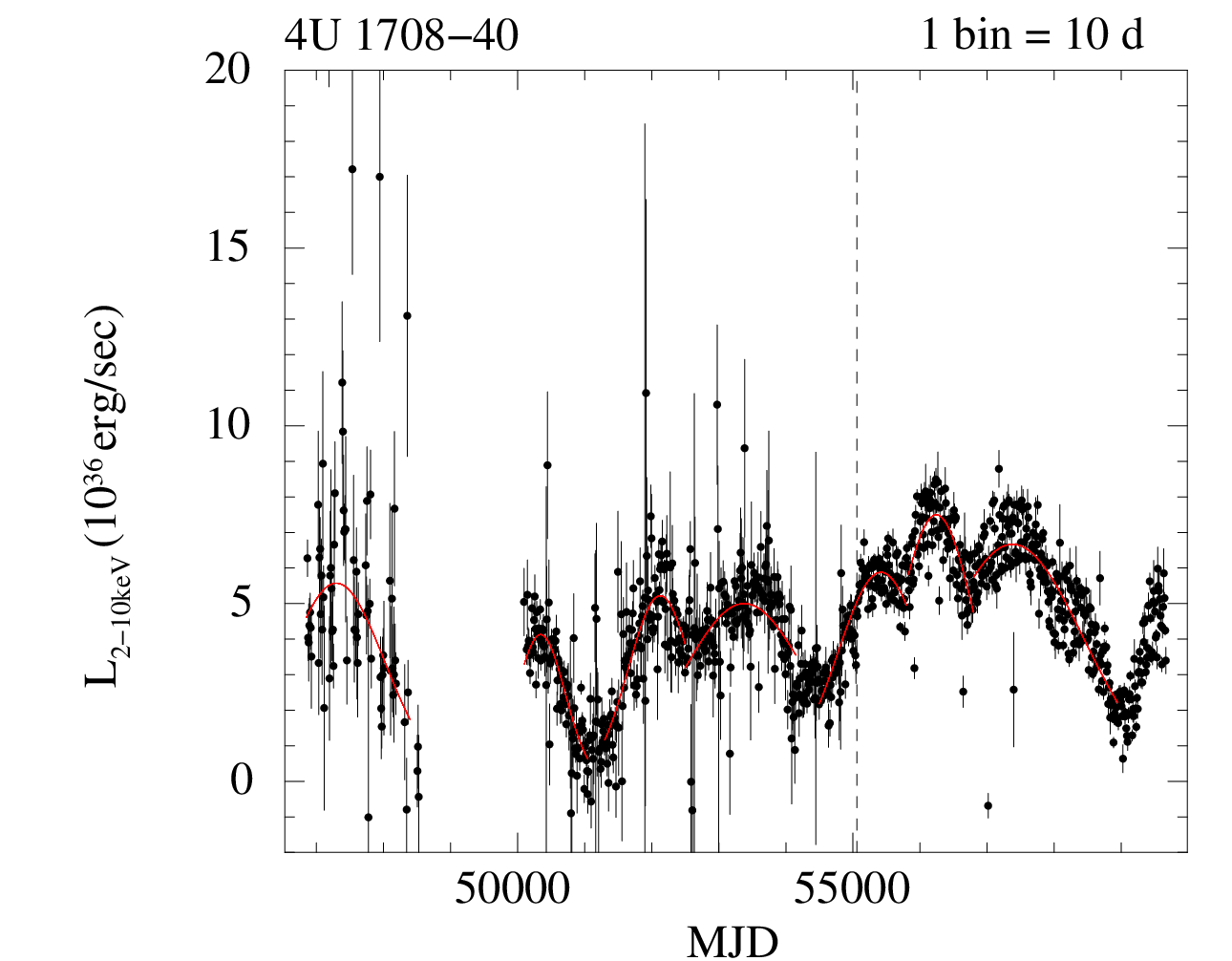}
  \includegraphics[width=75mm]{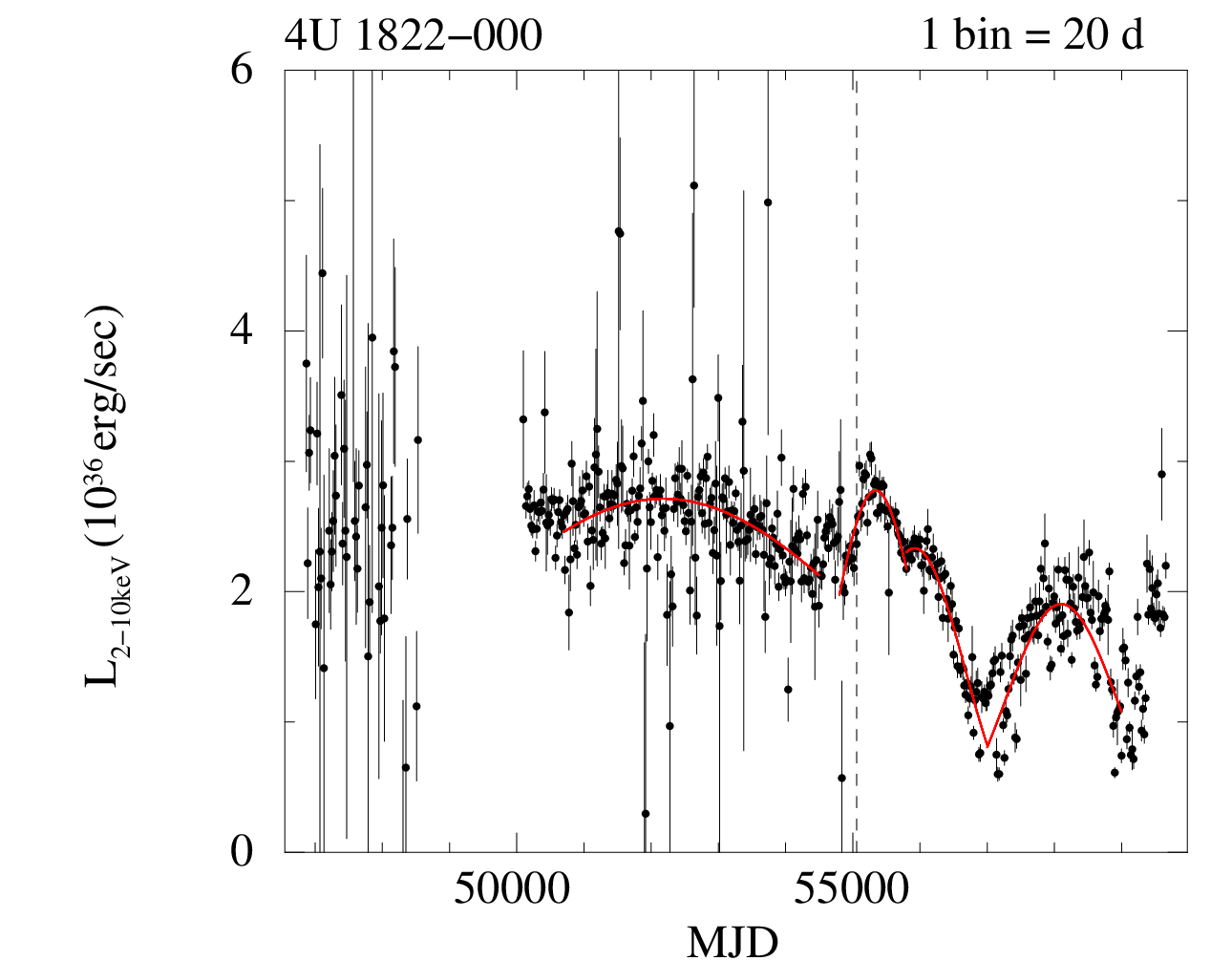}
  \includegraphics[width=75mm]{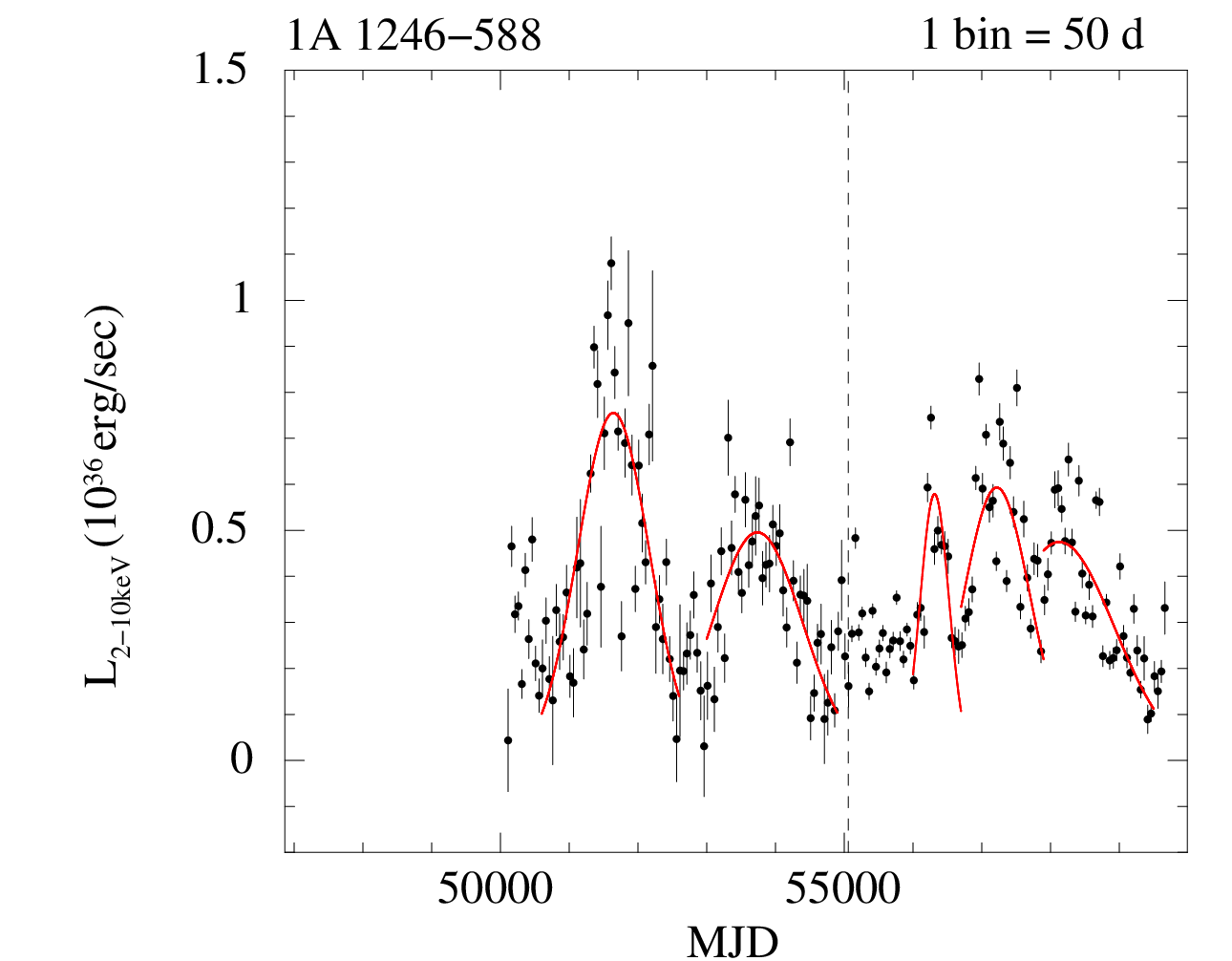}
\caption{Light curves of seven MP sources fitted the gaussian peaks. Parameters are shown in table~3.
The fitting results were shown in red color curve.
For 1A1246$-$588, there were no data of Ginga/ASM.  
(Color online)}
\label{fig4}
\end{figure*}  

\renewcommand{\arraystretch}{0.75}
\begin{table*}
  \tbl{Fitting results of gaussian components. Plots in figure~3 and 4.}
{
 \begin{tabular}{lccccc}
      \hline
       Name & Type & MJD 
        & GC\footnotemark[$*$]
        & GW\footnotemark[$*$]
        & GN\footnotemark[$*$] \\
    \hline
GX~3$+$1 & CP & 47000--48600 & 
$47580\pm28$ & $985^{+70}_{-58}$ & $15.4\pm0.3$ \\
& & 50000--51000 &
$50438\pm4$ & $546\pm7$ & $14.59\pm0.05$ \\
& &  51500--53300 & 
$52282\pm3$ & $864\pm5$ & $16.44\pm0.03$ \\
& & 53700--55550 & 
$54524\pm2$ & $678\pm2$ & $16.33\pm0.04$ \\
& & 55550--57500 & 
$56347\pm3$ & $784\pm3$ & $15.21\pm0.04$ \\
& & 57400--59400 & 
$58410\pm3$ & $907\pm4$ & $9.98\pm0.02$ \\
\hline
GX~9$+$1 & CP & 50000--53600 &
$51595\pm5$ & $2718\pm11$ & $30.25\pm0.03$ \\
& & 53600--57000 & 
$55490\pm4$ & $2141\pm8$ & $30.94\pm0.03$ \\
& & 57000--59660 & 
$58418\pm5$ & $2058\pm13$ & $31.20\pm0.04$ \\
\hline
Ser~X-1 & MP & 46870--47700 &
$47220\pm14$ & $544\pm34$ & $36.1\pm0.5$ \\
& & 51000--52300 &
$51484\pm6$ & $1172\pm17$ & $34.73\pm0.06$ \\
& & 52600--55000 &
$53611\pm5$ & $1486\pm9$ & $37.42\pm0.05$ \\
& & 55000--59650 &
$58778\pm37$ & $4132\pm55$ & $36.92\pm0.06$ \\
\hline
4U~1735$-$444 & MP
& 51000--55000 &
$52884\pm3$ & $1545\pm5$ & $36.46\pm0.06$ \\
& & 55000--56800 &
$55847\pm3$ & $708\pm4$ & $27.7\pm0.1$ \\
& & 56900--58300 &
$57575\pm3$ & $618\pm5$ & $24.4\pm0.1$ \\
\hline
GX~9$+$9 & MP & 51100--51900 &
$51383\pm13$ & $825\pm36$ & $13.91\pm0.05$ \\
& & 52400--53900 &
$53260\pm8$ & $1304\pm21$ & $15.77\pm0.04$ \\
& & 54100--55500 &
$54374\pm12$ & $1103\pm14$ & $17.70\pm0.04$ \\
& & 55500--56600 &
$56181\pm6$ & $846\pm14$ & $11.79\pm0.03$ \\
& & 56600--5960 &
$57589\pm39$ & $4212\pm140$ & $12.02\pm0.02$ \\
\hline
4U~1746$-$37 & MP & 50400--51080 &
$50824\pm10$ & $322\pm18$ & $9.1\pm0.2$ \\
& & 51080--51250 &
$51157\pm5$ & $92\pm13$ & $15\pm1$ \\
& & 51330--52050 &
$51597\pm8$ & $338\pm14$ & $13.0\pm0.2$ \\
& & 52050--52440 &
$52248\pm6$ & $180\pm13$ & $8.8\pm0.3$ \\
& & 52470--53500 &
$52876\pm6$ & $375\pm9$ & $14.4\pm0.3$ \\
& & 53500--54800 &
$54164\pm13$ & $557\pm22$ & $8.4\pm0.2$ \\
& & 55250--57400 &
$56216\pm6$ & $634\pm8$ & $11.5\pm0.1$ \\
& & 57500--58000 &
$57680\pm5$ & $183\pm7$ & $10.7\pm0.2$ \\
\hline
4U~1708$-$40 & MP & 46850--48400 &
$47295^{+67}_{-101}$ & $721^{+156}_{-98}$ & $5.6\pm0.4$ \\
& & 50100--51050 &
$50343\pm24$ & $363\pm21$ & $4.1\pm0.1$ \\
& & 51300--52500 &
$52127\pm13$ & $478\pm16$ & $5.2\pm0.1$ \\
& & 52500--54150 &
$53380\pm22$ & $929^{+49}_{-43}$ & $5.0\pm0.1$ \\
& & 54500--55800 &
$55418\pm13$ & $652\pm19$ & $5.87\pm0.05$ \\
& & 55830--56800 &
$56243\pm8$ & $582\pm18$ & $7.5\pm0.1$ \\
& & 56800--58950 &
$57379\pm17$ & $1057\pm17$ & $6.67\pm0.04$ \\
\hline
4U~1822$-$000 & MP & 50700--54500 &
$52177^{+59}_{-67}$ & $3322^{+191}_{-164}$ & $2.71\pm0.02$ \\
& & 54800--55800 &
$55340\pm13$ & $665^{+35}_{-30}$ & $2.78\pm0.03$ \\
& & 55800--57000 &
$55920^{+41}_{-50}$ & $748^{+42}_{-36}$ & $2.33\pm0.04$ \\
& & 57000--59000 &
$58103\pm11$ & $841\pm18$ & $1.90\pm0.03$ \\
\hline
1A1246$-$588 & MP & 50600--52600 &
$53737^{+42}_{-46}$ & $657^{+62}_{-52}$ & $0.50\pm0.03$ \\
& & 56000--56700 &
$56315\pm9$ & $209\pm11$ & $0.58\pm0.02$ \\
& & 56700--57900 &
$57218\pm17$ & $483\pm26$ & $0.59\pm0.01$ \\
& & 57900--59500 &
$58128^{+74}_{-98}$ & $807^{+59}_{-63}$ & $0.47\pm0.02$ \\
\hline
    \end{tabular}
    }
    \label{tab3}
\begin{tabnote}
\footnotemark[$*$] GC is the center MJD of gaussian component, 
GW is the width (sigma) of gaussian component,
and GN is the center luminosity of gaussian component.\\ 
\end{tabnote}
\end{table*}
\renewcommand{\arraystretch}{1}

\begin{figure*}
\includegraphics[width=170mm]{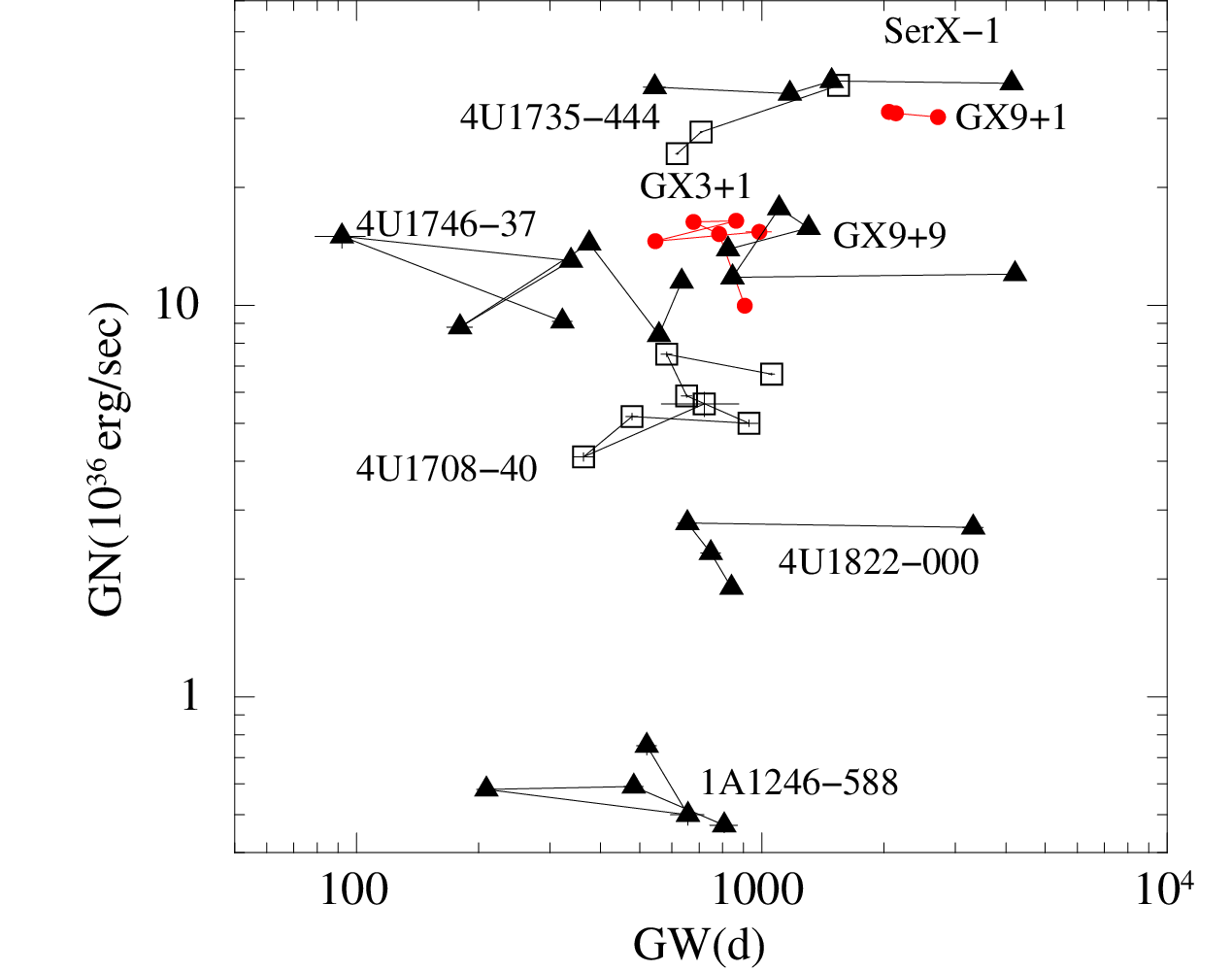}
\caption{
The center luminosity of Gaussian model (GN)
as a function of the width of Gaussian model (GW).
Filled circles represent data of CP sources.
Filled triangles and open squares represent 
data of MP sources.
To avoid the confusion of data points, we use two kind of
marks.
}
\label{fig5}
\end{figure*}  

The individual properties in the nine sources are as follows:

\begin{description}
\item[GX~3$+$1 (figure~2a and 3a):]
The sinusoidal variation is clear,
although the shapes of sinusoidal profiles are complex with some peaks.
The baseline decreases.
The long-term period is $\sim 6$ yr, which is similar 
to the results of KC10.
The long-term periodic variation turned to be stable over the twice time-length in KC10.

\item[GX~9$+$1 (figure~2b and 3b):]
The sinusoidal variation is also clear
but with a little longer period ($\sim 11$ yr) than that of GX~3$+$1.
The baseline has gradually increased in $\sim 25$~yr,
which is against the KC10 report that the baseline was constant for $\sim 13$~yr observed by RXTE/ASM.
The period of long-term is similar to the results ($\sim 12$ yr) of KC10.

\item[Ser~X-1 (figure~4a):]
Quasi periodic modulation is clear.
The period is not constant,
but seems to become longer.
KC10 reported the period of $\sim 7.3$ yr
observed by RXTE/ASM. 
In the MAXI/GSC era, 
the period is longer and amplitude may be larger.

\item[4U~1735$-$444 (figure~4b):]
The sinusoidal variation is clear.
However, the period is not constant, but seems to become shorter.
KC10 reported the period of $\sim 10$ yr
from the large hump around MJD=53000. 
In the MAXI/GSC era, 
the period is shorter, about a half, and amplitude may be smaller.

\item[GX~9$+$9 (figure~4c):]
Although KC10 reported that the baseline was increasing,
it cannot be extrapolated in the MAXI/GSC era. 
The light curve dropped after RXTE/ASM. 
Since then the baseline has stayed almost constant. 
However, the long-term period is
similar to the results ($\sim 4$ yr) of KC10 throughout
by RXTE/ASM and MAXI/GSC.
The amplitude of the sinusoidal variation seems to be smaller in the MAXI/GSC era.

\item[4U~1746$-$37 (figure~4d):]
Quasi periodic modulation is clear.
However, the period and the amplitude is not constant.
KC10 reported the period of $\sim 4.36$ yr
observed by RXTE/ASM.

\item[4U~1708$-$40 (figure~4e):]
Quasi periodic modulation is clear.
However, the period and the amplitude is not constant.
Although we divided the data into small peak with
short period, the data of MAXI/GSC era can also
be fitted with one gaussian profile for
large one peak.
KC10 did not report this source,
because they focused on the Z sources and Atoll sources.

\item[4U~1822$-$000 (figure~4f):]
Quasi periodic modulation is clear.
However, the period and the amplitude is not constant.
Although we divided the data between MJD=54800 and 57000
into two small peak with short period,
the data can also be fitted with one gaussian profile for
large one peak.
KC10 did not report this source.

\item[1A~1246$-$588 (figure~4g):]
Quasi periodic modulation is seen.
However, the period and the amplitude is not constant.
KC10 did not report this source.

\end{description}

\section{Discussion}

The long-term sinusoidal variations were presented 
for the two CP sources (GX~3$+$1 and GX~9$+$1).
The long-term periods range in $\sim 5$~yr and $\sim 10$~yr
as shown in table~2--3 and figure~2--3.
We also investigated the seven MP sources with
modified periodic variation, and estimated
periods of quasi-periodic modulation
(table~3 and figure~4).
The rage of each average period was approximately
from $\sim$~2.5~yr to $\sim$~10~yr.
These time scales are much longer than the orbital period
($\sim$ 2--5~hr), although, strictly speaking,
those of the four sources
(CP: GX~3$+$1, GX~9$+$1 and MP:4U~1708$-$40, 1A~1246$-$588) are unknown.

We discuss mechanisms of long-term variation.
First, we focus on the physical motion of the accretion disk.
Precession of accretion disk may occur 
due to excitation of resonances 
in case of mass ratio of
$q = M_{\rm c} / M_{\rm NS} \sim$ 0.25--0.33 \citep{Whitehurst1991}.
It is likely to occur in the case of LMXBs
with donor mass of 0.35$\Mo$--0.46$\Mo$.
Although the donor mass of our CP and MP sources
are not known, precession of accretion disk
could be possible.
In general, according to \citet{Inoue2012},
precession occurs when 
the ratio of the precession period $P_{\rm p}$
to the binary orbital period $P_{\rm B}$
is $P_{\rm p}/P_{\rm B}$ = 10--100. 
The ratios of our results are
$P_{\rm p}/P_{\rm B} \geq 10^{4}$.
There is no such case in figure~2 of \citet{Inoue2012}.
However, if we extrapolate one line for
$P_{\rm p}/P_{\rm B} = 10^{4}$,
the range of donor mass for 0.14$\Mo$--0.42$\Mo$ ($q\sim$0.1--0.3) would be possible.
Since donor mass is not known,
the disk precession scenario could be possible.

On the other hand,
another possibility of radiation-induced warping is excluded. 
Kotze and Charles (2012) explicitly discuss
the stability of radiation-induced warping.
In the region below the bottom solid line in their figure~1,
a radiation-induced warping of the disk is unlikely,
and there is in fact no super-orbital cycle observed in this region.
Our four sources
lie below the bottom solid line for 
the typical values of
mass ratio: $q = M_{\rm c} / M_{\rm NS} \sim 0.3$
and orbital period $\sim 2-5$ hr. 
Thus the radiation-induced warping of the accretion disk is unlikely.

Next, we consider the possibility of
the variation of mass transfer rate from the donor.
KC10 suggested that  they are consequence of the solar-like magnetic
cycles seen in the late-type star
(\cite{Applegate1987}, \cite{Warner1988}).
They pointed out that the flux modulation of a sin wave of $\leq 30$ percent
is plausible by the magnetic cycles.
In our result, the flux modulation (see the amplitude of table~2)
is $\leq 30$ percent.
Thus the solar cycle-like variations can be responsible for its long-term variations.

Here, we discuss another possibility of
the variation of mass transfer rate from the donor.
It is the irradiation by the central X-ray source to the donor star.
\citet{Ritter2008} showed that irradiation-driven mass
transfer cycles could only
occur when the irradiation is sustained
for a sufficiently long time.
B{\"u}ning and  Ritter (2004) shows the numerical results
in terms of 
irradiation-driven mass transfer cycles and then indicates
the possibility that NS-LMXB undergoes
those cycles.
Their numerical results show that
the stability of the mass transfer rate
from the donor star is a function of the orbital period.

In figure~1, filled marks represent quasi periodic modulation (MP: Ser~X-1, 4U~1735$-$444, GX~9$+$9, 4U~1746$-$37, 4U~1822$-$000) in our analysis.
In figure~1b, the filled marked five MP sources
are located in the region
which is middle in the binary separation
and high irradiation average flux.
The region may tend to show a periodic modulation.

However, in the region,
there are also three sources
(GS~1826$-$238, 4U~1636$-$536, and 4U~1254$-$690)
of non-periodic modulation.
The 2 LV sources (GS~1826$-$238 and 4U~1636$-$536)
show the large luminosity change of 1–2 orders of magnitude, and 
it is difficult to see the flux modulation of the amplitude (
$\leq 30$ percent ) 
which a change of the mass transfer rate induces.
The one NP source (4U~1254$-$690) has similar binary properties to that of GX~9$+$9 (CP).
The cause of no clear periodic variation is unclear.

We discuss the features of the region in which a long-term variation trend to occur.
One feature is high irradiation average flux
($\geq 1\times10^{13}$ erg s$^{-1}$ cm$^{-2}$).
Even in the high irradiation-flux, 
there are four NP sources
(Sco~X-1, GX~349$+$2, LMC~X-2 and 4U1624$-$490)
without a long-term variation in the
larger binary-separation region.
Here, the three sources 
(Sco~X-1, GX~349$+$2, and LMC~X-2) are Z sources.
KC10 reported that the amplitudes of long-term variation of the 
Z sources are small and 
noted that it is because
their luminosity is close to the Eddington.
Although the luminosity of 4U~1624$-$490 is not
close to Eddington luminosity, the intrinsic
luminosity is uncertain because it is an ADC source.
Here, we focus on the donor star.
The three sources (Sco~X-1, GX~349$+$2 and LMC~X-2)
are Z sources, and the donor stars are suggested to
be evolved stars 
(\cite{Hasinger1989} for Z sources, \cite{Cherepashchuk2021} for Sco~X-1).
4U~1642$-$490 did not belong to Z sources. However,
\citet{Jones1989} reported that the flaring behavior
of the source shows similarity to the flaring blackbody
component (in temperature and radius) of Sco~X-1 and other
Z sources.
Although the donor star of 4U~1624$-$490 is not identified,
it is possible to be an evolved star, similar
to the Z sources.
In this case, the mass transfer rate on to the neutron star 
may be above the Eddington luminosity.

On the other hand,
in the high irradiation-flux, but in the smaller binary-separation region,
there are four sources (FV: 4U~1820$-$303 and
three NPs: 4U~1543$-$624, 4U~2127$+$119, and 4U~1916$-$053). 
The sources are Ultra Compact X-ray Binaries (UCXBs).
The donors in UCXBs may be white dwarfs (WDs) or He stars 
(4U~1820$-$303: \cite{Rappaport1987},
4U~1543$-$624: \cite{Nelemans2004},
4U~2127$+$119: \cite{Dieball2005}, and 4U~1916$-$053: \cite{Joss1978}, \cite{Nelemans2006}).
\citet{Lu2017} suggests that, if the donor star is a WD, 
the irradiation flux can only penetrate into a very thin
layer of the WD surface, and the irradiation hardly affects the evolution of the persistent UCXB.
The irradiation flux would not affect 
the mass transfer rate from the donor star.

In summary,
the intense irradiation may induce variation of the mass transfer rate,
and a sinusoidal periodic variation may appear.
Our results also seem to show a dependence on the donor.
A WD donor would not 
be affected by the irradiation.
The periodicity might be
related to the mass transfer cycles caused by the irradiation
pointed out by \citet{Ritter2008}.

\appendix 
\section*{Light curves of 41 NS-LMXB observed with MAXI/GSC and RXTE/ASM}

The light curve used to analyze the long-term variation
are presented in this Appendix (figure~6--14).
The energy band is 2--10~keV band, and th
e period of data
is from 1996 February to 
\textcolor{black}{2021} 
December.
The left side of the vertical dash line is
the data of RXTE/ASM, and the right is MAXI/GSC. 

\begin{figure*}
\begin{center}
  \includegraphics[width=75mm]{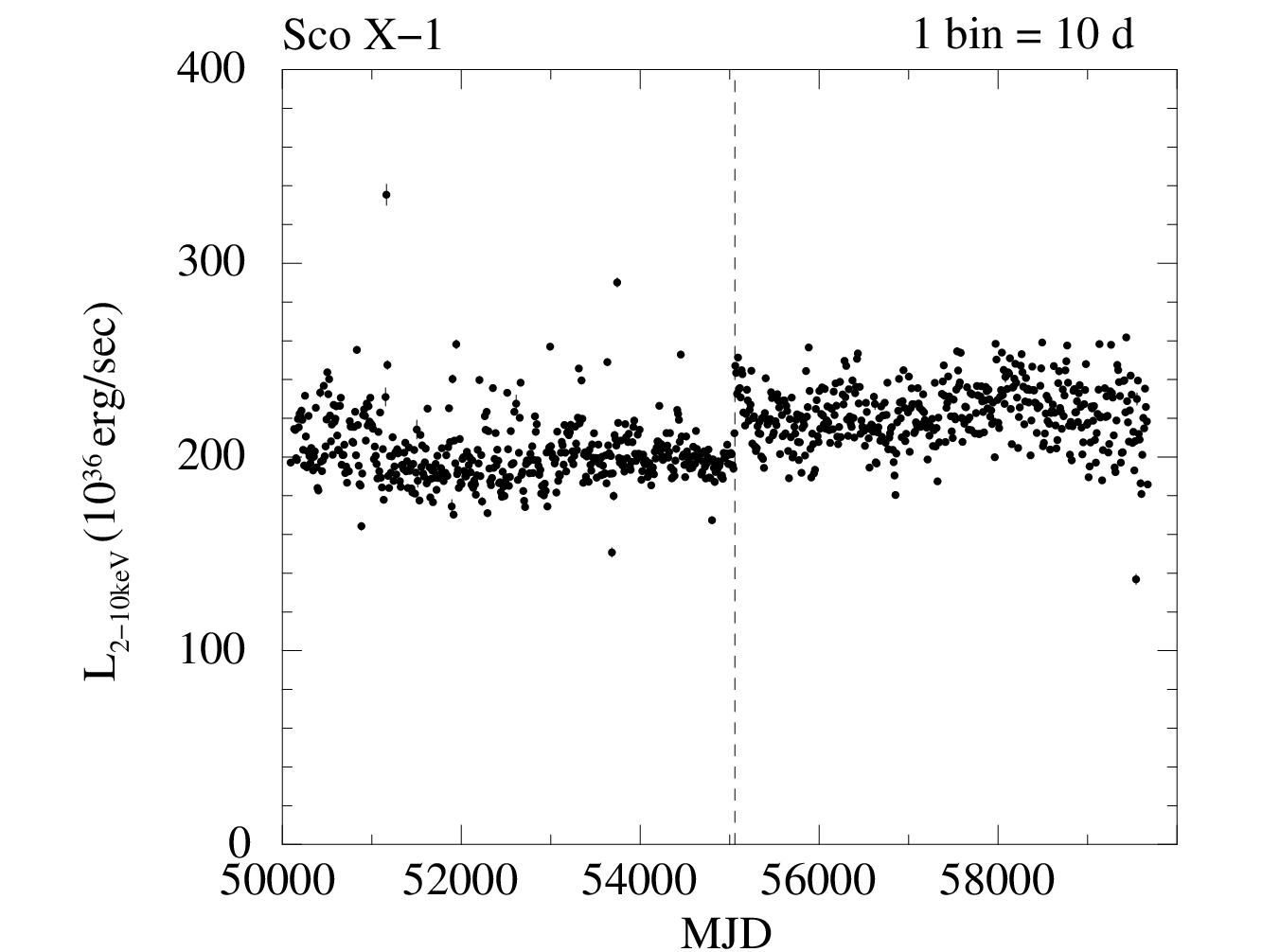}
  \includegraphics[width=75mm]{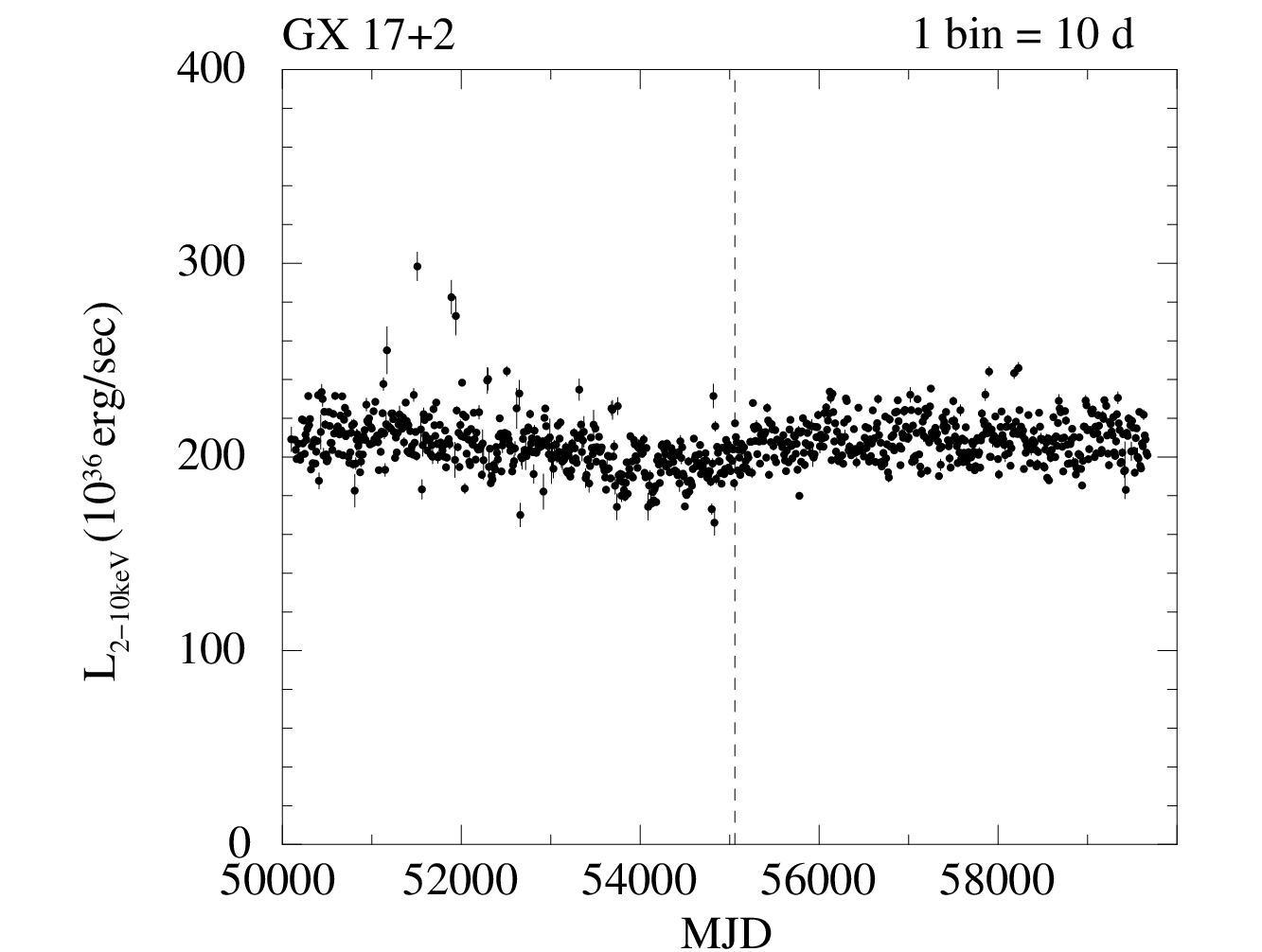}
  \includegraphics[width=75mm]{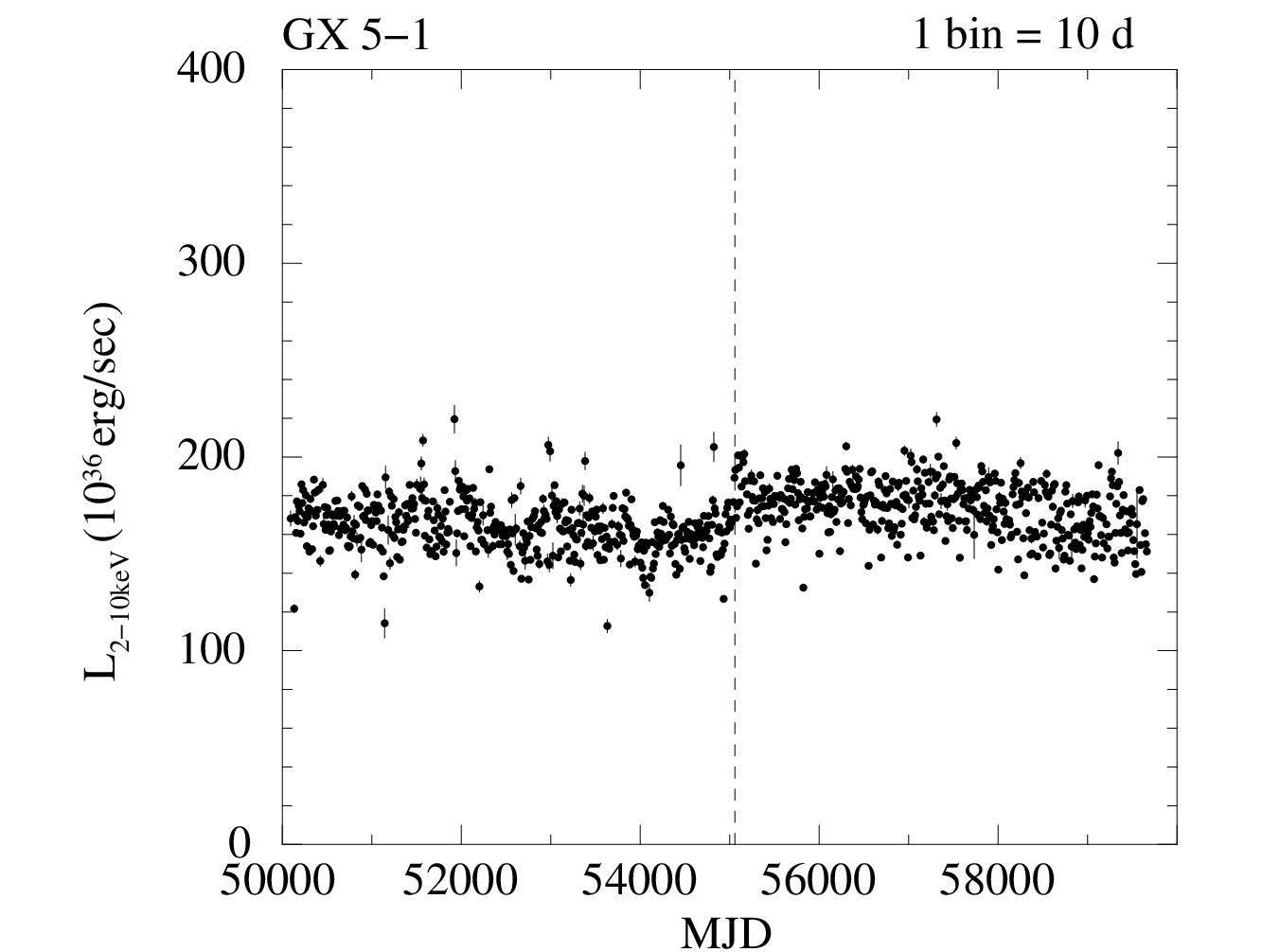}
  \includegraphics[width=75mm]{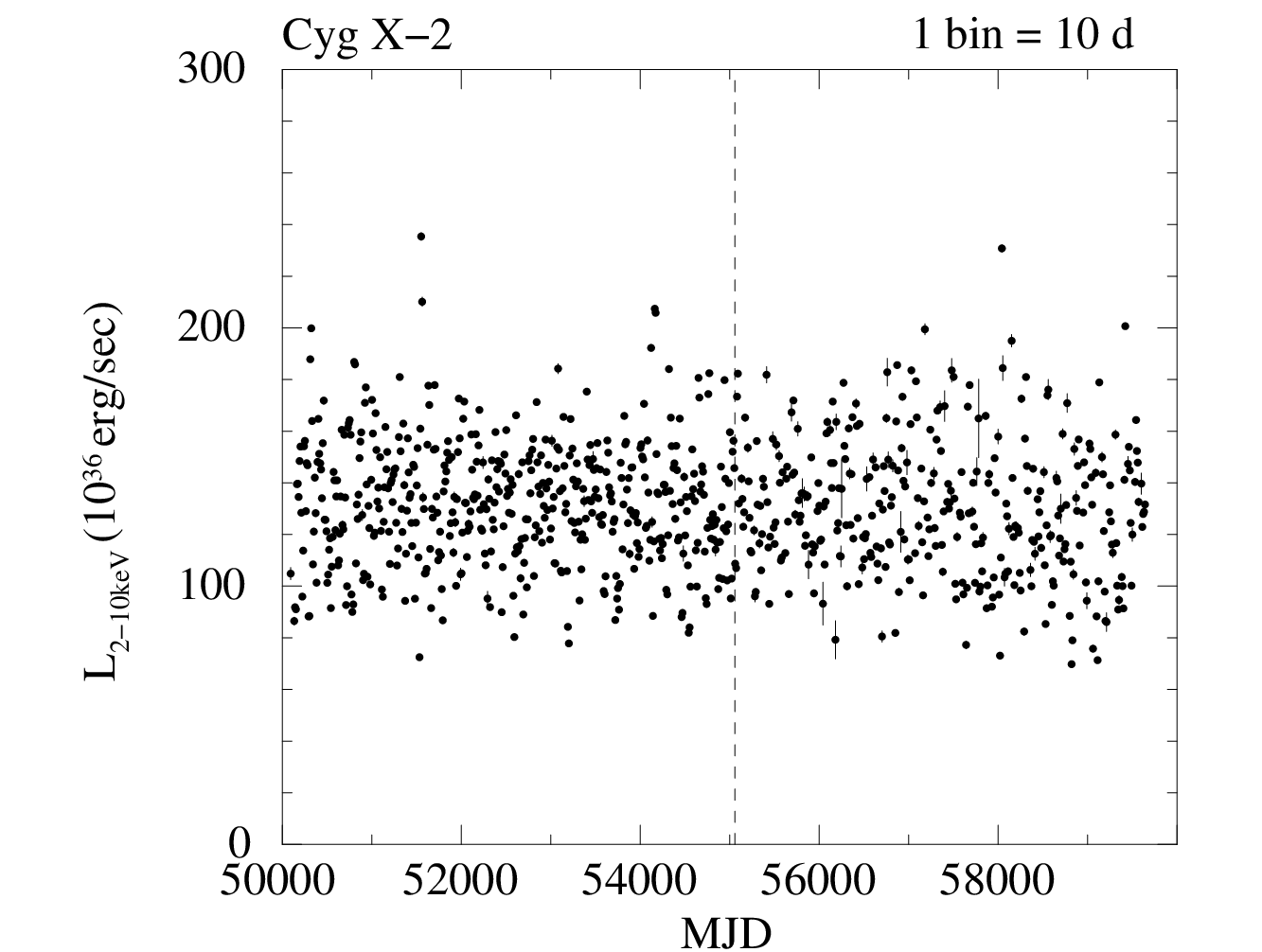}
  \includegraphics[width=75mm]{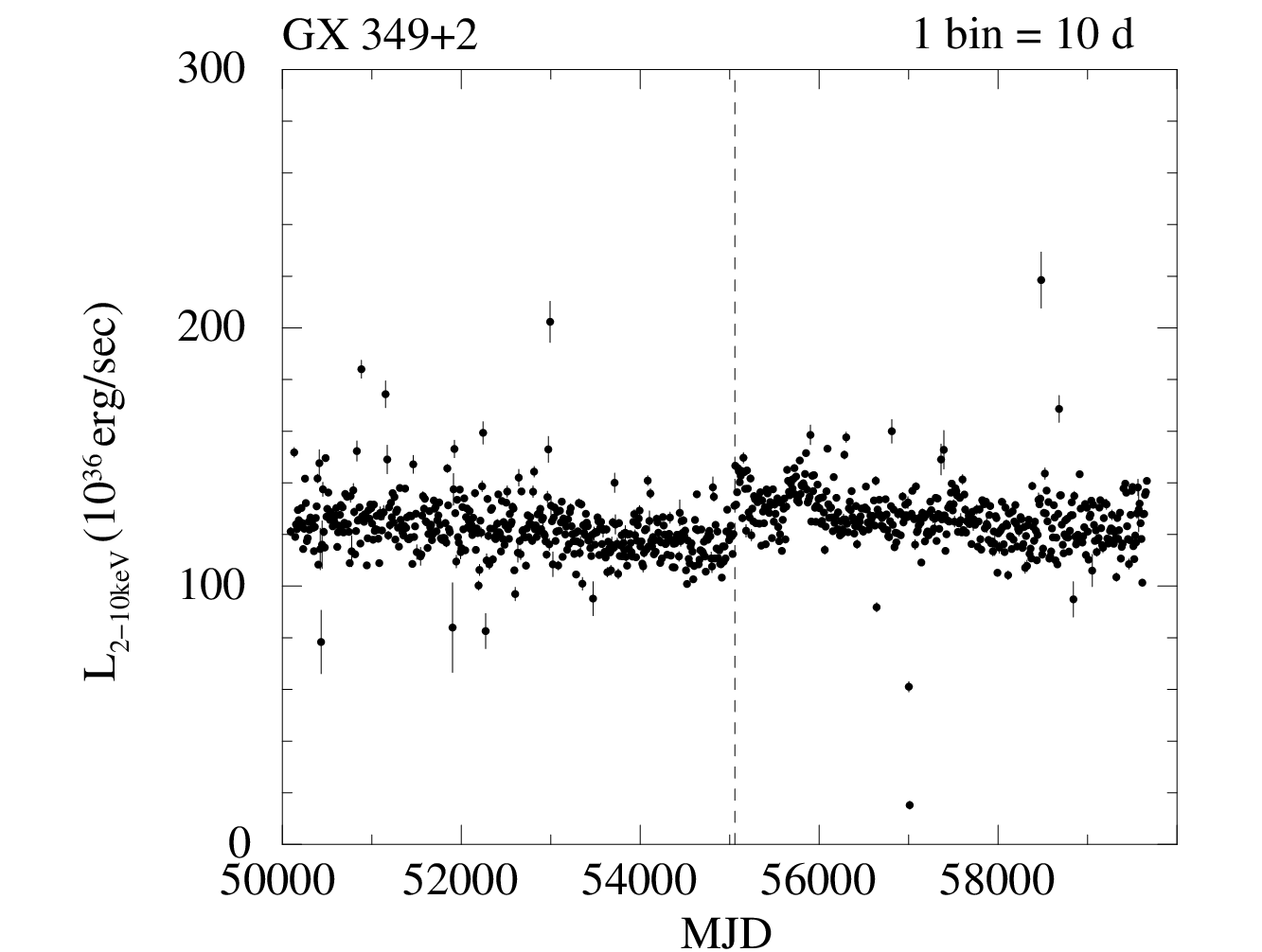}
  \includegraphics[width=75mm]{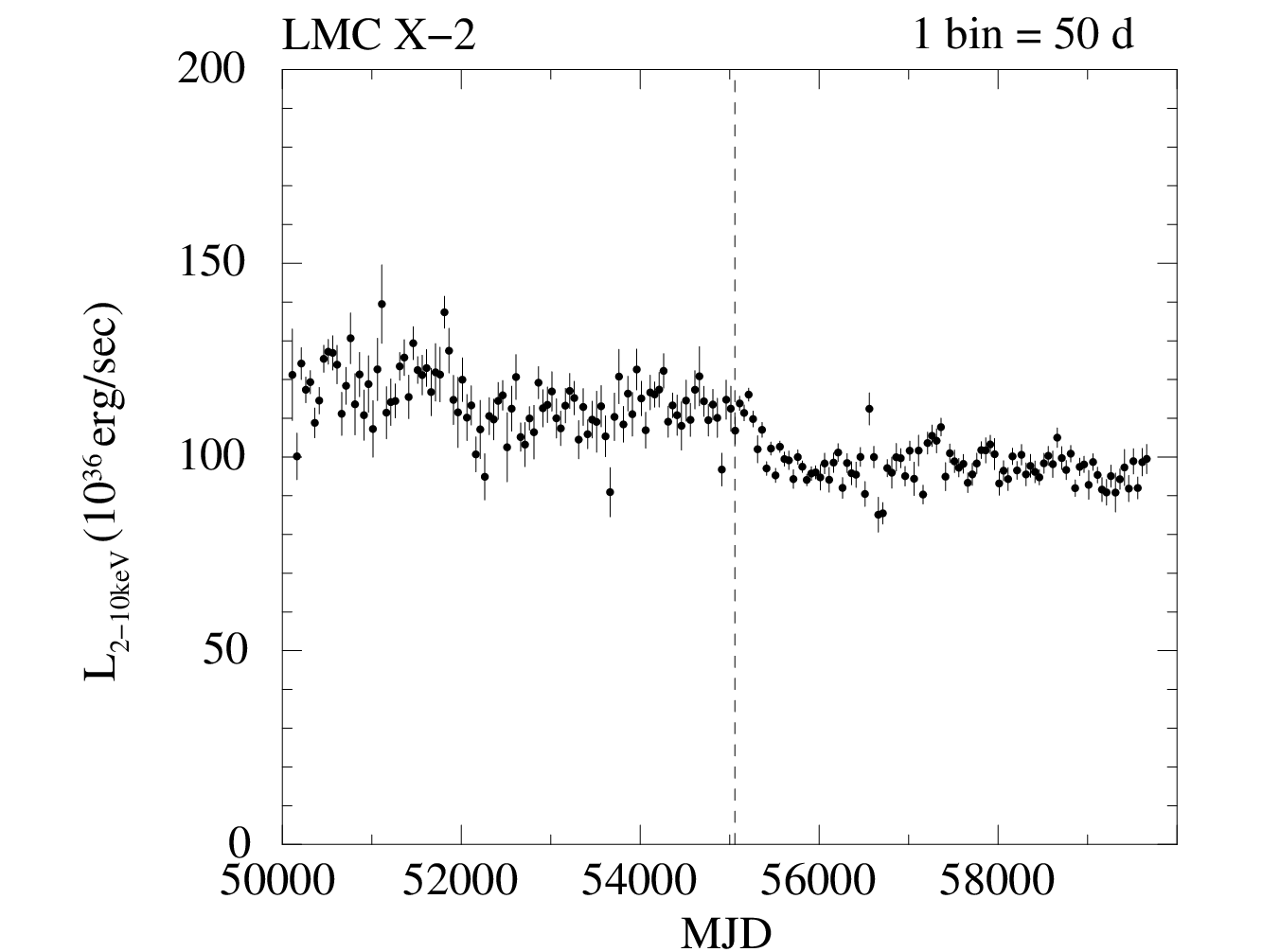}
 \includegraphics[width=75mm]{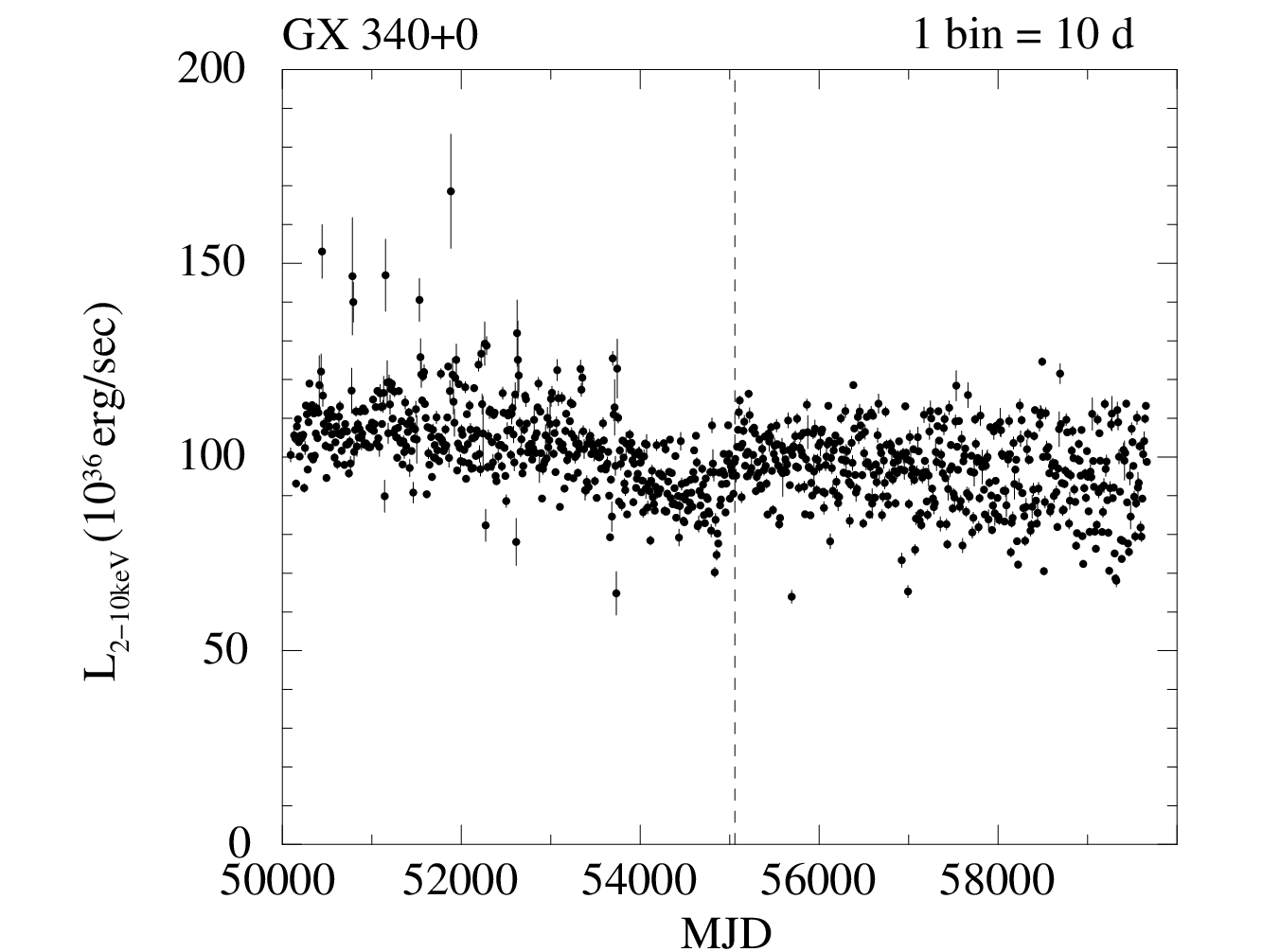}
 \includegraphics[width=75mm]{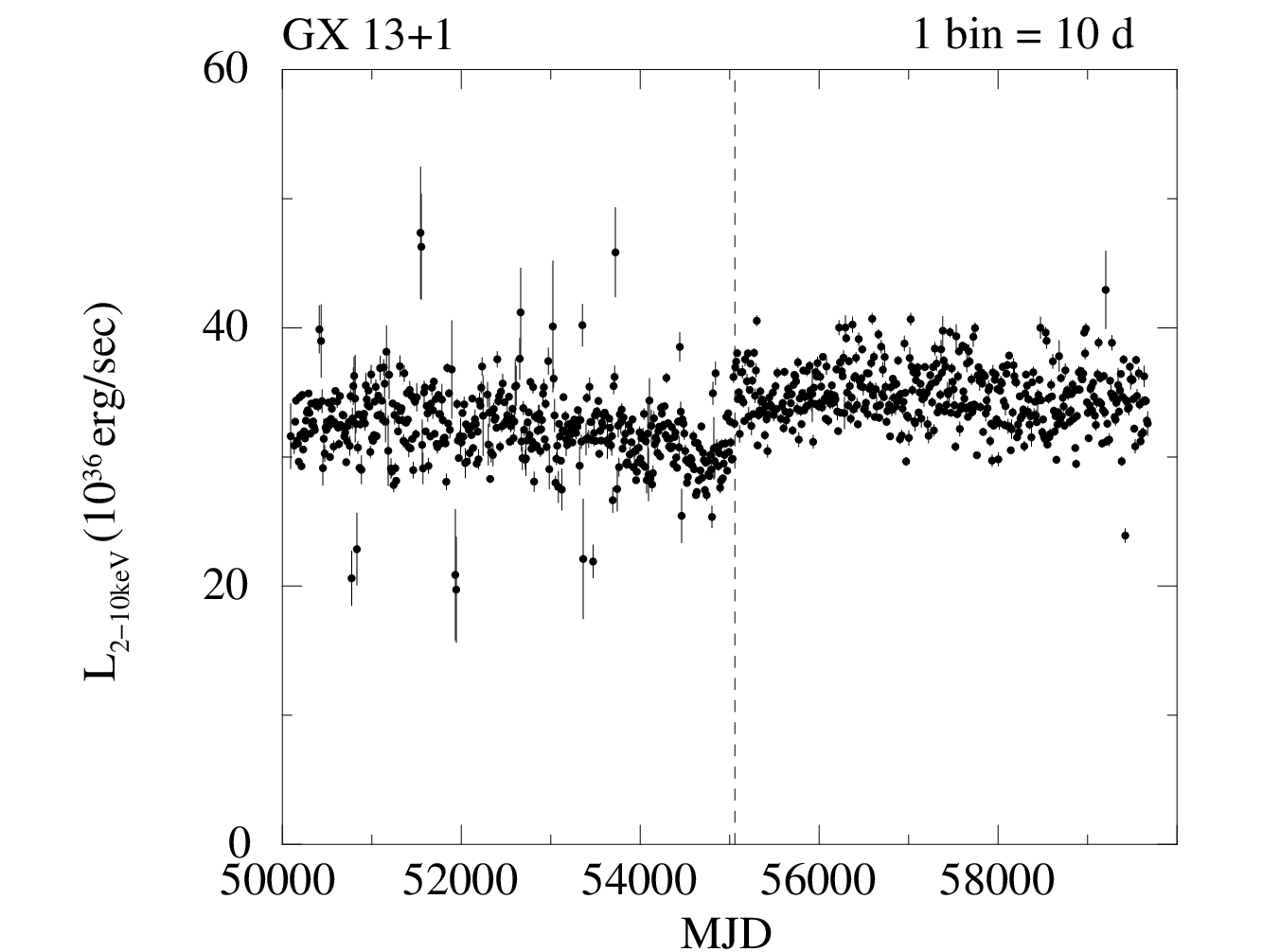}
 \end{center}
 \caption{Light curve in 2--10~keV of sources with
 Z sources and GX~13$+$1 (NP: no periodic variation).
The left side of the vertical dash line was the data of RXTE/ASM, and the right was MAXI/GSC. 
The MAXI/ASM flux ratio is not adjusted.}
\label{LP}
 \end{figure*}

\begin{figure*}
\begin{center}
 \includegraphics[width=75mm]{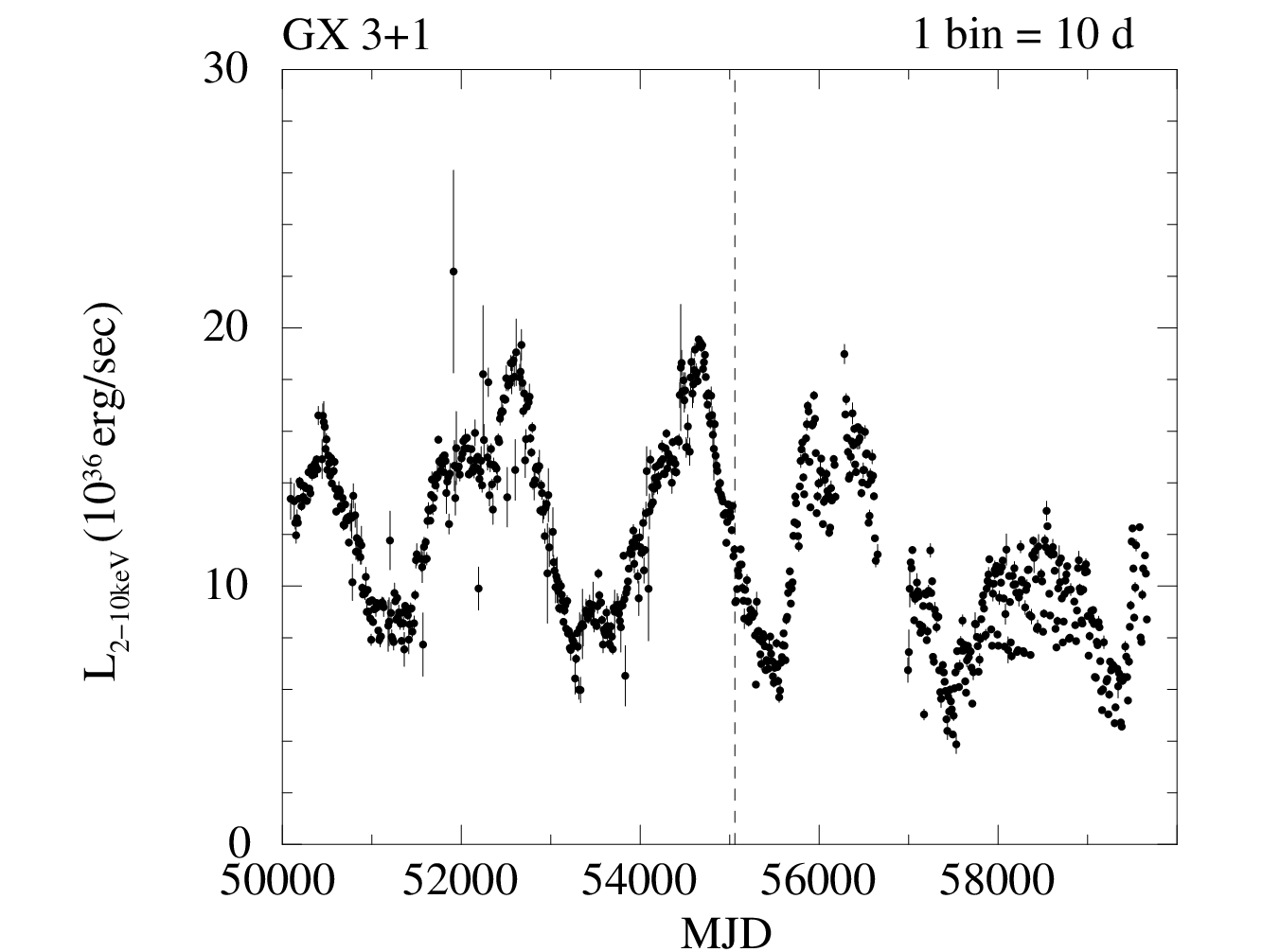}
 \includegraphics[width=75mm]{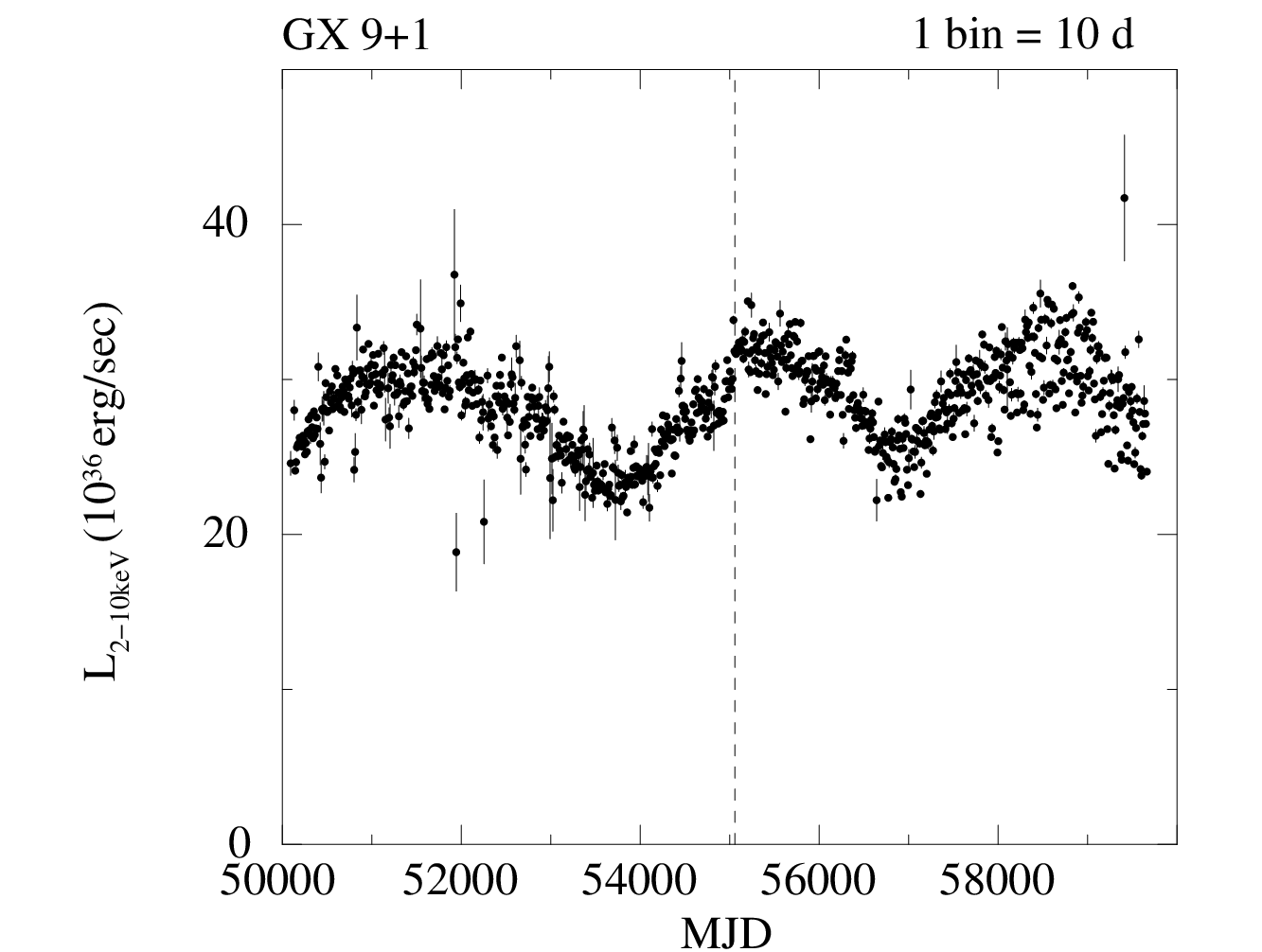}
  \end{center}
 \caption{As figure~\ref{LP}, but for CP (clear periodic variation) sources.}
 \label{CP}
 \end{figure*}
 
 \begin{figure*}
 \begin{center}
 \includegraphics[width=75mm]{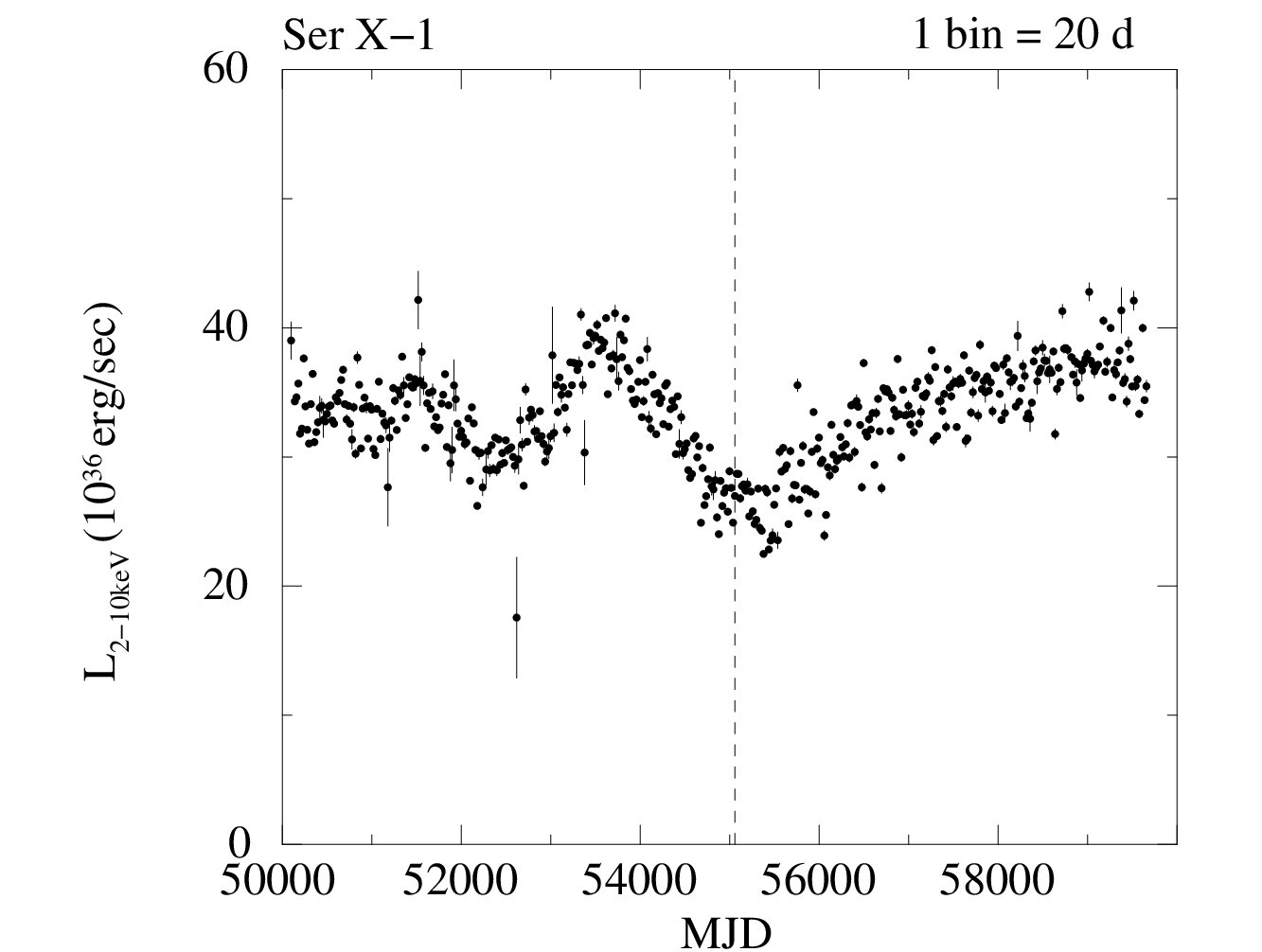}
 \includegraphics[width=75mm]{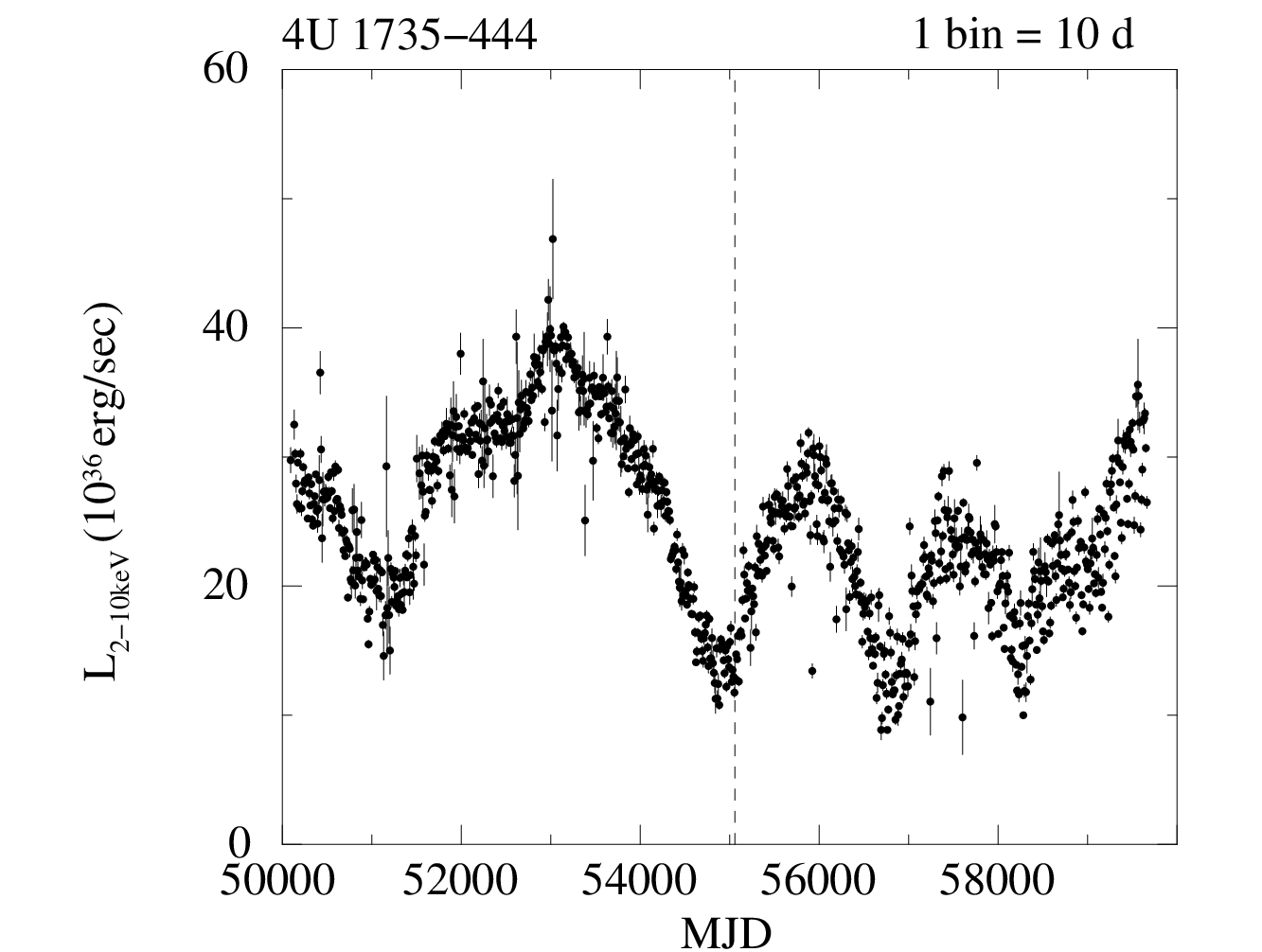}
 \includegraphics[width=75mm]{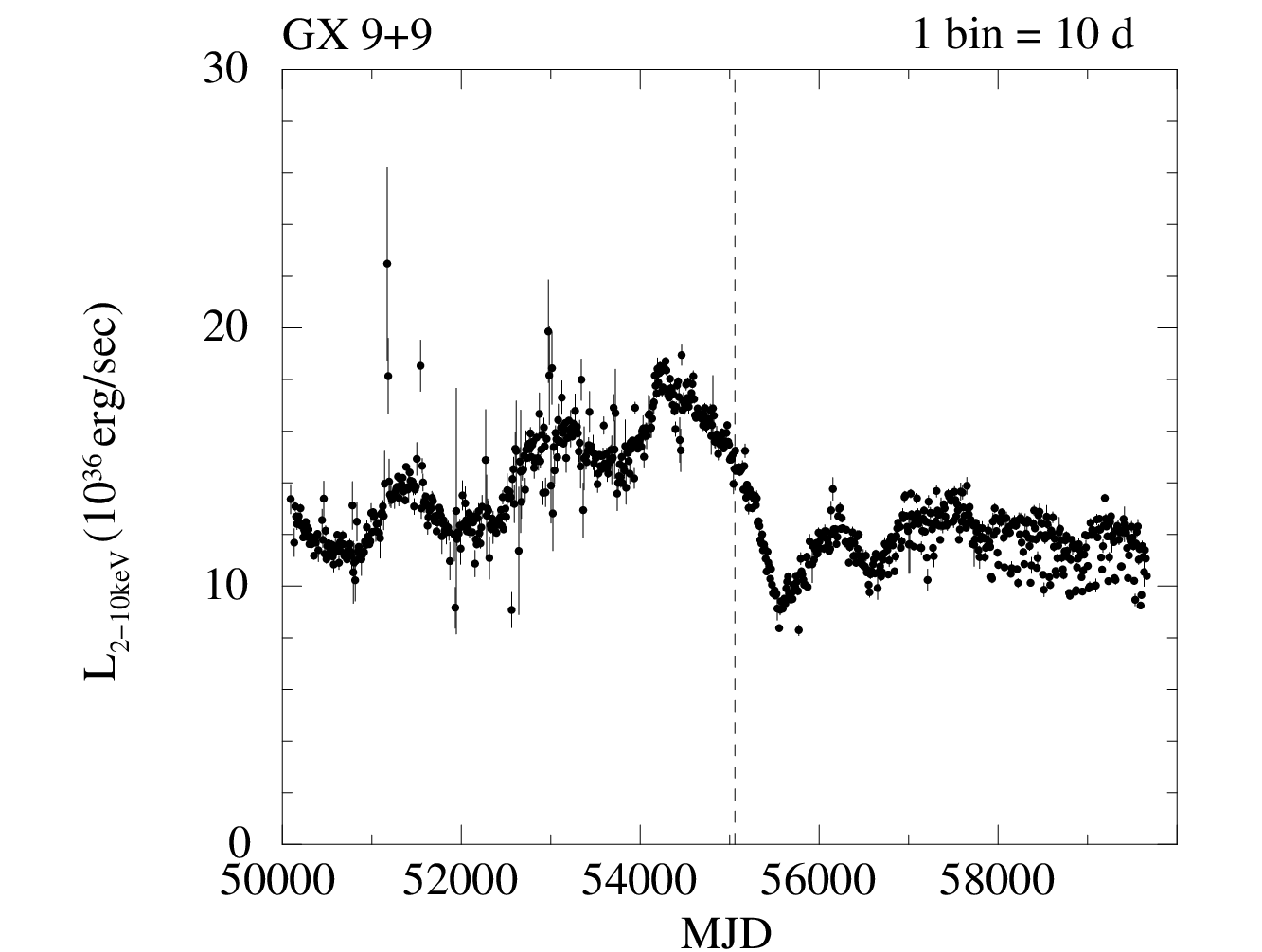}
 \includegraphics[width=75mm]{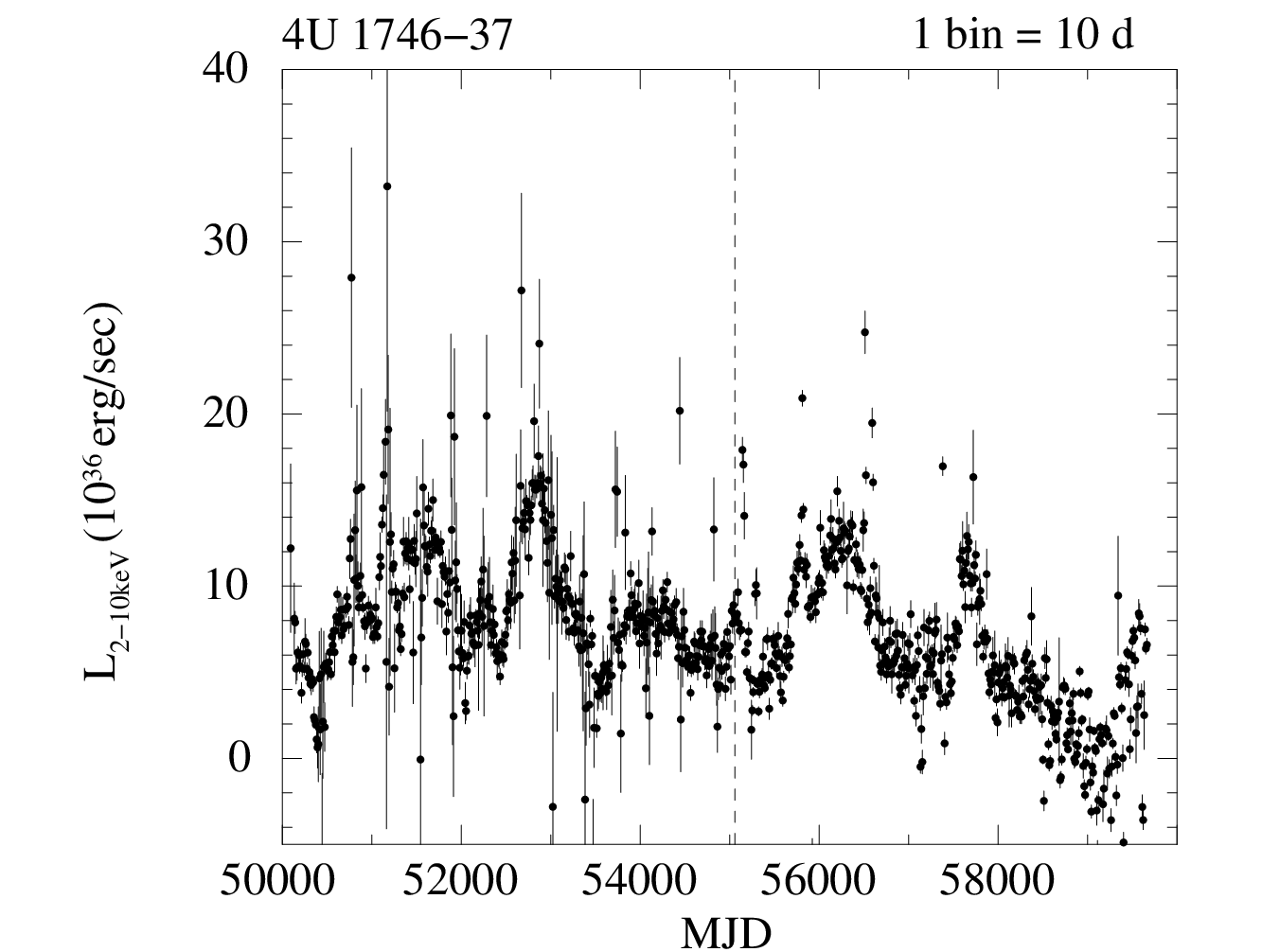}
 \includegraphics[width=75mm]{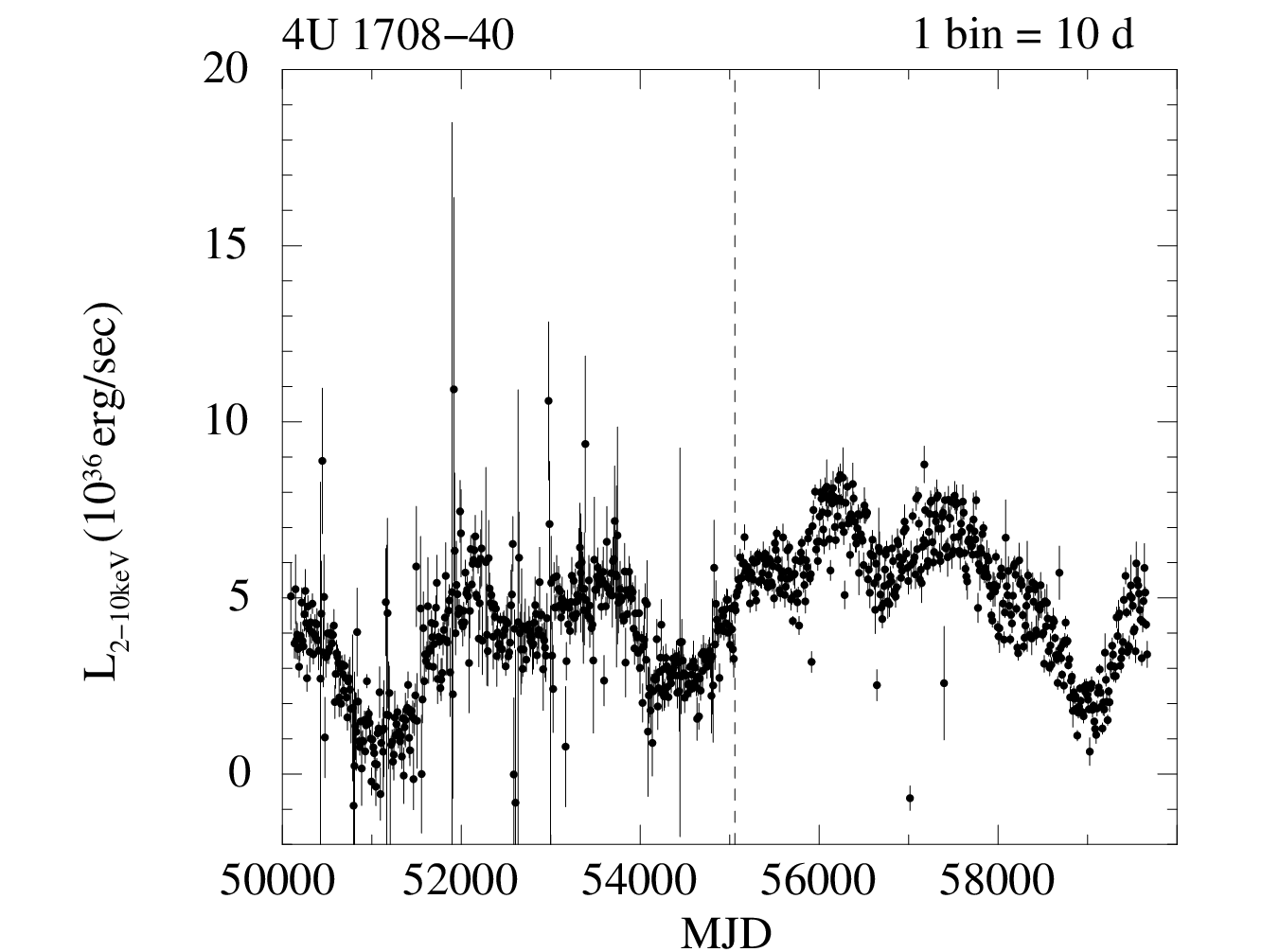}  \includegraphics[width=75mm]{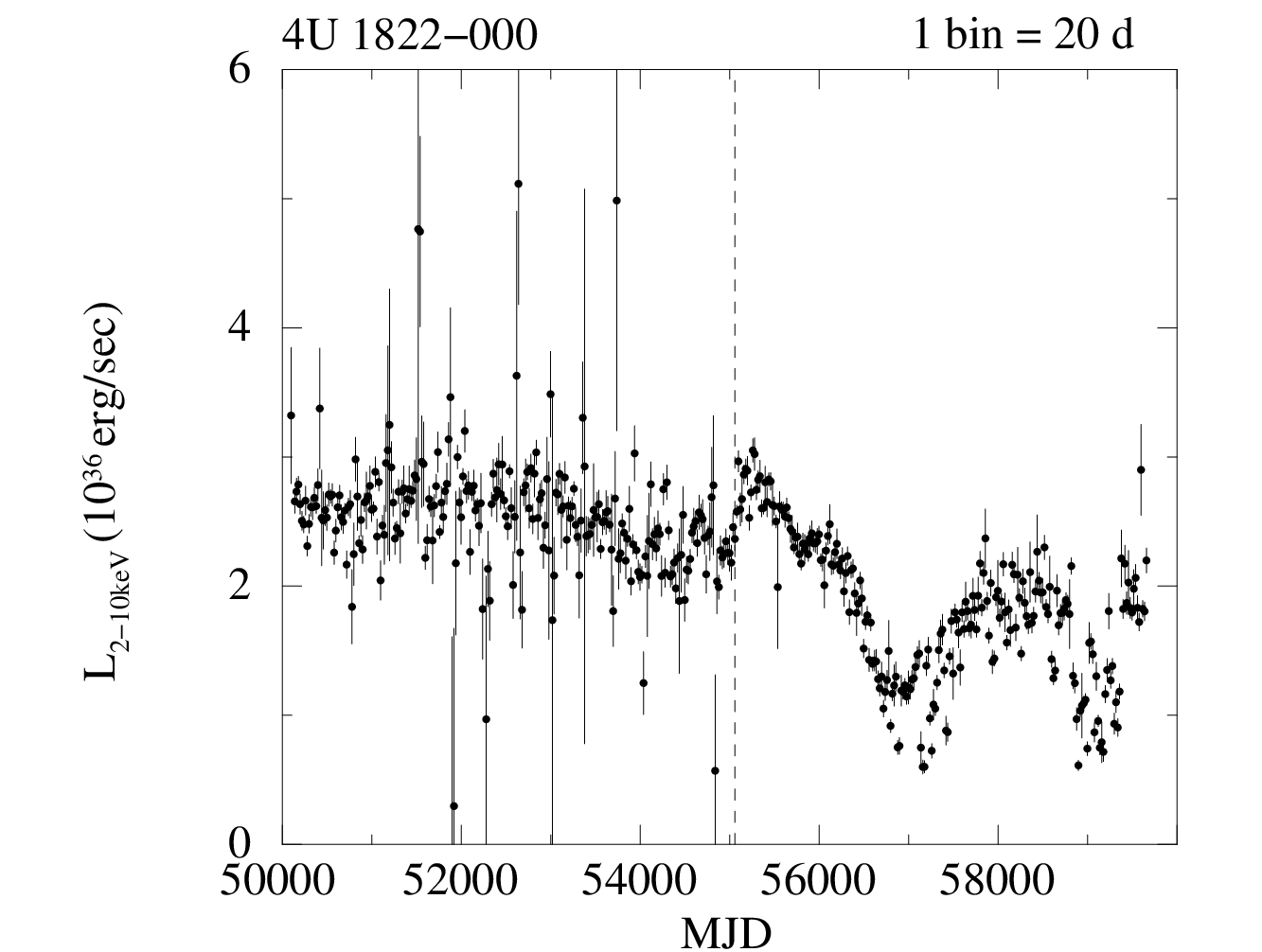}  \includegraphics[width=75mm]{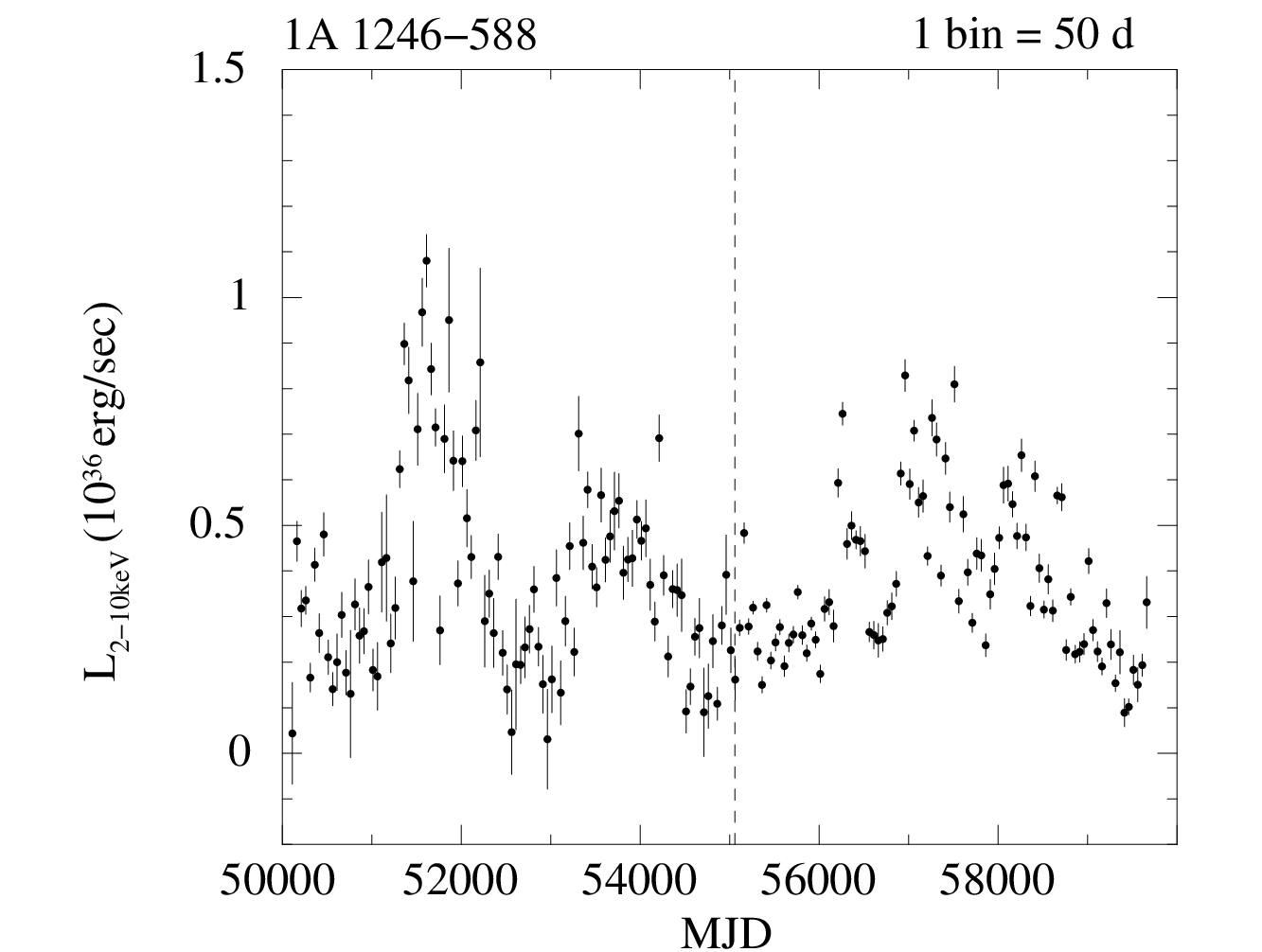}
 \end{center}
 \caption{As figure~\ref{LP}, but for MP (modified periodic variation) sources.}
 \label{MP}
\end{figure*}

\begin{figure*}
\begin{center}
 \includegraphics[width=75mm]{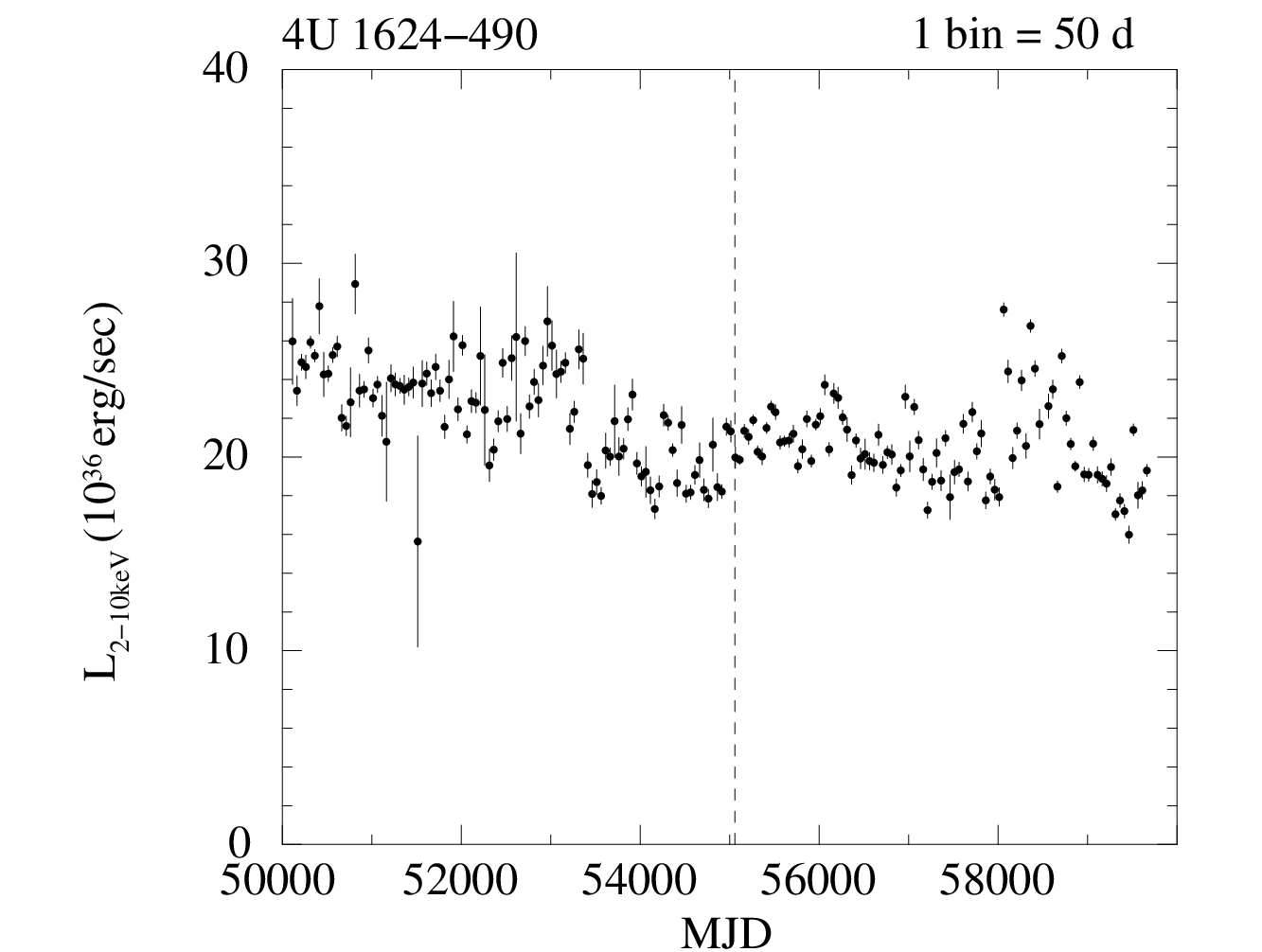}  \includegraphics[width=75mm]{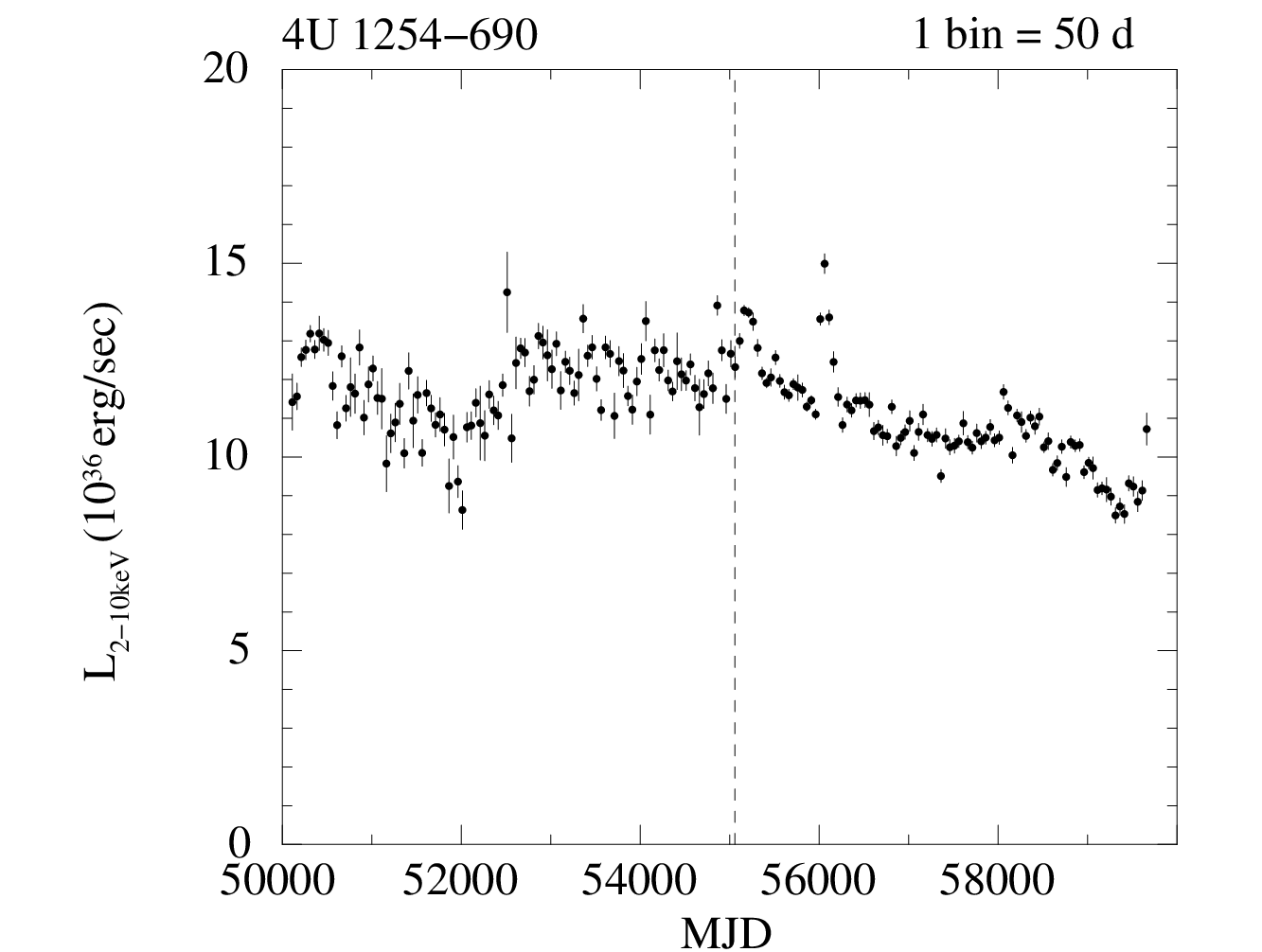}  \includegraphics[width=75mm]{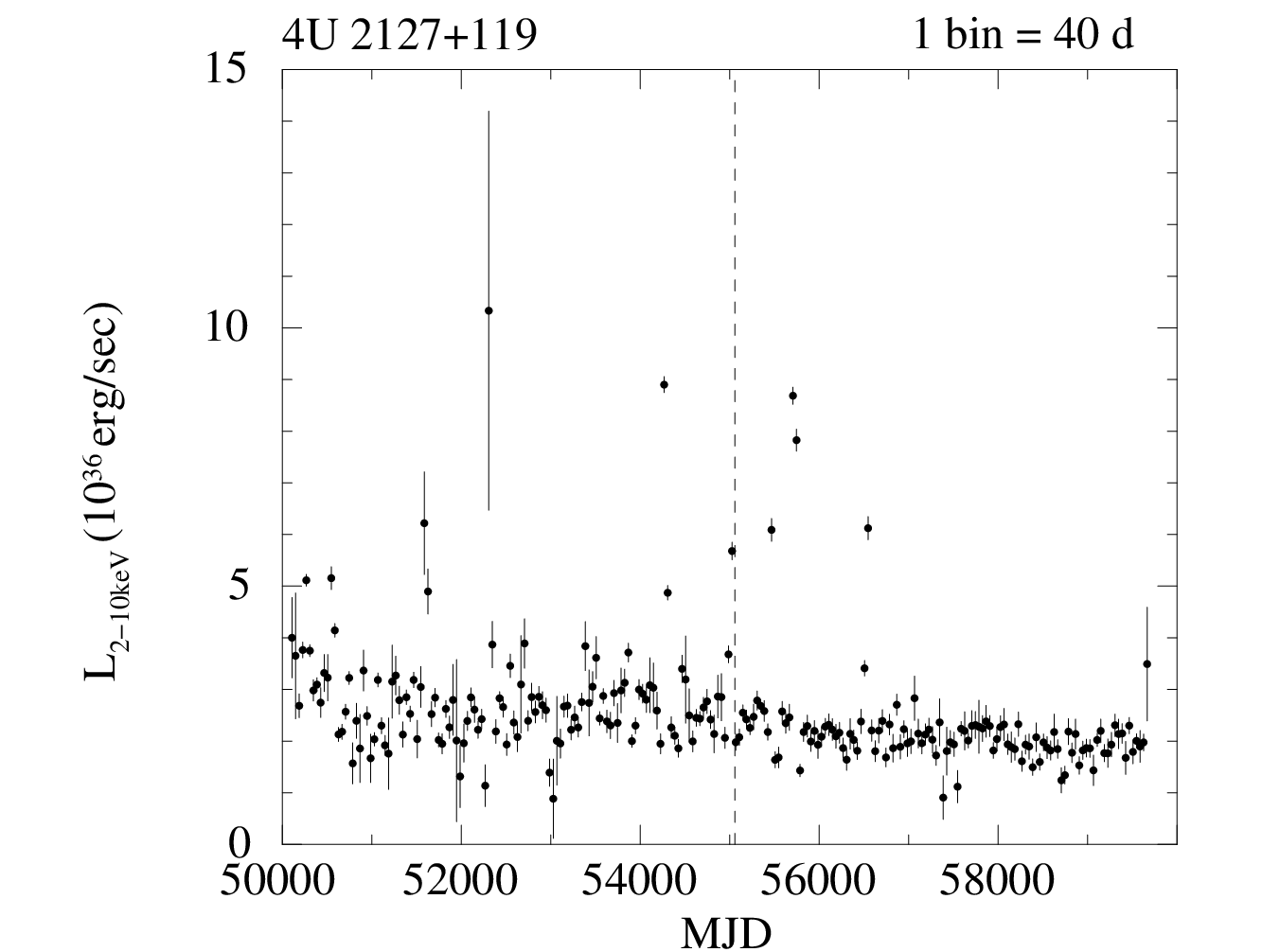}
 \includegraphics[width=75mm]{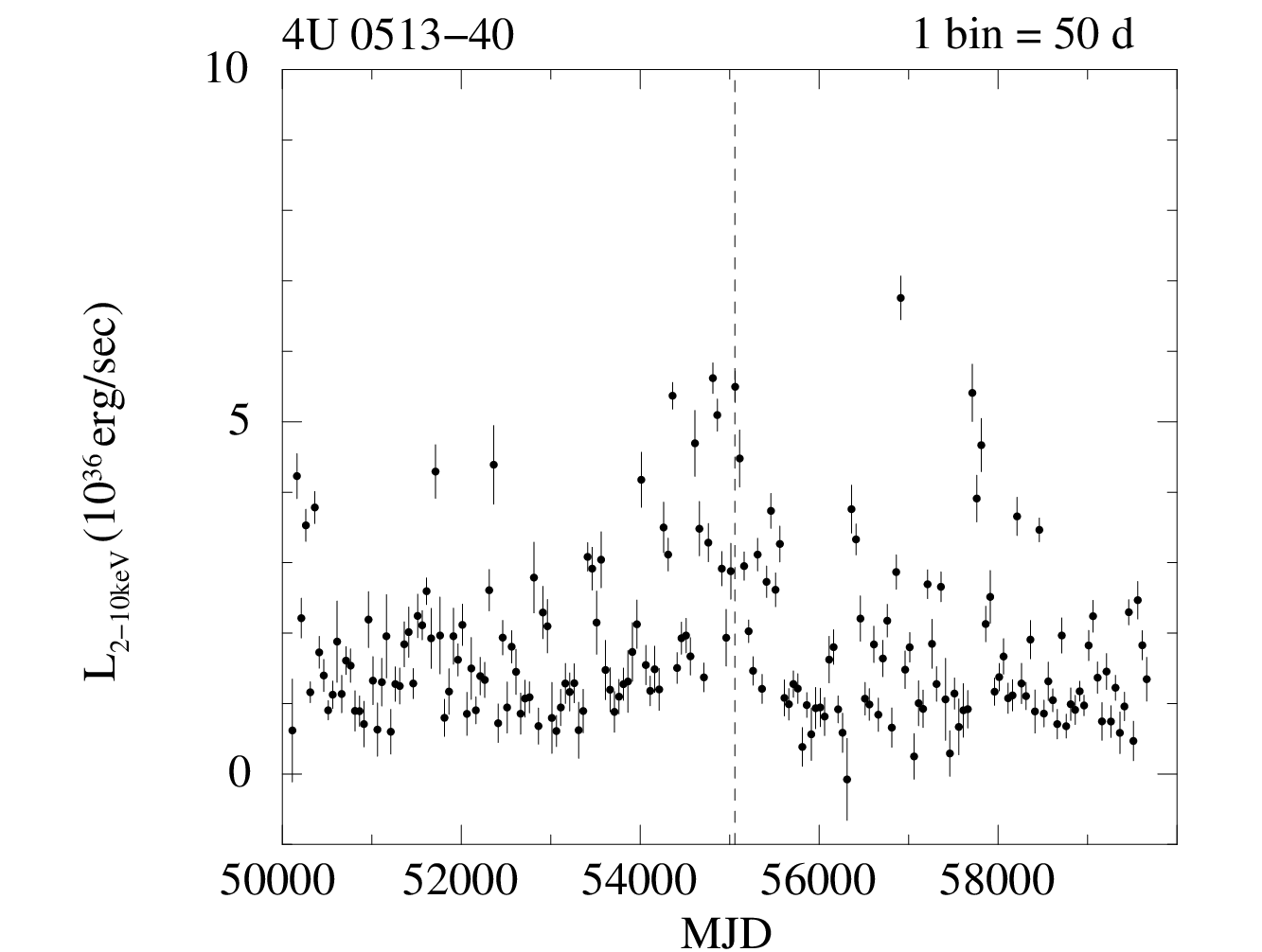}  \includegraphics[width=75mm]{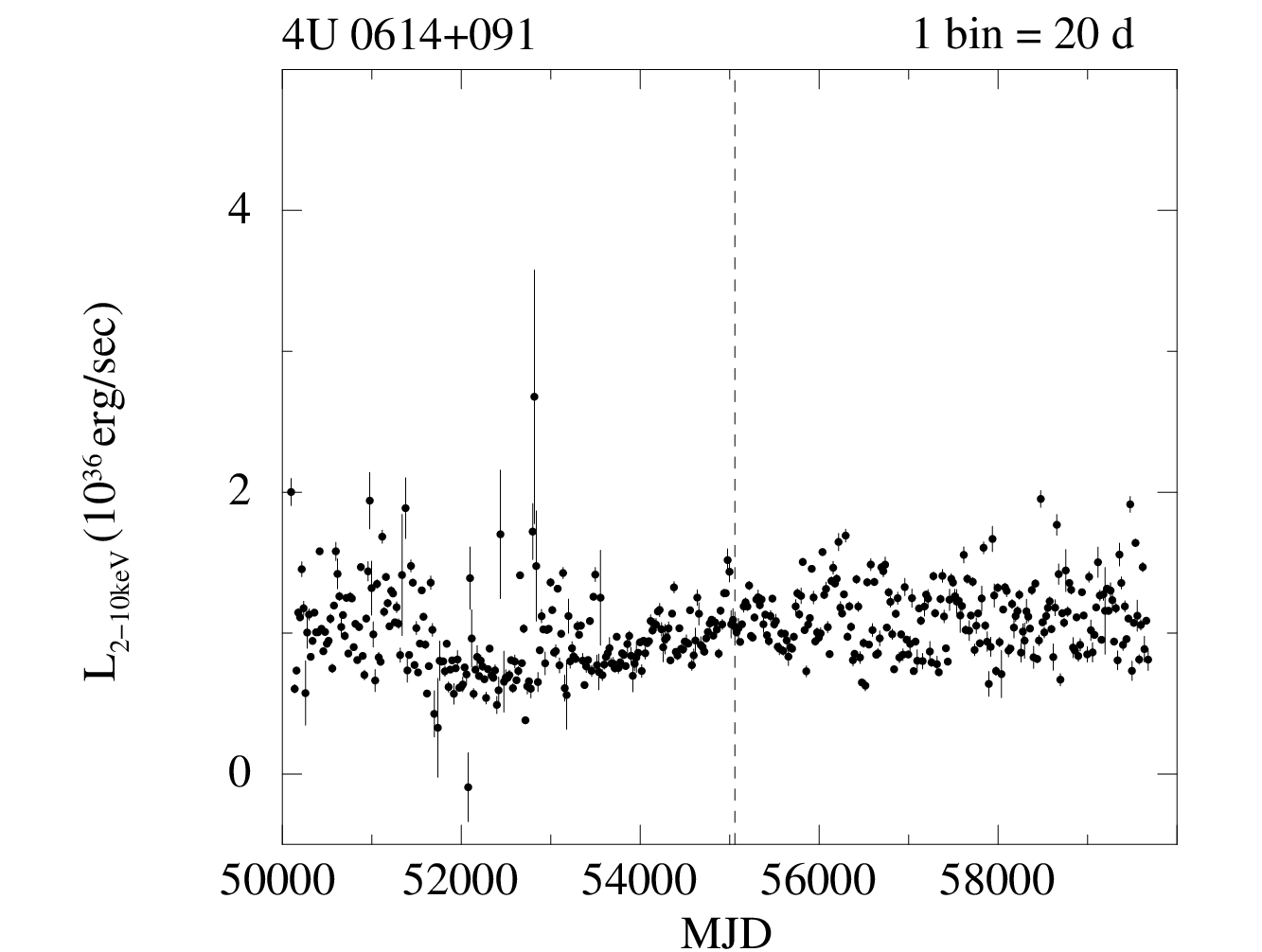} 
 \includegraphics[width=75mm]{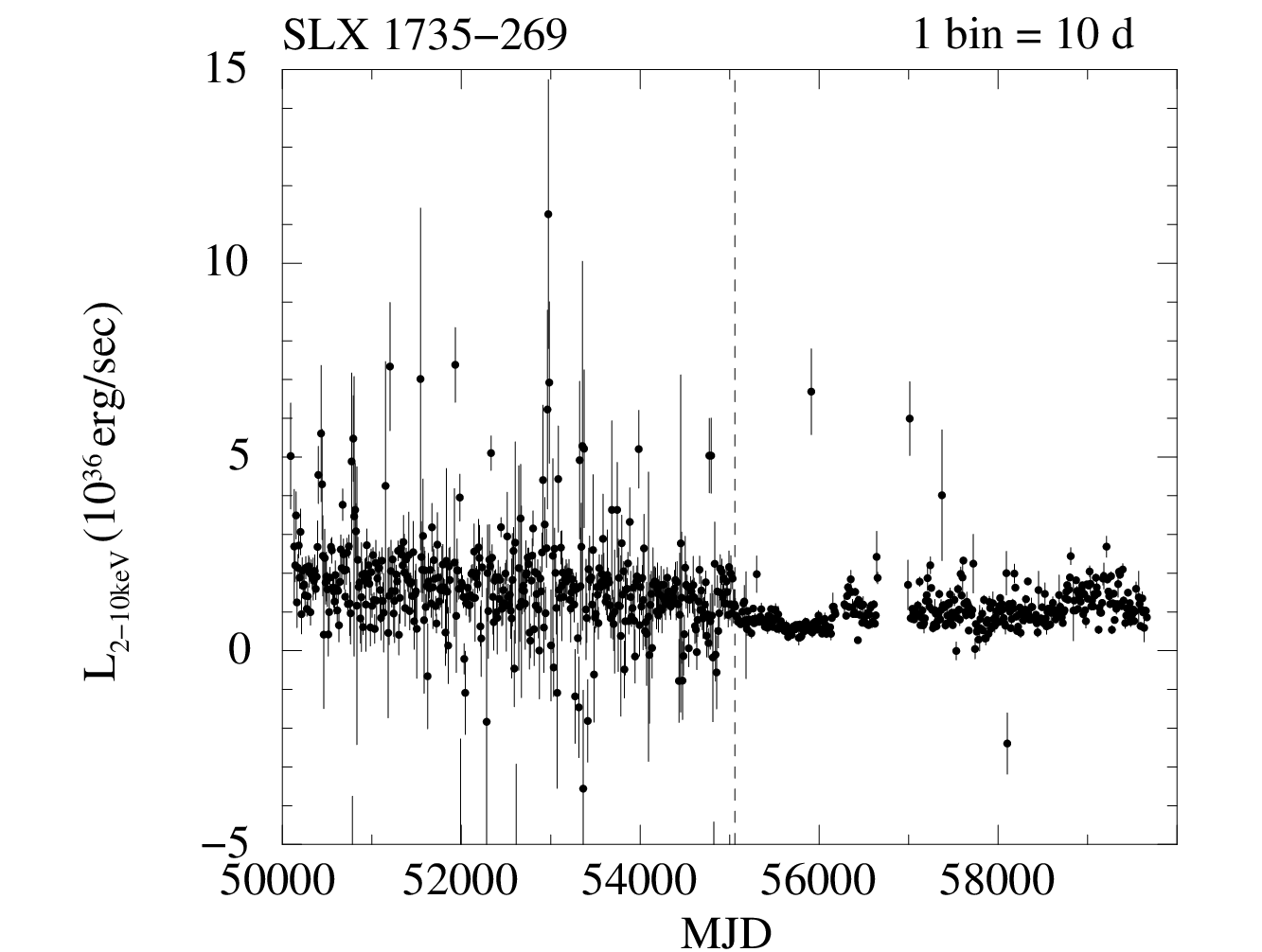} \includegraphics[width=75mm]{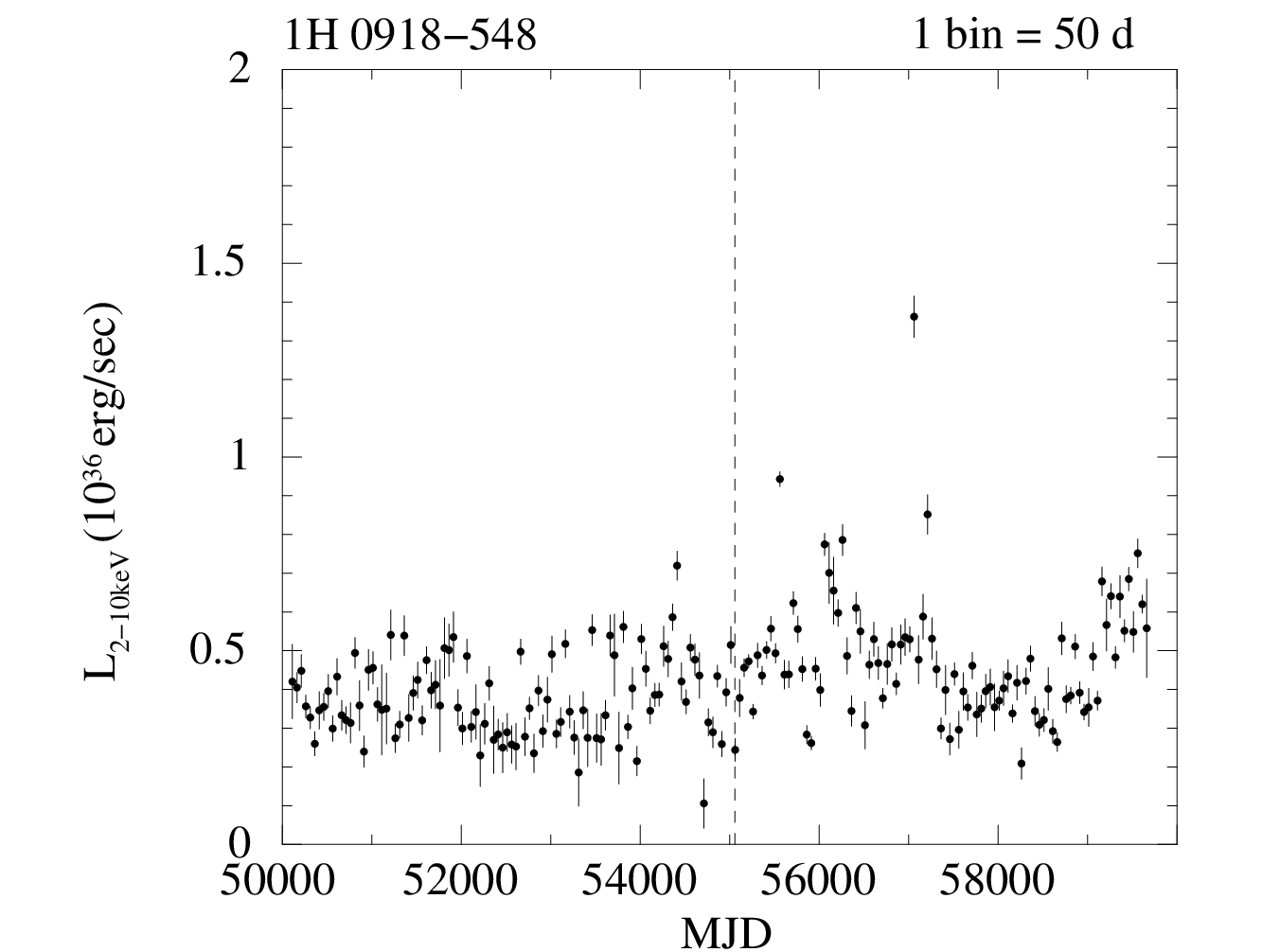} \includegraphics[width=75mm]{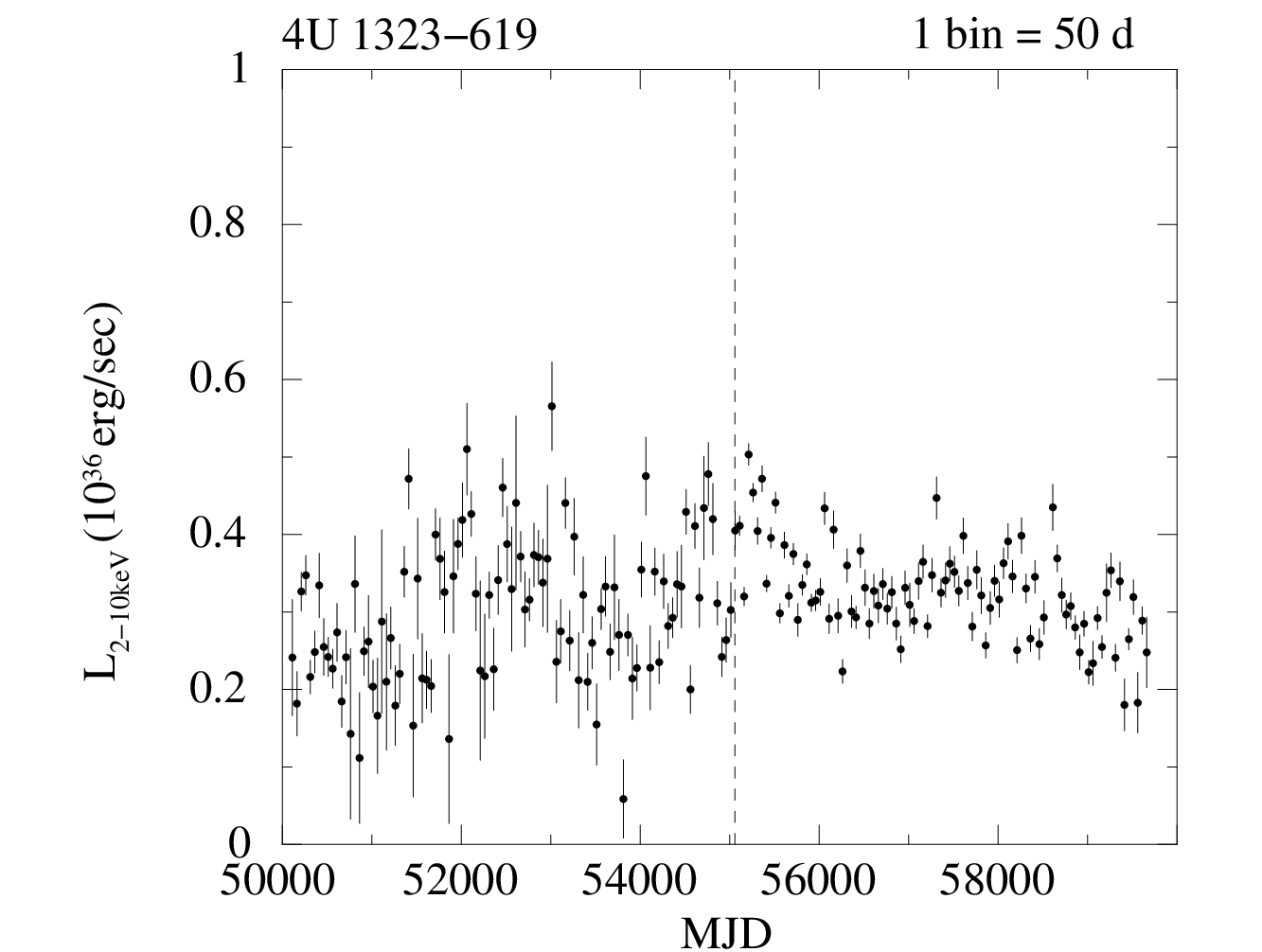} 
 \end{center}
 \caption{As figure~\ref{LP}, but for NP (no periodic variation) and almost constant sources.}
 \label{NP-cons}
 \end{figure*}
 
\begin{figure*}
\begin{center}
 \includegraphics[width=75mm]{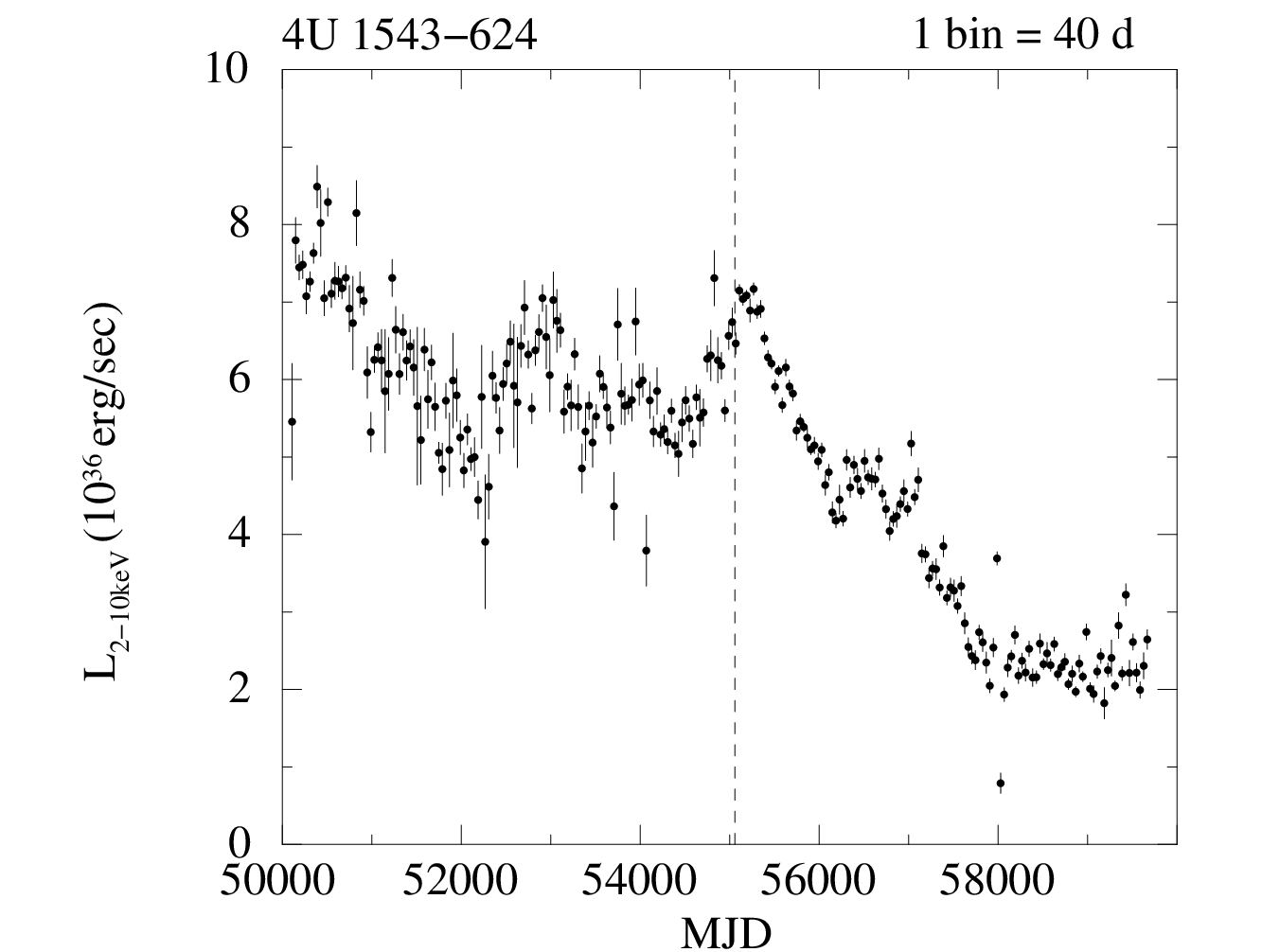}
 \includegraphics[width=75mm]{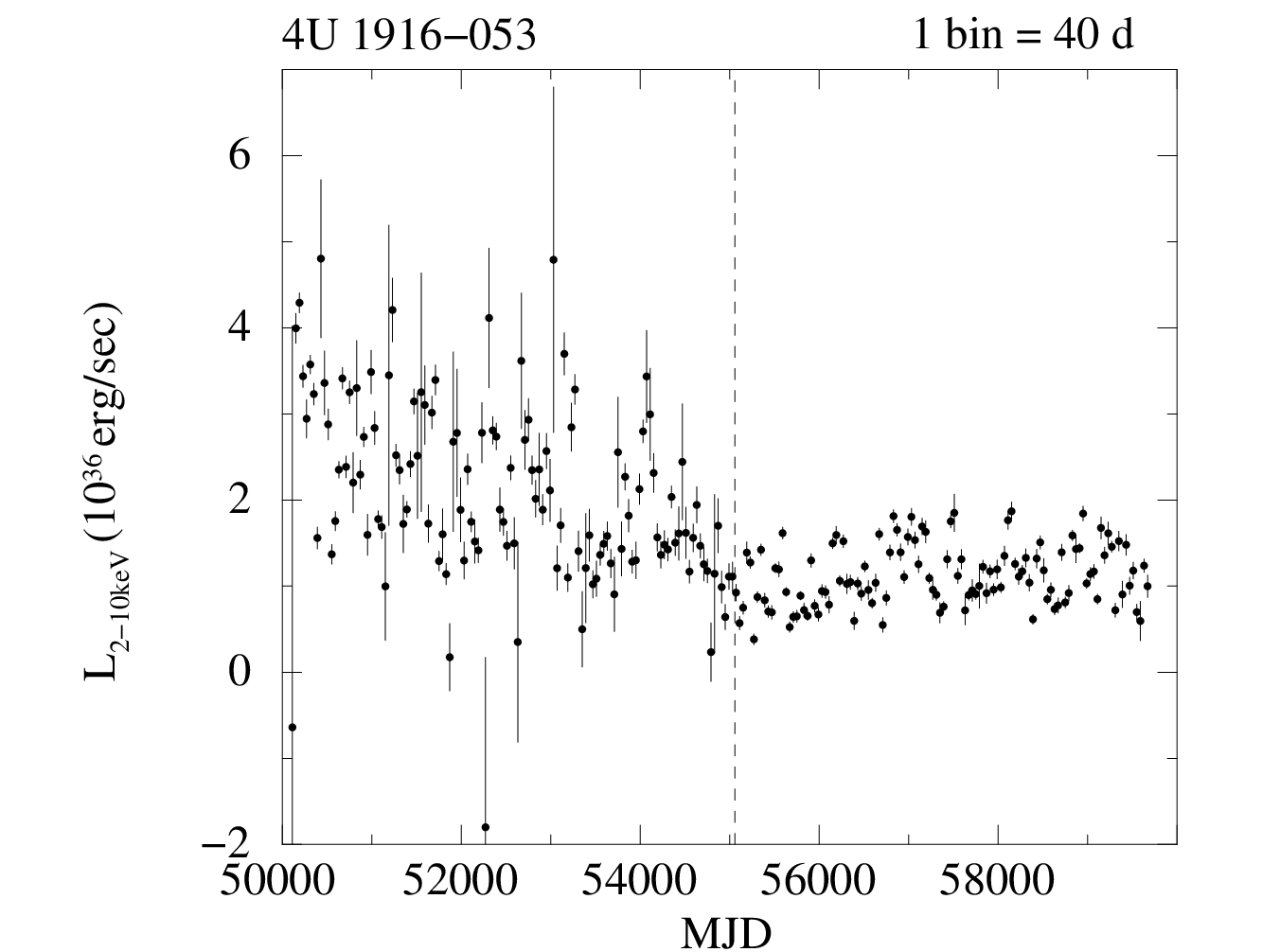} 
 \includegraphics[width=75mm]{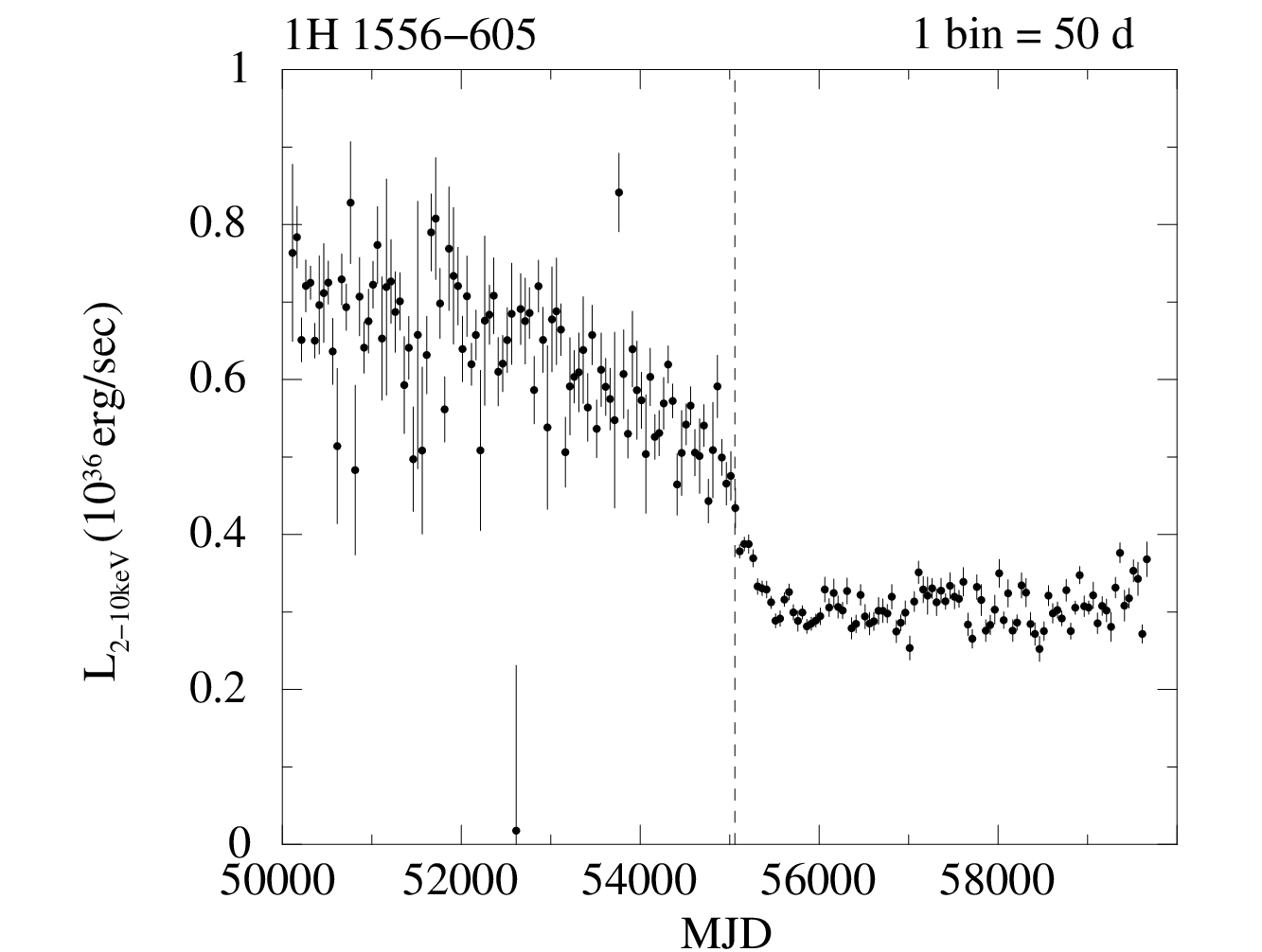}
 \end{center}
 \caption{As figure~\ref{LP}, but for NP (no periodic variation) and decrease sources.}
 \label{NP-dec}
 \end{figure*}
 
\begin{figure*}
\begin{center}
 \includegraphics[width=75mm]{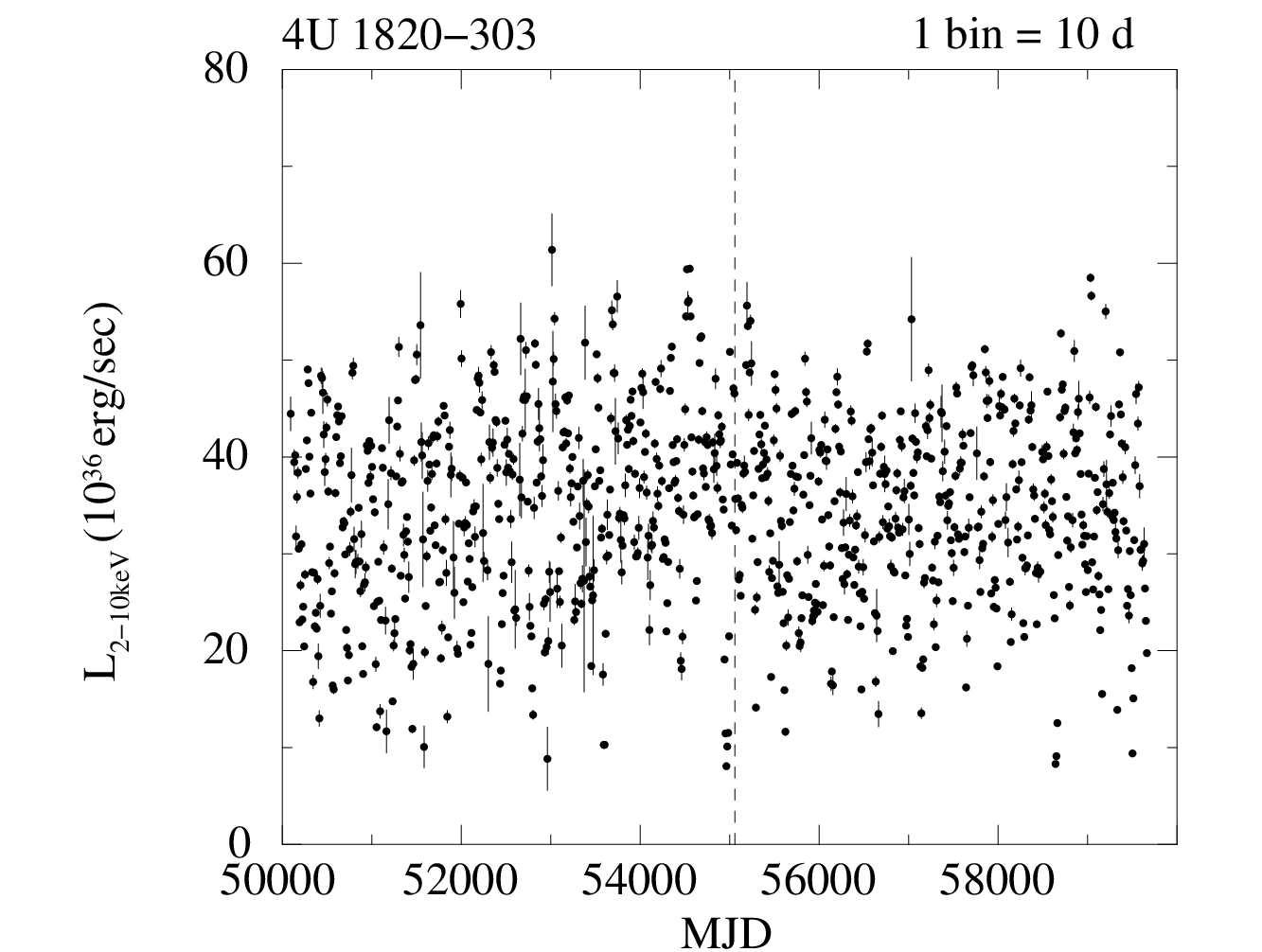}
 \includegraphics[width=75mm]{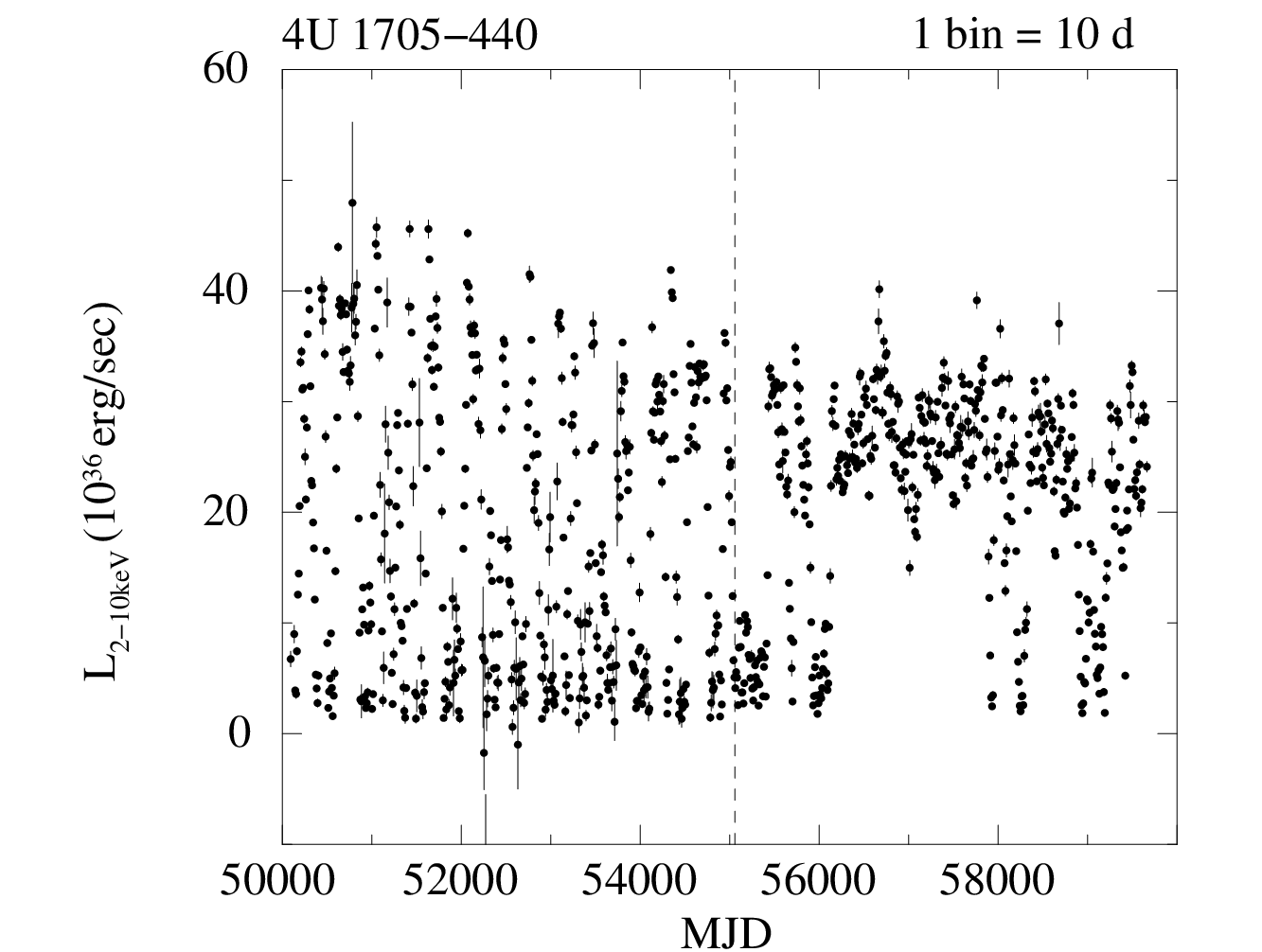}
 \end{center}
 \caption{As figure~\ref{LP}, but for FV (fast variability) sources.}
 \label{FV}
\end{figure*}

\begin{figure*}
\begin{center}
  \includegraphics[width=75mm]{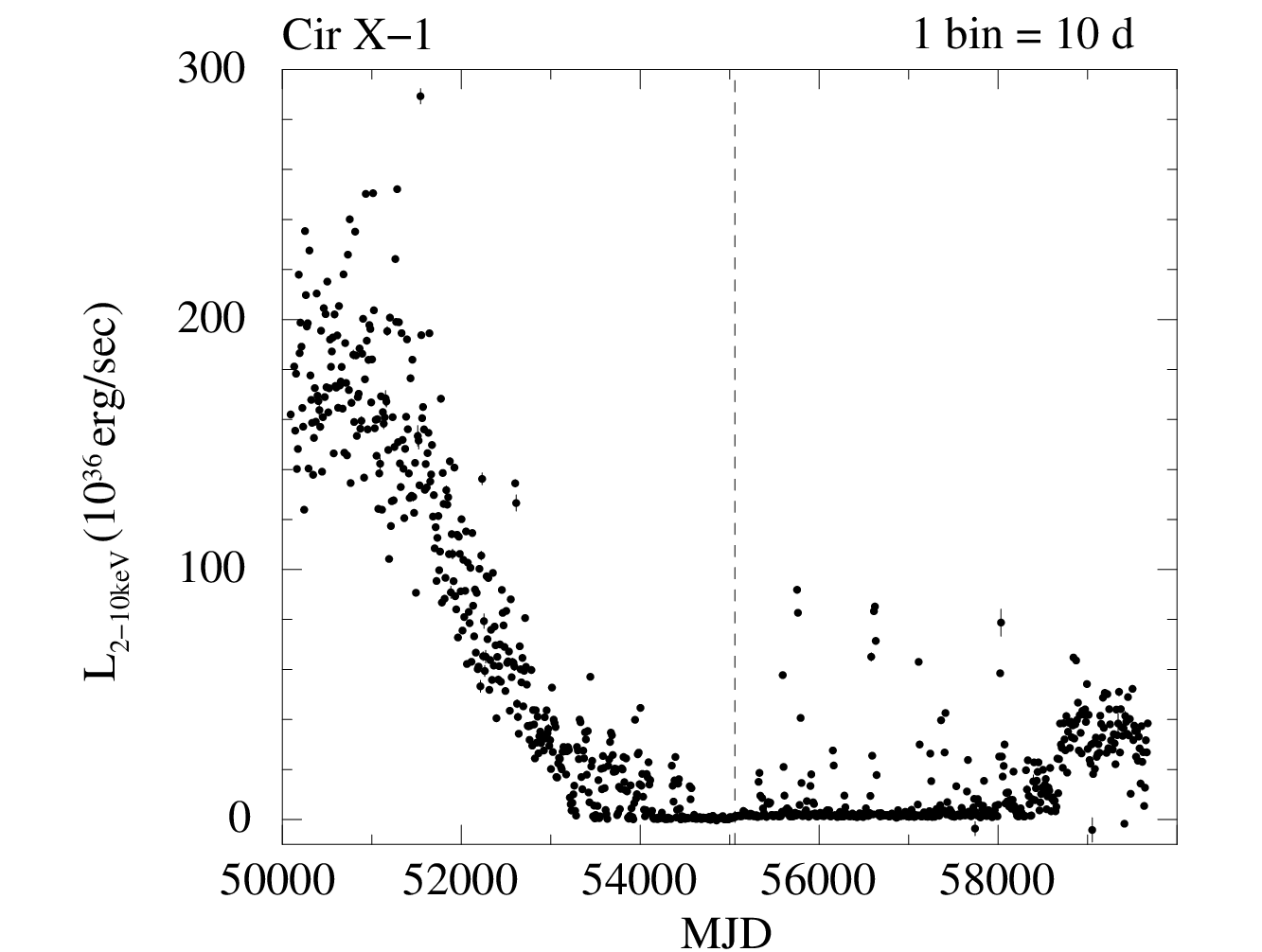} 
  \includegraphics[width=75mm]{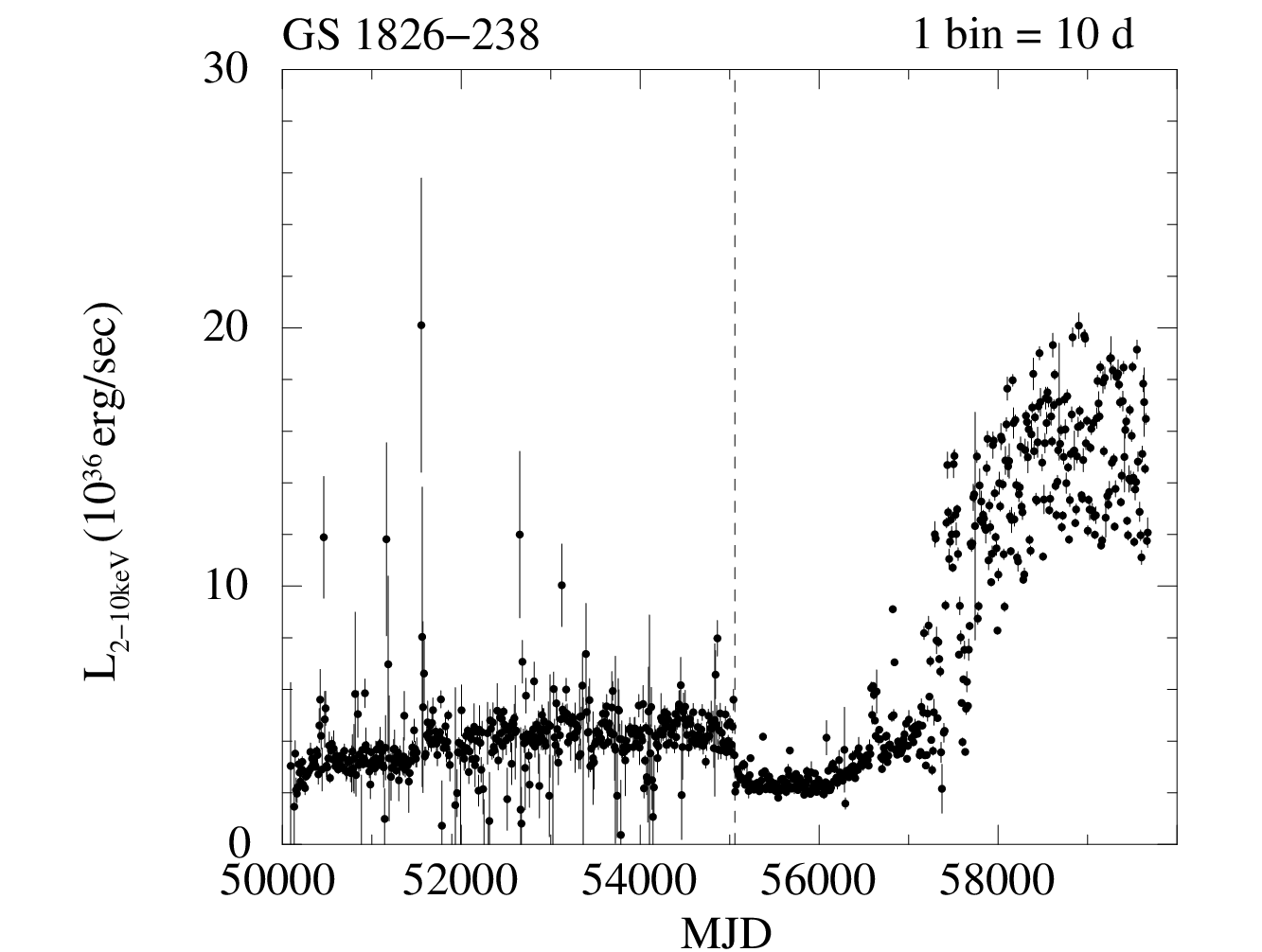} 
  \includegraphics[width=75mm]{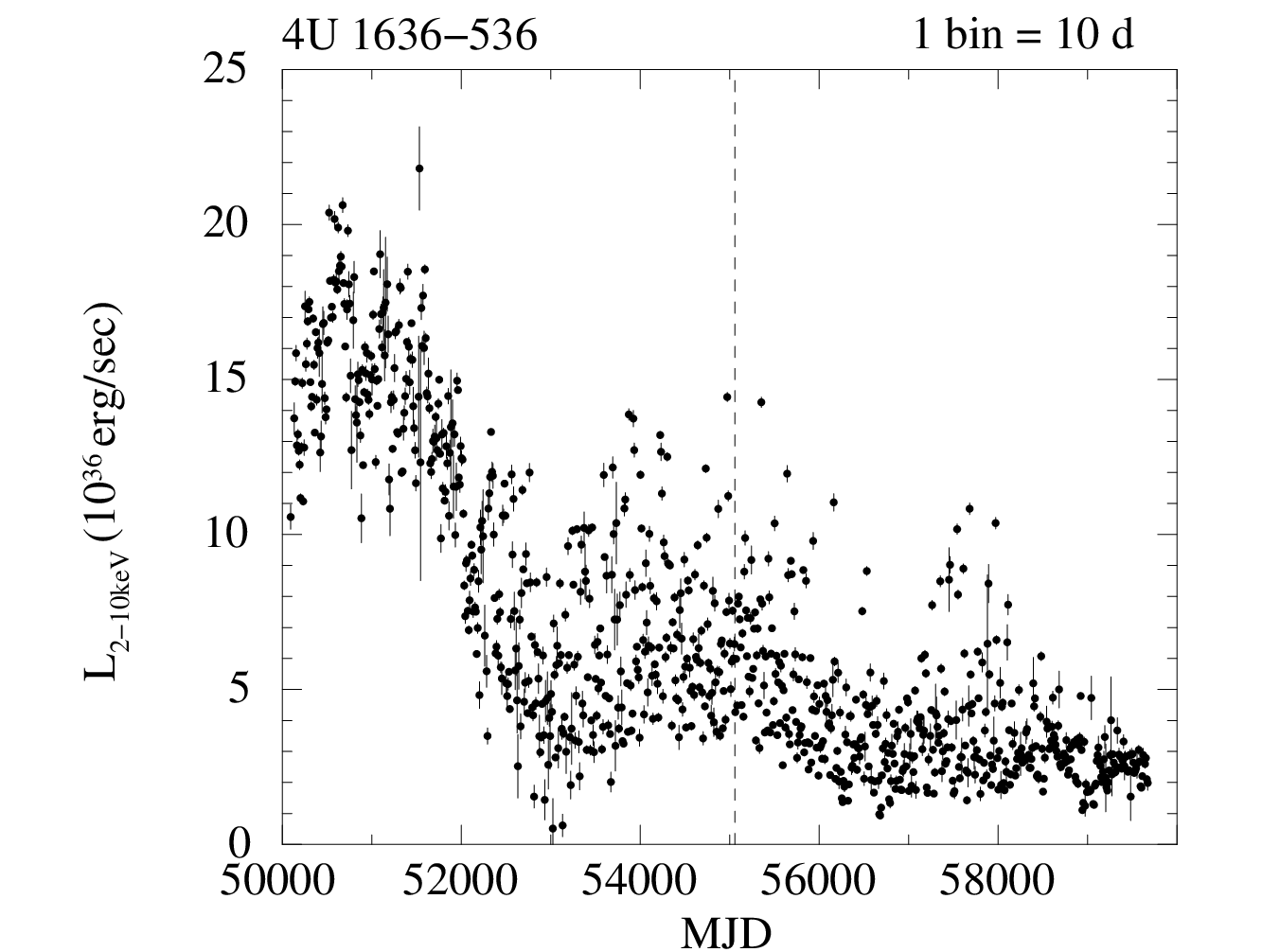}
\end{center}
\caption{As figure~\ref{LP}, but for LV (large variability) sources.}
\label{LV}
\end{figure*}

\begin{figure*}
\begin{center}
\includegraphics[width=75mm]{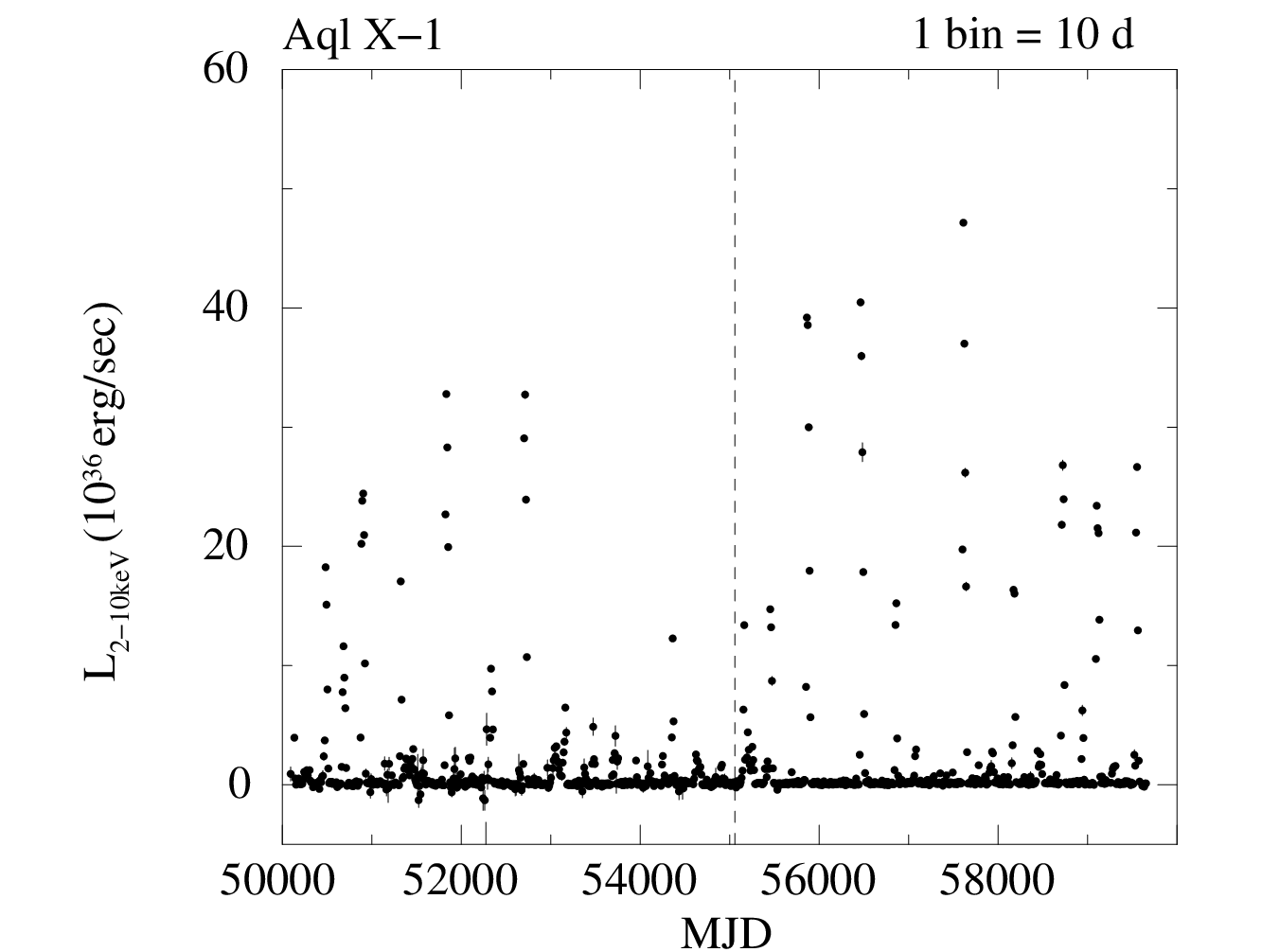}
 \includegraphics[width=75mm]{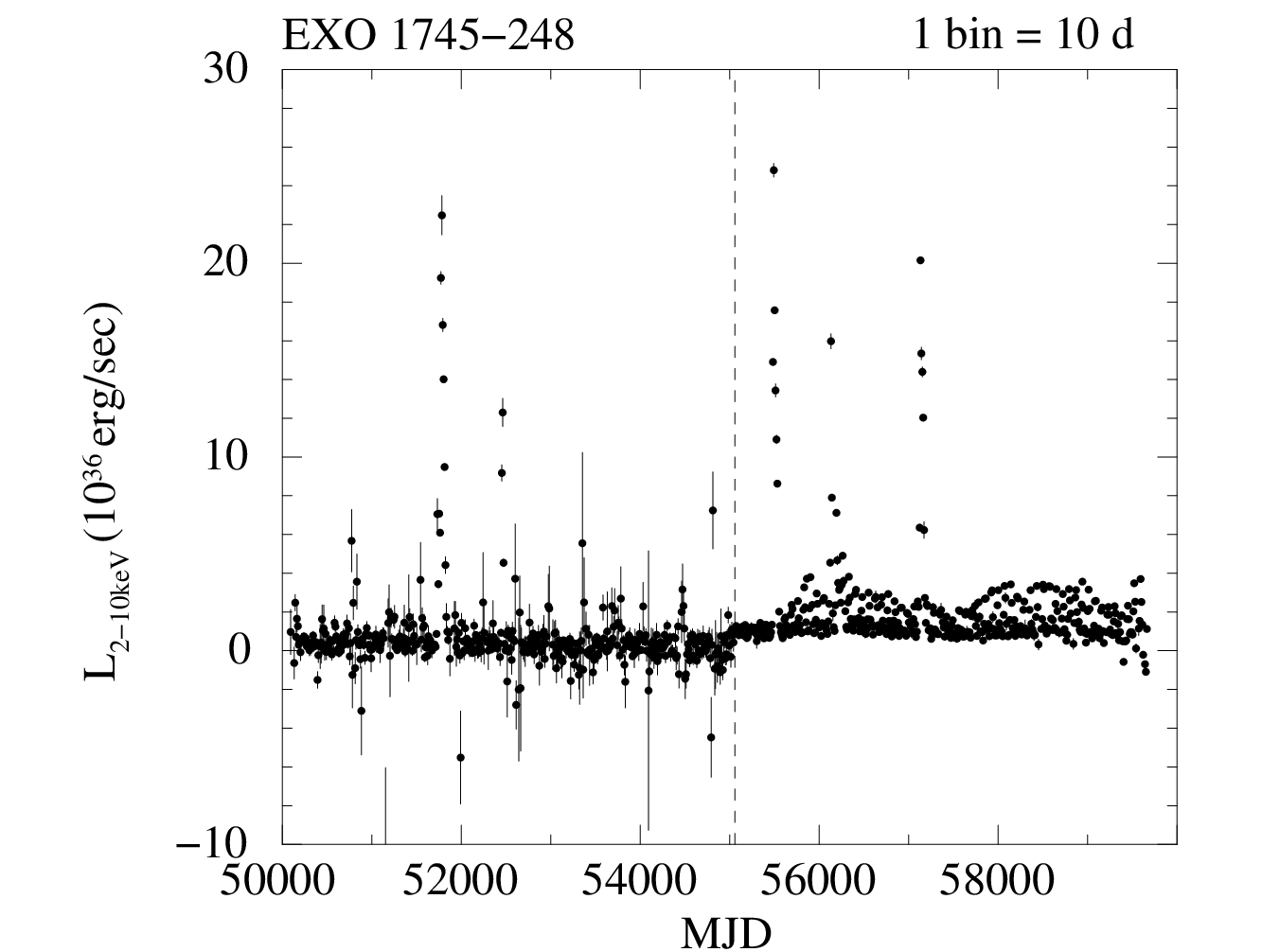}
 \includegraphics[width=75mm]{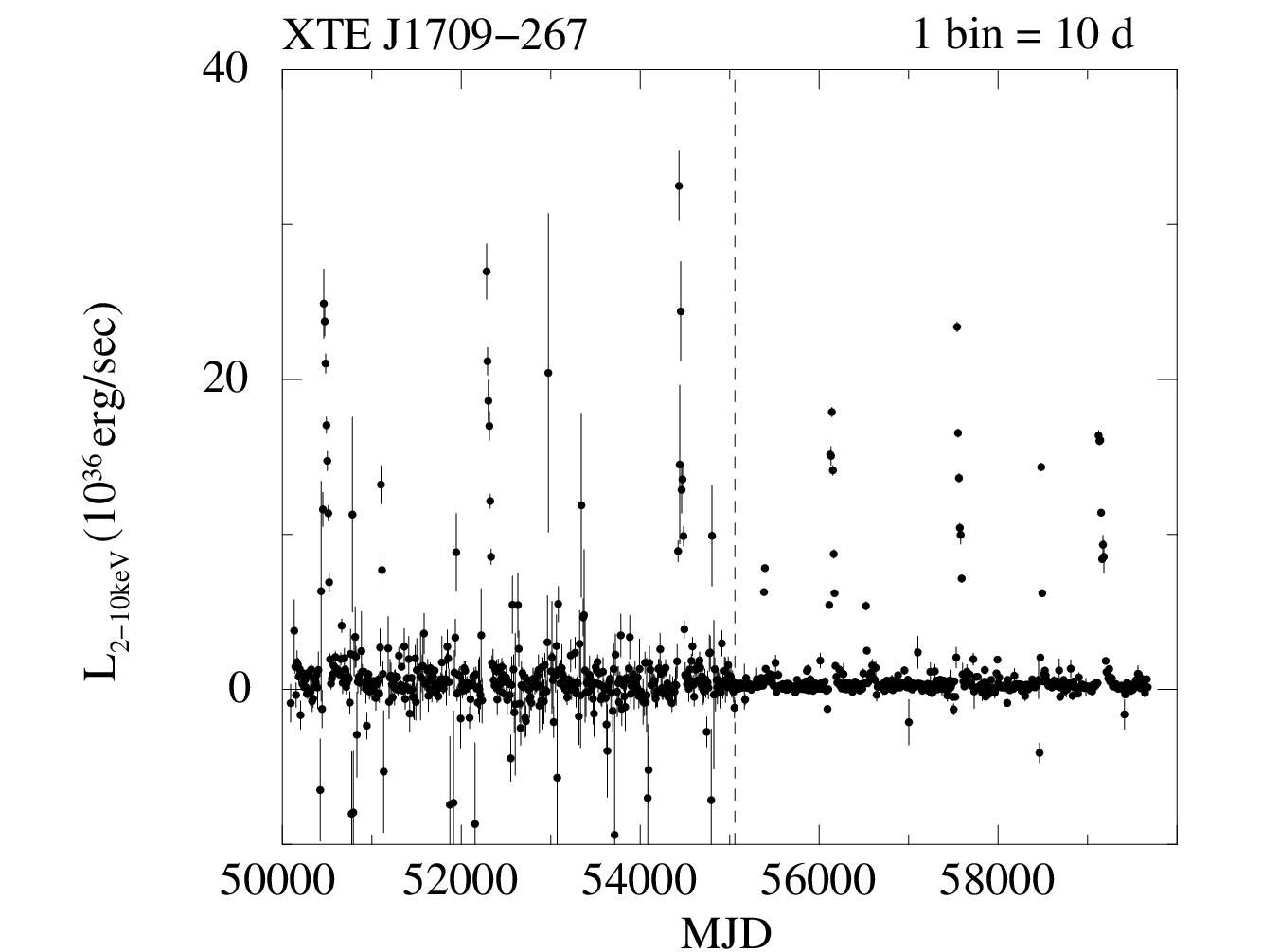}
 \includegraphics[width=75mm]{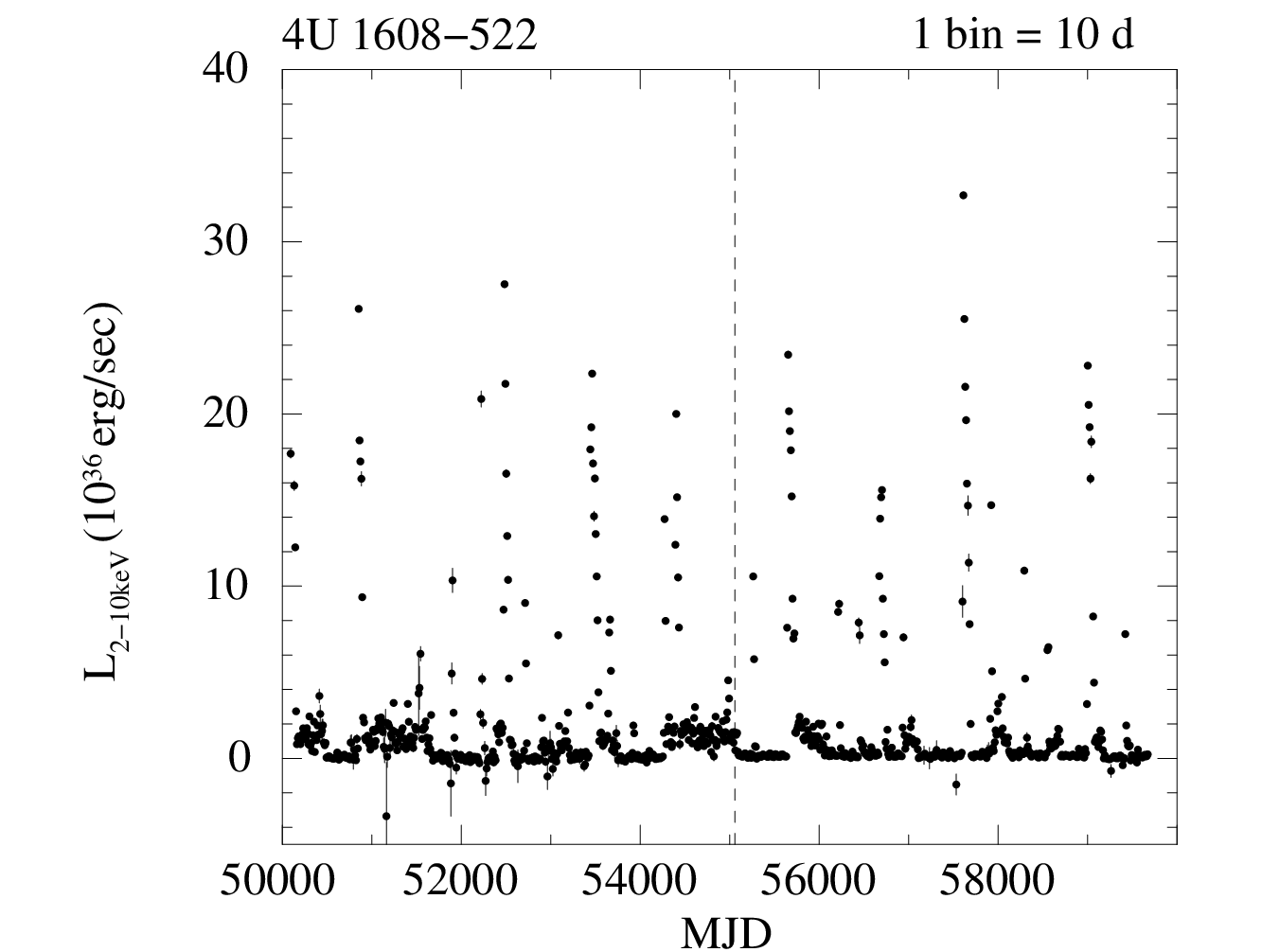}
  \includegraphics[width=75mm]{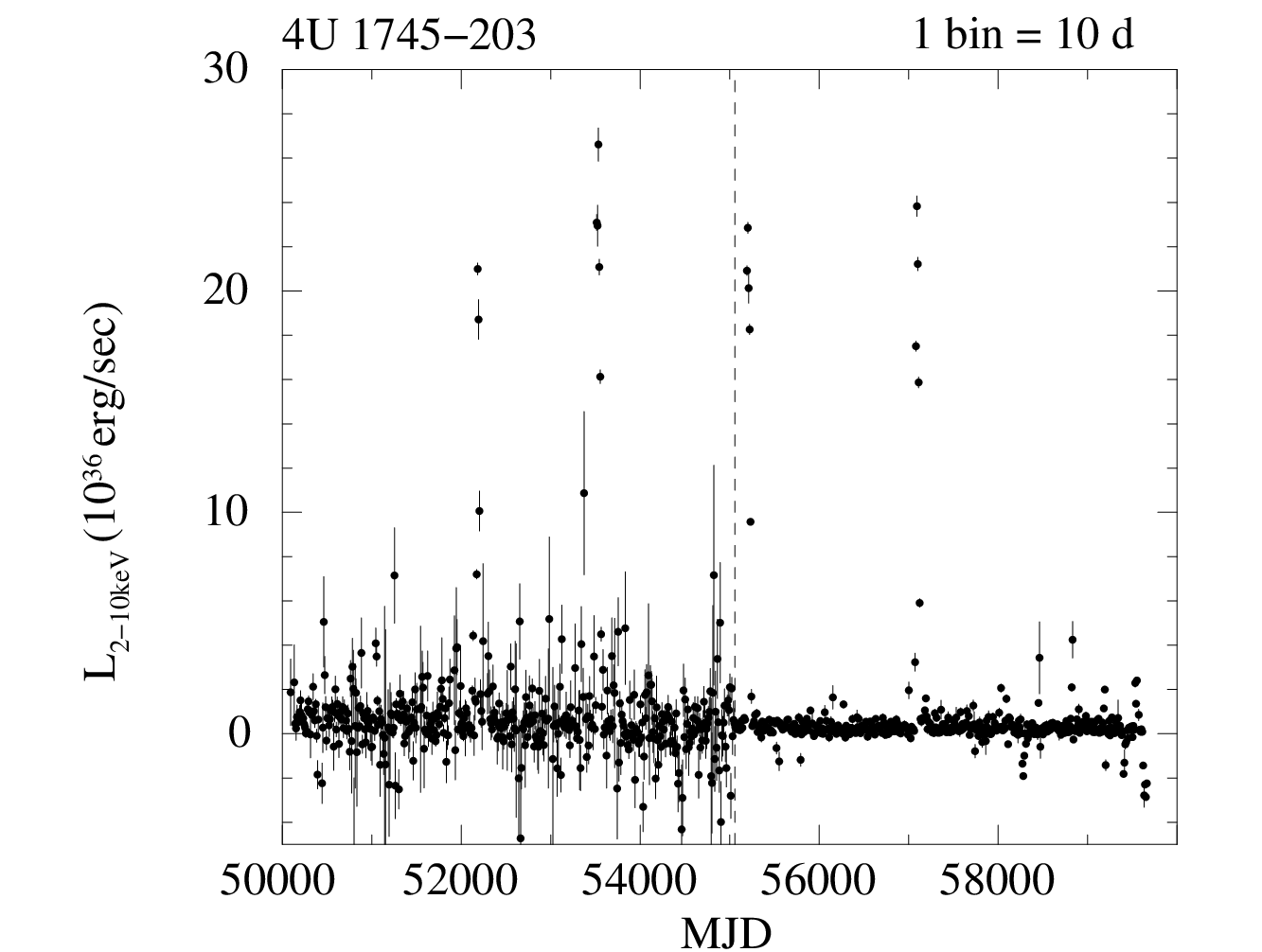} \includegraphics[width=75mm]{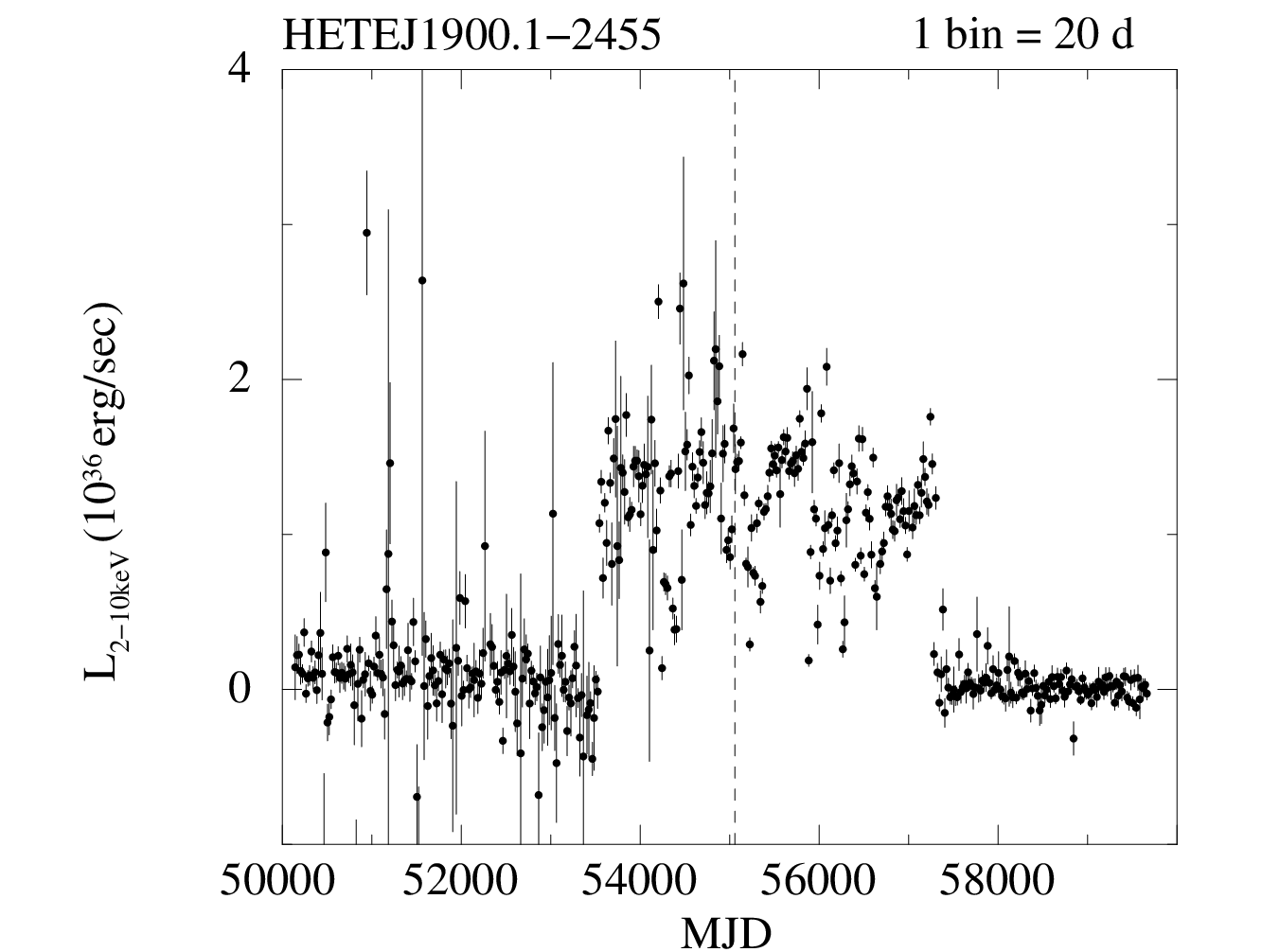}
\end{center}
\caption{As figure~\ref{LP}, but for Transient sources.}
\label{Tran}
\end{figure*}

\begin{figure*}
\begin{center}
 \includegraphics[width=75mm]{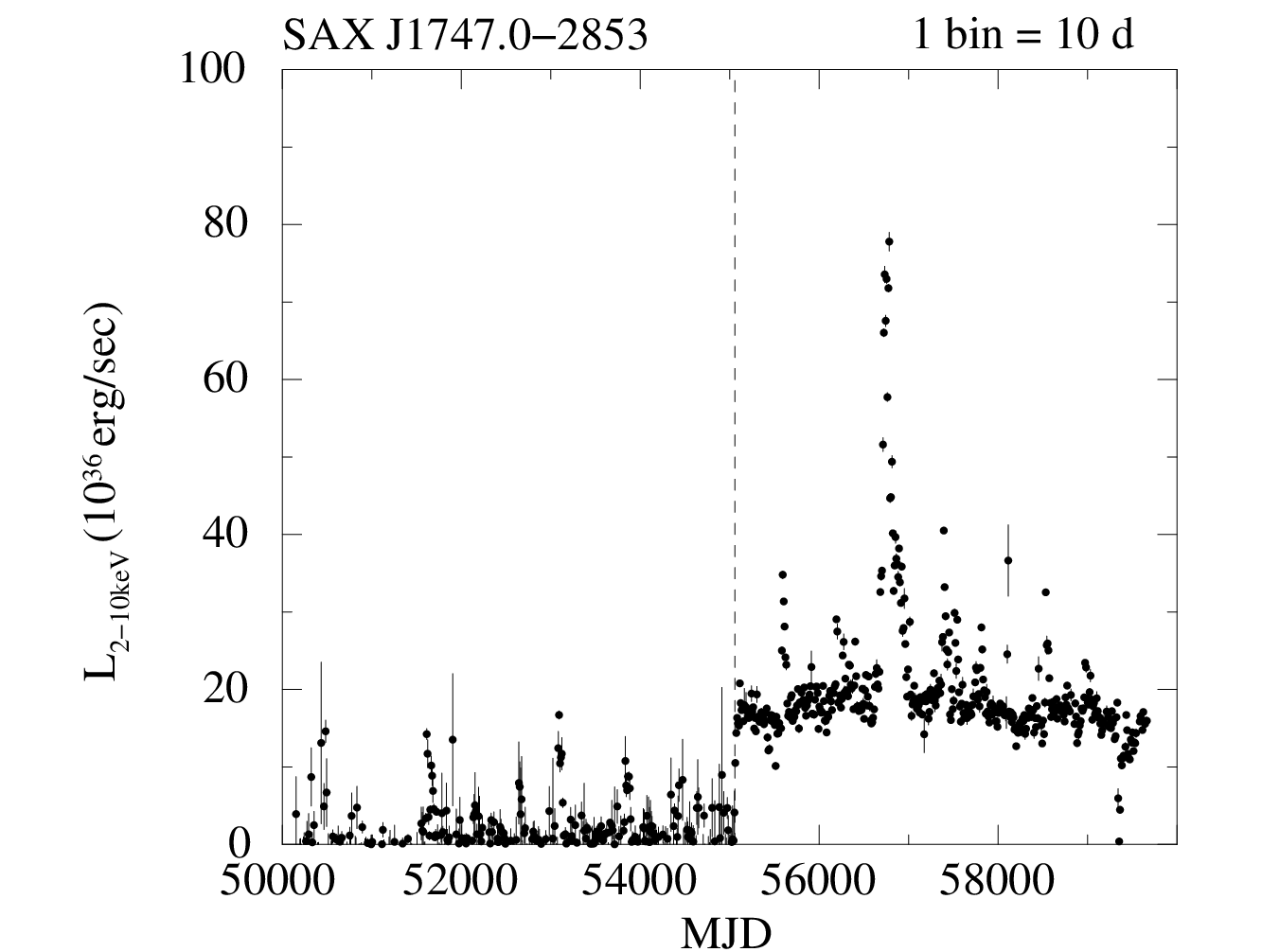}
 \includegraphics[width=75mm]{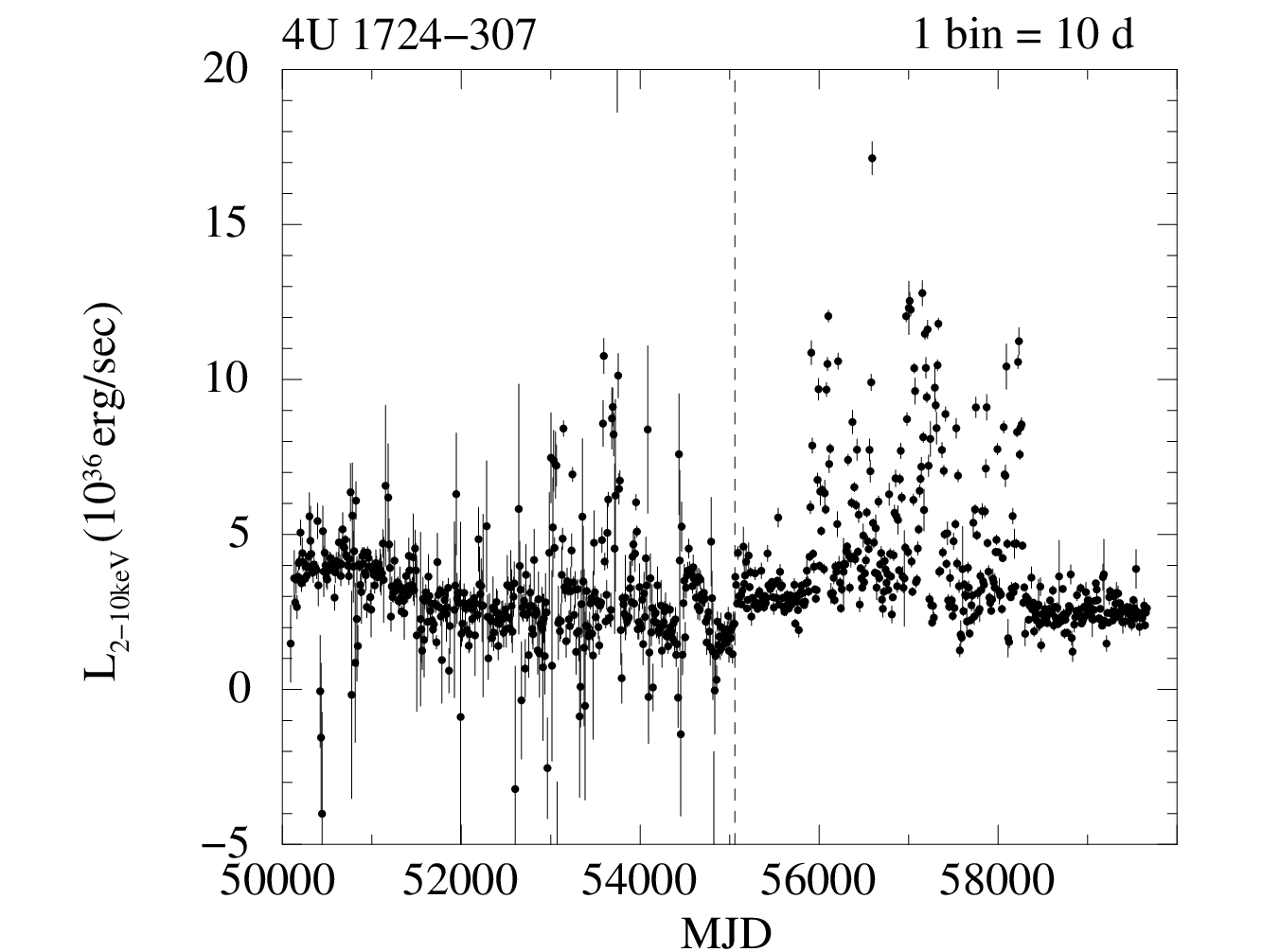}
 \end{center}
 \caption{As figure~\ref{LP}, but for the source with contamination.}
 \label{Con}
\end{figure*}

\clearpage
%%%

\end{document}